\documentclass[tightenlines,twocolumn,superscriptaddress,preprintnumbers,amssymb,nofootinbib]{revtex4-1}
\usepackage{amsmath, amsthm, amssymb, mathrsfs, bm, mathtools} 
\usepackage{graphics}
\usepackage{epsfig}
\usepackage{graphicx}
\usepackage{wrapfig}
\usepackage{color}
\usepackage{tikz}

\usepackage[hidelinks,colorlinks=true,linkcolor=blue,citecolor=blue]{hyperref}

\RequirePackage{subfigure}

\newcommand{\beq}{\begin{equation}}
\newcommand{\eeq}{\end{equation}}
\newcommand{\bqa}{\begin{eqnarray}}
\newcommand{\eqa}{\end{eqnarray}}

\newcommand{\ms}{\overline{\text{\tiny MS}}}

\newcommand{\cep}{\text{\tiny CEP}}
\newcommand{\tcp}{\text{\tiny TCP}}

\newcommand{\x}{x}
\newcommand{\y}{y}

\begin{document}

\title{Improved  results of chiral limit study with the large $N_c$ standard U(3) ChPT inputs in the on-shell renormalized  quark-meson model}
\author{Vivek Kumar Tiwari}
\email{vivekkrt@gmail.com}
\affiliation{Department of Physics, University of Allahabad, Prayagraj, India-211002}
\date{\today}

\begin{abstract}

~When the $f_{\pi}, f_{K} \text{ and } M_{\eta}^2$ given by the $ (m_{\pi},m_{K}) $ dependent scaling relations of the large $N_{c}$ standard U(3)  Chiral perturbation theory (ChPT) and the infrared regularized U(3) ChPT,~are used in the on-shell renormalized 2+1 flavor quark-meson (RQM) model to find its parameters in the path to the chiral limit $(m_{\pi},m_{K}) \rightarrow 0$ away from the physical point $ (m_{\pi}^{\text{\tiny{Phys}}},m_{K}^{\text{\tiny{Phys}}})=(138,496) $ MeV,~one gets the respective framework of RQM-S model and RQM-I model.~Computing and comprehensively comparing the RQM-S and I model Columbia plots for the $m_{\sigma}=400,500\text{ and }600$ MeV,~it has been shown that the use of large $N_c$ standard U(3) ChPT inputs give the better and improved framework for the Chiral limit studies as the RQM-S model tricritical lines show the expected saturation pattern after becoming flat in the vertical meson (quark) mass $\mu-m_{K}$ ($\mu-m_{s}$) plane for all the cases of $m_{\sigma}$ whereas the divergent RQM-I model tricritical line for the $m_{\sigma}=500$ MeV becomes strongly divergent when the $m_{\sigma}=600$ MeV.
~The  relative position $m_{K}^{\tcp}/m_{K}^{\text{\tiny{Phys}}} (m_{s}^{\tcp}/m_{s}^{\text{\tiny{Phys}}})$ of the RQM-S model tricritical point $\tcp$ on the $m_{K}(m_{s})$ axis for the $m_{\pi}(m_{ud})=0$,~shifts from the 0.5016 to 0.4405(0.2927 to 0.2280) whereas the critical pion (light quark) mass $m_{\pi}^{c}(m_{ud}^{c})$ changes from the 134.51(3.8) to 117.26(2.9) MeV when the $m_{\sigma}=400\rightarrow600$ MeV.~In contrast,~the shift of $m_{K}^{\tcp}/m_{K}^{\text{\tiny{Phys}}}(m_{s}^{\tcp}/m_{s}^{\text{\tiny{Phys}}})$  from the  0.4405 to 0.2988(0.2280 to 0.1068 ) and the change of $m_{\pi}^{c}(m_{ud}^{c})$ from the 117.26(2.9) to 82.07(1.43) MeV is significantly large when the $m_{\sigma}=600\rightarrow800$ MeV.~The tricritical line shrinks to very small regions for the high $m_{\sigma}=750$ MeV.~Even though the first order regions in the horizontal $m_{\pi}-m_{K}$ plane of the RQM-S model Columbia plot,~get much reduced for the $m_{\sigma}=750 $ MeV,~its spread is comparable to what has been found in the QM model Columbia plot drawn by the functional renormalization group methods for the $m_{\sigma}=530$ MeV.

\end{abstract}

\maketitle
\section{Introduction}
\label{secI}

The color charge neutral hadrons,~which are the units of Quantum Chromodynamics (QCD) at low energies,~get dissolved into color charge carrying quark gluon units for extremely high temperatures and densities and such a QCD phase transition~\cite{Cabibbo75,SveLer,Mull,Ortms,Riske} gives the color conducting medium of quark gluon plasma (QGP).~The QCD phase transition studies are very important for understanding the early universe evolution~\cite{EuniI,EuniII},~the structure of compact stars~\cite{BubaNs,FukuNs,TypeNs},~and the nature of the strong interaction in heavy-ion collision experiments~\cite{AgarExp,OdyExp,APandav,MishaConf}.~The chiral symmetry and color confinement are two important and deeply interlinked properties of QCD.~The phase transition is theoretically constructed by exploiting the approximate $ SU_{A}(3)$ chiral symmetry of the QCD Lagrangian which becomes exact for the massless light u, d and strange s quarks.~The exact $ SU_{A}(3)$ chiral symmetry gets spontaneously broken for the low temperatures and densities in the non-perturbative low energy QCD vacuum giving the eight massless Goldstone bosons which constitute the observed octet of light pions with a relatively heavier $K\text{ and } \eta$ mesons as pseudo-Goldstone modes in nature because the chiral symmetry is broken explicitly also by the very small (small) mass of light (strange) quark.~The first principle lattice QCD (LQCD) simulations  have settled that the chiral symmetry restoring phase transition at $\mu=0$ for the observed $\pi\text{ and }K$ meson masses corresponding to the real light and strange quark masses at the physical point,~is a crossover transition with pseudo-critical temperature $T_{\chi}\equiv 155\pm2$ MeV \cite{Wupertal2010, WB2010II, HotQCD2012, WB2014, HotQCD2014}.

~The chiral transition is first order in the massless chiral limit $m_{u}=m_{d}=m_{ud}=0= m_{s}$ according to the universality \cite{rob} arguments and it becomes second order of the $O(4)$ universality class if $m_{u,d}=0$, $m_{s}=\infty$ with the strong $U_A(1)$ axial anomaly \cite{tHooft:76prl} ($\eta^{\prime}$ mass $m_{\eta^{\prime}}(T_{c})>>T_{c}$) at the chiral symmetry restoring  critical temperature $T_{c}$.~The second chiral order transition becomes first order at a tricritical point $\tcp$ for some finite $m_{s}=m_{s}^{\tcp}$($m_{K}=m_{K}^{\tcp}$) in the light chiral limit $m_{u,d}=0$ ($m_{\pi}=0$).~The chiral crossover at $\mu=0$ turns first order for the critical quark (pion) mass $m_{q} \ (m_{\pi})<m_{q}^{c} (m_{\pi}^{c})$ for three flavors  $N_{f}=3$ of degenerate quarks.~The mass dependence of the order of chiral phase transitions and structures of the critical lines which separate the first from the second order and crossover chiral transition regions,~are presented in the $m_{\pi}-m_{K}$ or $m_{ud}-m_{s}$ planes of the Columbia plot \cite{columb} which is well understood in the heavy quark mass limit  both from the continuum and lattice QCD (LQCD) studies \cite{from, saito, reinosa, fischrA, mael}.~Since the LQCD results are prone to large variations on account of the strong cut-off and discretization effects,~the exact mapping of the critical lines for the small masses in the $m_{u,d}-m_{s}$ ($m_{\pi}-m_{K}$) plane,~becomes quite tough and challenging in the LQCD \cite{karsch1, karsch2, karsch3, forcrd, jin, jin2}.~Confirming the first order region close to the chiral limit,~different LQCD studies between 2001 to 2017 have found that the  $m_{\pi}^{c}$ lies in the range 290-67 MeV \cite{karsch2, karsch3, forcrd,jin, jin2,forcrnd2,Resch}.~On the other hand,~no evidence of first-order transition has been found for small pion mass as the LQCD study with improved Wilson fermions constrains $m_{\pi}^{c}\le100$ MeV ~\cite{Kura} while the study with  improved $N_{\tau}=6$ staggered fermions~\cite{Bazav} give $m_{\pi}^{c}\le50$ MeV whereas the LQCD study with highly improved $N_{\tau}=8$ staggered quarks~\cite{dini21} finds no first order region for $m_{\pi}^{c}\ge80$ MeV and another study~\cite{varnho} suggests that $m_{\pi}^{c}$ could even be zero.~Considering the flavor dependence of the position of  $\tcp$,~the Ref.~\cite{Cute} finds that the chiral transition is of second order in the chiral limit for $N_{f}=3$.~With the Möbius domain wall fermions,~the Ref.\cite {zhang24} finds small $m_{q}^{c}\le 4$ MeV.~Since the implementation of chiral fermions on the lattice is a notoriously difficult problem,~the order of chiral transition in the small mass regions of the Columbia plot,~remains an unsettled issue even if the improved LQCD studies are suggesting a small or no first order regions in some cases.

The Dyson-Schwinger approach $N_{f}=3$ study in Ref.~\cite{bernhardt23} and another study in Ref. \cite{kousvos22} apart from some LQCD studies,~find that the chiral transition is of second order in the small mass regions of  the Columbia plot.~Using the functional renormalization group (FRG) methods in the local potential approximation (LPA),~the very recent studies in Refs.~\cite{fejos22, fejoHastuda} have discussed their findings that instead of the $\epsilon$ expansion predictions of first order regions in the Ref.~\cite{rob},~the chiral transition can be of second order for $N_{f}\ge2$ if the $U_A(1)$ symmetry gets restored at the $T_{\chi}$.~However they have added a caveat note that the effect of wave function renormalization that gets completely neglected under the LPA,~might affect the final results.~Furthermore since some studies~\cite{dick15, ding21, kaczmarek21, bazazov12, bha14, kaczmarek23} find that the axial $U_A(1)$ anomaly remains finite and relevant at the critical point $T_{\chi}$ while others \cite{dini21, brandt16, tomiya16, aoki21, aoki22} claim that it vanishes,~the thermal fate of the $U_A(1)$ anomaly remains unsettled~\cite{lahiri21}.~The chiral limit behavior of the $U_A(1)$  symmetry restoring observables,~has been studied in Refs.~\cite{nico13,nicojhep,nico18I,nico18II,solnico}.~The recent study in the Ref.~\cite{pisarski24},~has conjectured some novel signals if anomaly strength becomes very weak at the $T_{\chi}$.~The Ref.~\cite{Giacosa} recently proposed a scenario where the decreasing first order regions completely vanish to give second order transition in the Columbia plot when the the sixth order coupling for the $U_A(1)$ anomaly term,~becomes relevant and strong.

The lattice QCD studies have a long history of getting support and insights 
from the effective theory model studies ~like the linear sigma model (LSM) \cite{Ortman, Lenagh, Rischke:00, Roder, jakobi, Herpay:05, Herpay:06, Kovacs:2007,Jakovac:2010uy, marko, Fejos},~the quark-meson (QM) model \cite{scav, mocsy, bj, Schaefer:2006ds, SchaPQM2F, Bowman:2008kc,  SchaeferPNP, Schaefer:09,  Schaefer:09wspax, SchaPQM3F, Mao, TiPQM3F,  koch, zacchi1, zacchi2} or the Nambu-Jona-Lasinio \cite{costaA, costaB, fuku08} (NJL) model  and the  non-perturbative FRG framework models~\cite{berges, bergesRep, Gies, braunii, pawlanal, fuku11, Herbst, grahl, mitter, Weise1, FejosI, Renke2, brauniii, FejosII,FbRenk, Weise2, Tripol, Fejos3, Weise3, Fejos4, fejos5, BrauPRL, BrauPRD}.~The 2+1 flavor QM model FRG method study under the LPA in the Ref.~\cite{Resch} finds that,~the critical pion mass $m_{\pi}^{c}\equiv 86$ MeV,~obtained using the extended mean field approximation (e-MFA) in the QM model where the thermal and vacuum quantum fluctuations of only the quark loops are considered,~becomes quite low $m_{\pi}^{c}\equiv 17$ MeV indicating much reduced first order regions in the Columbia plot when the thermal and vacuum quantum fluctuations of meson loops are also included in their study with the full FRG flow.~However the Ref.~\cite{pisarski24} has pointed out that the approximation used in the above work is known to overestimate the mesonic fluctuations~\cite{PawlRen} that tend to soften the chiral transition.~Including the nonperturbative mesonic loop corrections in the Hartree approximation of LSM using the Cornwall-Jackiw-Tomboulis (CJT)  two-particle (2PI) irreducible effective action formalism,~the beyond perturbation method studies~\cite{Rischke:00,Lenagh} reported a large first order region in the Columbia plot when the $U_A(1)$ anomaly is absent and the results show drastic $\sigma$ mass dependence in the presence of the $U_A(1)$ anomaly where the heavier $m_{\sigma}$ values give the first order regions which disappear for the realistic  $m_{\sigma}$.~Note that since it is not clear,~how the $\sigma$ meson of the Lagrangian is connected to the physical $\sigma/f_{0}(500)$ state~\cite{Pelaez},~one computes the $m_\sigma$ dependence of the results.~Using the symmetry improved CJT called the SICJT formalism~\cite{Sicjt} in the LSM,~the very recent study in the Ref.~\cite{Tomiya},~finds a stable first-order regime with a definite $\tcp$ which shows small variation on increasing the $m_{\sigma}$.~They clarified that the  enhanced first order regions found with the conventional CJT formalism,~are merely an artifact due to the lack of manifest invariance of Nambu Goldstone and low energy theorems at finite temperature.

The exact chiral effective potential of the renormalized QM (RQM) model was calculated recently after treating the quark one-loop vacuum fluctuations consistently in the Refs.~\cite{Adhiand1,Adhiand2,Adhiand3,asmuAnd,RaiTiw22,raiti23} for the two flavor and the Refs.~\cite{vkkr23,skrvkt24,vkt25I,vkt25II,Ander25,Gholami} for the three (2+1) flavor of the QM model.~The parameters of the RQM model are renormalized on-shell by matching the counter terms in the on-shell scheme with those in the modified minimal subtraction $\overline{\text{MS}}$ scheme when the mass parameter and the running couplings are put into the relation of the pole masses of the pseudoscalar $\pi, \  K, \  \eta, \  \eta'$ and the scalar  $\sigma$ mesons.~Since the condensate dependent part of the $U_{A}(1)$ anomaly term gets modified when the meson self energies due to quark loops are calculated using the pole masses of mesons,~the renormalized ’t Hooft coupling $c$ gets significantly stronger.~Further,~the renormalized explicit chiral symmetry breaking strength $h_{x} (h_{y})$ gets weaker  by a small (relatively large) amount.~The RQM model with the above novel features,~got developed recently to give very fruitful frameworks  of chiral limit studies by enriching the model with the chiral perturbation theory (ChPT) inputs~~\cite{vkt25I,vkt25II}.~Its parameter fixing away from the physical point  for the reduced $\pi \text{ and } K $ masses as the $m_{\pi},m_{K} \rightarrow  0$,~was refined by using the $ (m_{\pi},m_{K}) $ dependent scaling relations for the $f_{\pi}, f_{K} \text{ and } M_{\eta}^2$ given by the $ \mathcal{O}(\frac{1}{f^2})$  accurate results of the large $N_{c}$ standard U(3) chiral perturbation theory (ChPT)~\cite{gasser, LeutI,  KaisI, herrNPB, herrPLB, Escribano} in the RQM-S model framework~\cite{vkt25I} and the infrared regularized U(3) ChPT~\cite{Herpay:05, borasoyI, borasoyII, Beisert, Becher} in the RQM-I model framework~\cite{vkt25II}.~The enrichment of the RQM model with the ChPT inputs,~cured the problem of the loss of the spontaneous chiral symmetry breaking ( SCSB) for the mass range $m_{\sigma}=400-800$ MeV when the $m_{\pi},m_{K} \rightarrow  0$  in the commonly used method \cite{Ortman, Lenagh, Schaefer:09,  fuku08, berges, Herbst} of chiral limit studies in the QM model called the fixed-ultraviolet (UV) scheme \cite{Resch} where   the light (strange) explicit chiral symmetry breaking strengths  $ h_{x}(h_{y}) $ are changed while all other parameters are kept same as the ones at the physical point.~It is emphasized that RQM model frameworks do not have any ambiguity when parameters get fixed towards the chiral limit for the $m_{\sigma }=400\rightarrow800$ MeV  whereas the e-MFA:FRG QM model chiral limit study in the Ref.~\cite{Resch},~where the FRG flow allows for the scalar $\sigma$ masses only in the range $m_{\sigma} \in [400,600]$ MeV,~could avoid the loss of SCSB only by heuristically adjusting the initial effective action successively to larger scales $\Lambda^{\prime} > \Lambda $ at each step of reducing the strengths $ h_{x} \text{ and } h_{y} $  such that the $f_{\pi}$ does not change in their ChPT motivated fixed $f_{\pi}$ scheme.

The very recent work of the Ref.~\cite{vkt25I} showed that,~for the $m_{\sigma}=530$ MeV,~the chiral limit study results in the RQM-S model are approximately similar to those found in the RQM-I model  but the RQM-I model tricritical line,~that separates the first and second order  transitions  in the vertical $\mu-m_{K}$ plane of the Columbia plot at $m_{\pi}=0$,~looks divergent after the $m_{K}\ge500$ MeV  while the RQM-S model tricritical line became flat showing proper saturation that is consistent and desirable on the physical grounds because the 2+1 flavor tricritical line is expected to be connected to the tricritical point of the two flavor chiral limit~\cite{hjss} at higher $\mu$ and $m_{K}$.~The first motivation of the present work is to compute the Columbia plots  of the RQM-S model for the $m_{\sigma}=400,500,600$ MeV and the RQM-I model for the $m_{\sigma}=500\text{ and }600$ MeV.~Bringing out similarities and differences after a thorough and comprehensive comparison of the RQM-S and RQM-I model Columbia plots for the $m_{\sigma}=400,500\text{ and } 600$ MeV,~our first goal will let us  find if the large $N_c$ standard U(3) ChPT scaling relations for the ($m_{\pi},m_{K}$) dependence of the $f_{\pi}, f_{K} $ and $M_{\eta}^2$,~constitute the  improved and better prescription for the Chiral limit studies for all the $m_{\sigma}$.~Computing RQM-S model Columbia plots for the $m_{\sigma}=750\text{ and } 800$ MeV and  comparing them with those for the $m_{\sigma}=400\rightarrow 600$ MeV,~our second goal is to find the relative shifts of the $\tcp$  at $m_{K}^{\tcp}(m_{s}^{\tcp})$  and the  critical pion (quark) mass $m_{\pi}^{c}(m_{ud}^{c})$ for the shrinking first order regions when the $m_{\sigma}=600\rightarrow800$ MeV.~Furthermore the divergence in the shape of tricritical line,~that corresponds to significantly shrunk first order region in the $\mu-m_{K}$ ( $\mu-m_{s}$) plane,~will also be revealed when the $m_{\sigma}=750$ MeV is high.~Our third goal is to compare  the Columbia plots of the RQM-S model,~when the strength of chiral transition gets significantly diluted for the $m_{\sigma}=750\text{ and }800$ MeV,~with the Columbia plots of the e-MFA:FRG QM model study for the $m_{\sigma}=530$ MeV and the QM with fermionic vacuum term (QMVT) model study for  the $m_{\sigma}=400$ MeV where the strength of chiral transition gets strongly softened by the very large smoothing effects of quark one-loop vacuum fluctuations due to its treatment in the $\overline{\text{MS}}$ scheme of renormalization  and the use of curvature masses of mesons to fix the model parameters.

The paper is arranged as follows.~The 2+1 flavor RQM model is presented briefly in the  section~\ref{secII}.~
~The Chiral perturbation theory (ChPT) scaling of the $f_{\pi}, f_{K} $ and $M_{\eta}$ has been discussed in the section~\ref{subsec:Chpt} where the infrared regularized U(3) ChPT formulas are presented in the section~\ref{subsec:IRChpt}-1 whereas the large $N_{c}$ standard   U(3) ChPT formulas are described in the section~\ref{subsec:SChpt}.~Results and discussions are presented in the section~\ref{secIII}.~The temperature variations of the $\pi\text{ and }\sigma$ curvature masses in the RQM-S and RQM-I models for the $ m_{\sigma}=400,500,600,750 \text{ and }800$ MeV are presented in the section~\ref{secIIIA}.~The RQM-S and RQM-I model Columbia plots for the $m_{\sigma}=400,500,600\text{ MeV }$ are presented and compared with each other in the section~\ref{Colplotmeson} where the RQM-S model Columbia plot for the $m_{\sigma}=750,800\text{ MeV}$ are also compared.~The RQM-S and RQM-I model Columbia plots in the quark mass $ {m_{ud}-m_{s}}$ and $ {\mu-m_{s}}$ planes for the $m_{\sigma}=400,500,600\text{ MeV }$ are presented and compared with each other in the section~\ref{Colplotquark} where the RQM-S model Columbia plots for the $m_{\sigma}=750 \text{ and }800\text{ MeV}$ are also presented.~The whole work is summarized in  the section~\ref{secIV}. 

\section{The RQM Model }
\label{secII}
The RQM model effective potential as presented in the Ref.~\cite{vkt25I,vkt25II} is reproduced below in this section.~The model Lagrangian~\cite{Rischke:00,Schaefer:09,TiPQM3F} is written as :
\bqa
\label{lag}
{\cal L_{QM}}&=&\bar{\psi}[i\gamma^\mu D_\mu- g\; T_a\big( \sigma_a 
+ i\gamma_5 \pi_a\big) ] \psi+\cal{L(M)}\;. \\ 
\nonumber
\label{lagM}
\cal{L(M)}&=&\text{Tr} (\partial_\mu {\cal{M}}^{\dagger}\partial^\mu {\cal{M}}-m^{2}({\cal{M}}^{\dagger}{\cal{M}}))\\ \nonumber
&&-\lambda_1\left[\text{Tr}({\cal{M}}^{\dagger}{\cal{M}})\right]^2-\lambda_2\text{Tr}({\cal{M}}^{\dagger}{\cal{M}})^2\\ 
&&+c[\text{det}{\cal{M}}+\text{det}{\cal{M}}^\dagger]+\text{Tr}\left[H({\cal{M}}+{\cal{M}}^\dagger)\right]\;.
\eqa
The Yukawa coupling $g$ couples the flavor triplet quark fields $\psi$ (color $N_c$-plet  Dirac spinor) to the nine scalar(pseudo-scalar) meson $\xi$ fields $\sigma_a (\pi_a$) of $3\times3$ complex matrix ${\cal{M}}=T_{a} \xi_{a}=T_{a}(\sigma_{a}+i\pi_{a})$.~$T_{a}=\frac{\lambda_{a}}{2}$ where $\lambda_a$ ($a=0,1..8$) are Gell-Mann matrices with $\lambda_0=\sqrt{\frac{2}{3}}{\mathbb I}_{3\times3}$.~The field $\xi$ picks up the non-zero vacuum expectation value $\overline{\xi}$ in the $0$ and $8$ directions.~The condensates $\bar{\sigma_0}$ and $\bar{\sigma_8}$ break the $SU_L(3) \times SU_R(3)$ chiral symmetry spontaneously   while the external fields $H= T_{a} h_{a}$ with $h_0$, $h_8  \neq 0$ break it explicitly.~The change from the singlet octet $(0,8)$  to the light strange basis $(\x,\y)$ gives $\x ( h_{x})= \sqrt{\frac{2}{3}}\bar{\sigma}_0 (h_{0}) +\frac{1}{\sqrt{3}} \bar{\sigma}_8(h_{8})$ and $ \y (h_{y}) = \frac{1}{\sqrt{3}}\bar{\sigma}_0 (h_{0})-\sqrt{\frac{2}{3}}\bar{\sigma}_8 (h_{8})$.~Considering mesons at mean filed level  with the thermal and quantum fluctuations of the quarks/anti-quarks,~the grand potential \cite{Schaefer:09,TiPQM3F}, is the sum of the vacuum effective potential $ U(\x,\y) $ and the quark/anti-quark  contribution $\Omega_{q\bar{q}}$ at finite temperature $T$ and quark chemical potential $\mu_{f} (f=u,d,s)$.
\bqa
\label{Grandpxy}
&&\Omega_{\rm MF }(T,\mu)=U(\x,\y)+\Omega_{q\bar{q}} (T,\mu;\x,\y)\;. 
\eqa
\bqa
&&\Omega_{q\bar{q}}(T,\mu;\x,\y)=\Omega_{q\bar{q}}^{vac}+\Omega_{q\bar{q}}^{T,\mu}. \\
\nonumber
\label{eq:mesop}
&&U(\x,\y)=\frac{m^{2}}{2}\left(\x^{2} +
\y^{2}\right) -h_{x} \x -h_{y} \y
- \frac{c}{2 \sqrt{2}} \x^2 \y \\  
&&+ \frac{\lambda_{1}}{2} \x^{2} \y^{2}+
\frac{1}{8}\left(2 \lambda_{1} +
\lambda_{2}\right)\x^{4} 
+\frac{1}{8}\left(2 \lambda_{1} +
2\lambda_{2}\right) \y^{4}\;. \\ \nonumber \\
\label{vac1}
&&\Omega_{q\bar{q}}^{vac} =- 2 N_c\sum_f  \int \frac{d^3 p}{(2\pi)^3} \ E_q \  \theta( \Lambda_c^2 - \vec{p}^{2})\;.\\ \nonumber \\
\label{vac2}
&&\Omega_{q\bar{q}}^{T,\mu}=- 2 N_c \sum_{f=u,d,s} \int \frac{d^3 p}{(2\pi)^3} T \left[ \ln g_f^{+}+\ln g_f^{-}\right].\; \\ \nonumber \\
\label{GrandQM}
&&\Omega_{\rm QM }(T,\mu,x,y)=U(\x,\y)+\Omega_{q\bar{q}}^{T,\mu}\;. 
\eqa
The $ g^{\pm}_f = \left[1+e^{-E_{f}^{\pm}/T}\right] $ where $E_{f}^{\pm} =E_f \mp \mu_{f}$ and $E_f=\sqrt{p^2 + m{_f}{^2}}$ is the quark/anti-quark energy.~The light (strange) quark mass  $m_{u/d}=\frac{g\x}{2}$ ($m_{s}=\frac{g\y}{\sqrt{2}}$) and $\mu_{u}=\mu_{d}=\mu_{s}=\mu$.~The quark one-loop vacuum term with ultraviolet cut-off $\Lambda_c$ in Eq.~(\ref{vac1}) is dropped for the standard mean field approximation (s-MFA) of the quark meson (QM) model grand effective potential in Eq.~(\ref{GrandQM}).~Several studies have used the minimal subtraction scheme to regularize the  quark one-loop vacuum divergences after including the vacuum fluctuations in the extended mean field approximation (e-MFA) \cite{vac, lars, schafwag12,  guptiw, vkkr12, vkkt13, Rai} but the effective potential $\Omega_{vac} (\x,\y)=U(\x,\y)+\Omega_{q\bar{q}}^{vac}$
in their treatment,~turns inconsistent as they fix the model parameters using curvature masses of mesons which are obtained by taking the
double derivatives of the effective potential with respect
to the different fields at its minimum.~The above mentioned inconsistency  becomes apparent when one notes that the calculation of the curvature masses,~involves the evaluation of the meson self-energies at zero momentum because the effective potential is the generator of the n-point functions of the theory at vanishing external momenta \cite {Adhiand1, Adhiand2, Adhiand3, asmuAnd, RaiTiw22, raiti23,vkkr23,skrvkt24, laine, BubaCar, fix1}.~It has to be noted that the pole definition of the meson mass is the physical and gauge invariant \cite{Kobes, Rebhan} one.

~Here,~we will employ the consistent e-MFA RQM model effective potential   calculated in our very recent works \cite{vkkr23,skrvkt24}  after relating the counter-terms in the $\overline{\text{MS}}$ scheme to those in the on-shell (OS) scheme \cite{Adhiand1, Adhiand2, Adhiand3, asmuAnd, RaiTiw22, raiti23}.~The relations between the renormalized parameters of both the schemes are determined when the physical quantities ( the on-shell pole masses of the $m_{\pi}, m_{K}, m_{\eta},m_{\eta^{\prime}} \ \text{and} \ m_{\sigma}$,~the pion and kaon decay  constants $f_{\pi}$ and $f_{K}$ )  are put into the relation of the $\overline{\text{MS}}$ running  couplings and mass parameter.~These relations are used as input when the effective potential is calculated using the modified minimal subtraction procedure.~After the cancellation of the  $ 1 / \epsilon$ divergences,~the vacuum effective potential 
$ \Omega_{vac}=U(x_{\ms},y_{\ms})+\Omega^{q,vac}_{\ms}+\delta U(x_{\ms},y_{\ms})$ in the $\overline{\text{MS}}$ scheme has been rewritten in Ref. \cite{vkkr23} in terms of the scale $\Lambda$ independent constituent quark mass parameters $\Delta_{x}=\frac{g_{\ms} \ x_{\ms}}{2}$ and $\Delta_{y}=\frac{g_{\ms} \ y_{\ms}}{\sqrt{2}}$ as the following.
\begin{align}
\label{OmegDelxy}
\nonumber 
&\Omega_{\rm vac}(\Delta_{x},\Delta_{y})=\frac{m^2_0}{g^2_0}(2\Delta_{x}^2+\Delta_{y}^2)-2\frac{h_{x0}}{g_0}\Delta_{x}-\sqrt{2}\frac{h_{y0}}{g_0}\Delta_{y}\\ \nonumber 
&-2\frac{c_{0}}{g^3_{0}}\Delta_{x}^2 \ \Delta_{y}+4\frac{\lambda_{10}}{g^4_{0}}\Delta_{x}^2 \ \Delta_{y}^2+2\frac{(2 \lambda_{10}+\lambda_{20})}{g^4_{0}}  \Delta_{x}^{4} \ \\ \nonumber 
&+\frac{(\lambda_{10}+\lambda_{20})}{g^4_{0}}  \Delta_{y}^{4}+
\frac{2N_c\Delta_{x}^4}{(4\pi)^2}\left[\frac{3}{2}+\ln\left(\frac{\Lambda^2}{m_{u}^2}\right)+\ln\left(\frac{m_{u}^2}{\Delta_{x}^2}\right)\right] \\ 
&+\frac{N_c\Delta_{y}^4}{(4\pi)^2}\left[\frac{3}{2}+\ln\left(\frac{\Lambda^2}{m_{u}^2}\right)+\ln\left(\frac{m_{u}^2}{\Delta_{y}^2}\right)\right]\;.  
\end{align}
The condition that the minimum of the effective potential of the RQM model does not shift from that of the QM model fixes the scale $\Lambda_0$ \cite{vkkr23,skrvkt24} as :
\bqa
\label{Sclcond}
&\ln\left(\frac{\Lambda^2_0}{m_u^2}\right)+\mathcal{C}(m^2_\pi)+m^2_\pi \mathcal{C}^{\prime}(m^2_\pi)=0\;. 
\eqa
The terms $\mathcal{C}(m^2_\pi) $ and $ \mathcal{C}^{\prime}(m^2_\pi) $ and the derivations for the renormalized parameters  $m^{2}_{0}=(m^{2}+m^2_{\text{\tiny{FIN}}})$,~$h_{x0}=(h_{x}+h_{x\text{\tiny{FIN}}})$,~$h_{y0}=(h_{y}+h_{y\text{\tiny{FIN}}}) $,~$\lambda_{10}=(\lambda_{1}+\lambda_{1\text{\tiny{FIN}}})$,~$\lambda_{20}=(\lambda_{2} +\lambda_{2\text{\tiny{FIN}}})$ and $c_{0}=(c+c_{\text{\tiny{FINTOT}}})$ are given in detail in Refs. \cite{vkkr23,skrvkt24}.~The $m^2_{\text{\tiny{FIN}}}$,
$ h_{x\text{\tiny{FIN}}} $, $ h_{y\text{\tiny{FIN}}} $, $\lambda_{1\text{\tiny{FIN}}} $, $ \lambda_{2\text{\tiny{FIN}}}$ and 
$c_{\text{\tiny{FINTOT}}}$ are the finite on-shell corrections in the parameters at the scale $\Lambda_0$.~The experimental values of the pseudo-scalar meson masses $m_{\pi}$, $m_{K}$, $m_{\eta}$, $m_{\eta^{\prime}} $ ($m_{\eta}^2+m_{\eta^{\prime}}^2$), the scalar $\sigma$ mass $m_{\sigma}$ and the $f_{\pi}$, $f_K$ as input determine  the tree level QM model quartic couplings  $\lambda_1$, $\lambda_2$,~mass parameter  $m^2$,~$h_x$, $h_y$ and the coefficient $c$ of the t'Hooft determinant term for the $U_A(1)$ axial anomaly \cite{Rischke:00,Schaefer:09}.

~Although the $f_{\pi}$, $f_{K}$ and $g$ get renormalized due to the dressing of the meson propagator in the on-shell scheme,~they do not change as the $g_{\ms}=g_{ren}=g_{0}=g$, $x_{\ms}=x$, $y_{\ms}=y$ at $\Lambda_0$.~Further  $x_{\ms}=f_{\pi,ren}=f_\pi$ and $y_{\ms}=\frac{2f_{K,ren}-f_{\pi,ren}}{\sqrt{2}}$= $\frac{2f_K-f_\pi}{\sqrt{2}}$ at the minimum.~Using $\Delta_{x}=\frac{g \ x}{2}$ and $\Delta_{y}=\frac{g \ y}{\sqrt{2}}$,~the vacuum effective potential in the Eq.~(\ref{OmegDelxy}) can be  written in terms of the $x$ and $y$ as :
\begin{align}
\label{vacRQM}
\nonumber
&\Omega_{vac}^{\rm RQM}(\x,\y)=\frac{(m^{2}+m^2_{\text{\tiny{FIN}}})}{2} \ \left(\x^{2} +
\y^{2}\right)-(h_{x}+h_{x\text{\tiny{FIN}}}) \ \x 
\\ \nonumber 
&-(h_{y}+h_{y\text{\tiny{FIN}}}) \y -\frac{(c+c_{\text{\tiny{FINTOT}}})}{2 \sqrt{2}} \x^2 \y + \frac{(\lambda_{1}+\lambda_{1\text{\tiny{FIN}}})}{2} \x^{2} \y^{2}
\\ \nonumber
&+\frac{\left\{2(\lambda_{1} +\lambda_{1\text{\tiny{FIN}}})+
( \lambda_{2} +\lambda_{2\text{\tiny{FIN}}})\right\}\x^{4}}{8}+\left( \lambda_{1} +\lambda_{1\text{\tiny{FIN}}}+\lambda_{2}
 \right.  \\  \nonumber
& \left. +\lambda_{2\text{\tiny{FIN}}}\right)\frac{ \y^{4}}{4} +\frac{N_c g^4 (\x^4+2\y^4)}{8(4\pi)^2} \left[\frac{3}{2}-\mathcal{C}(m^2_\pi)-m^2_\pi \mathcal{C}^{\prime}(m^2_\pi)\right]\ \\
&-\frac{N_c g^4 }{8(4\pi)^2} \left[\x^4 \ln\left(\frac{\x^2}{f_{\pi}^2}\right)+2\y^4 \ln\left(\frac{2 \  \y^2}{f_{\pi}^2}\right) \right]. \\ \nonumber  \\ 
\label{grandRQM}
&\Omega_{\rm RQM }(T,\mu,x,y)=\Omega_{vac}^{\rm RQM}(\x,\y)
+\Omega_{q\bar{q}}^{T,\mu}\;.
\end{align}
The search of the grand potential minima  $\frac{\partial \Omega_{\rm RQM}}{\partial \x}= \frac{\partial \Omega_{\rm RQM}}{\partial \y}=0$ for the  Eq.~(\ref{grandRQM}) gives the  $T$ and  $\mu$ dependence of the $ \x$ and $ \y$.~The curvature masses of mesons are different from their pole masses in the RQM model due to the on-shell renormalization of the parameters \cite{BubaCar,fix1}.~The derivation of the RQM model  curvature mass expressions for the mesons are presented in the section (IIA) of the Ref.\citep{vkt25I}.~When one uses the Eq.~(\ref{vacRQM}) to evaluate the equations of motion $\frac{\partial \Omega_{vac}^{\rm RQM}}{\partial \x}=0= \frac{\partial \Omega_{vac}^{\rm RQM}}{\partial \y}$,~one gets the renormalized explicit chiral symmetry breaking strengths $h_{x0}$ and $h_{y0}$ as the following.

\bqa
\label{hx0}
&h_{x0}=m_{\pi,c}^2 \ f_{\pi}. \\
\label{hy0}
&h_{y0}=\biggl(\sqrt{2} \ f_K \ m^2_{K,c}-\frac{f_{\pi}}{\sqrt{2}} \ m^2_{\pi,c}\biggr).
\eqa 
The pion and kaon curvature masses $m_{\pi,c}$ and $m_{K,c}$ as derived in the Ref. \cite{vkkr23} have the following expressions.
\begin{align}
\label{mpicr}
&m_{\pi,c}^2=m_{\pi}^2\biggl\{ 1-\frac{N_{c}g^2}{4\pi^2} \ m_{\pi}^2 \ \mathcal{C}^{\prime}(m^2_\pi,m_u)  \biggr\}. \\ \nonumber
\label{kcr}
&m_{K,c}^2= m_{K}^2 \biggl[ 1-\frac{N_{c}g^2}{4\pi^2} \biggl\{  \mathcal{C}(m^2_\pi,m_u)+m^2_\pi \ \mathcal{C}^{\prime}(m^2_\pi,m_u)  \\ \nonumber
&\hspace*{.8cm}-\biggl(1-\frac{(m_s-m_u)^2}{m_K^2}\biggr)\ \mathcal{C}(m^2_K,m_u,m_s)+ \\  &\hspace*{.8cm}\biggl(1-\frac{f_\pi}{f_K}\biggr) \ \biggl(\frac{m^2_u-m_s^2+2m_s^2\ln(\frac{m_s}{m_u})}{m^2_K}\biggr)\biggr\}\biggr]. 
\end{align}

In order to see the consistency check of the above expressions,~one can  verify that when the $x=f_{\pi}$ and $y=\frac{(2f_{K}-f_{\pi})}{\sqrt{2}}$ in the vacuum,~the expression of $m_{\rm p,11\rm c}^{2}$ in the Eq.~(18) of Ref.~\citep{vkt25I} gives the same value as the $m_{\pi,\rm c}^2$ in the Eq.~(\ref{mpicr}) while the expression of $m_{\rm p,44 \rm c}^{2}$ in the Eq.~(19) of Ref.~\citep{vkt25I} gives the value that is equal to the kaon curvature mass square $m_{K,\rm c}^2$ given by the Eq.~(\ref{kcr}).~The pion curvature mass  $m_{\pi;c}= 135.95$ MeV in vacuum turns out to be 2.05 MeV smaller than its pole mass  $m_{\pi}= 138$ MeV whereas the vacuum kaon curvature mass $m_{K;c}= 467.99$ MeV is 28.01 MeV smaller than its pole mass of $m_{K}= 496$ MeV \citep{vkkr23,skrvkt24}.

\subsection{The ChPT scaling of  $\bf f_{\pi},f_{K} \ \text{and} \ M_{\eta}^2=m_{\eta}^2+m_{\eta^{\prime}}^2$}
\label{subsec:Chpt}

In order to investigate the current quark mass ($m_{u/d}-m_{s}$ plane) sensitivity of the chiral phase transition in the lower left corner of the Columbia plot,~one resorts to the equivalent description of the pion kaon mass $m_{\pi}-m_{K}$ plane because the low energy effective theory framework of the $2+1$ flavor QM model deals with scalar and pseudo-scalar mesons rather than the $u,d$ and $s$ quarks of the perturbative QCD regime.~The  explicit chiral symmetry breaking source strengths in the non-strange and strange directions  $h_{x}=f_{\pi} m^2_{\pi}$,~$h_{y}=\sqrt{2} f_{K} m^2_{K}-f_{\pi} m^2_{\pi}/\sqrt{2}$ as well as the effective theory model parameters can be determined only at the physical point because here the experimental information is in tune with the underlying theory of QCD.~Since it  is  not clear,~a priori,~how the parameters change as the system is tuned away from the physical point,~different strategies are adopted to fix the model parameters.~The fixed-ultraviolet (UV) scheme \cite{Resch} is the most commonly used strategy \cite{Ortman, Lenagh, Schaefer:09, fuku08, berges,Herbst} in the literature.~It relies on the assumption that the change in current quark masses of QCD can be directly mapped onto a change of symmetry breaking source strengths.~Hence one varies the $h_{x}(h_{y})$ while keeps all other parameters same as the ones at the physical point such that the initial effective action does not change in the UV \cite{Resch}.~Since the spontaneous chiral symmetry breaking (SCSB) gets lost in the chiral limit as the mass parameter $m^2$ becomes positive for the moderate $m_{\sigma}$,~the fixed-UV scheme does not work for the scalar $\sigma$ mass range $m_{\sigma}=400-800$ MeV~\cite{Schaefer:09,Resch}.~This problem for the chiral limit study in the QM model got circumvented in the Ref.~\cite{Schaefer:09} by choosing quite a large $m_{\sigma}\ge800$ MeV.

Resch et. al. computed Columbia plots for the QM model in the Ref.~\cite{Resch} using nonperturbative FRG methods under the LPA.~Motivated by the findings of ChPT,~they proposed a fixed-$f_{\pi}$ scheme  where the initial effective action is to be heuristically adjusted to the larger scales ($\Lambda^{\prime} > \Lambda $) at each step of the calculation when smaller masses are taken in the path to the chiral limit in the Columbia plot such that the pion decay constant  $f_{\pi}$ in the infra-red (IR) always remains fixed to its physical value.~In the fixed-$f_{\pi}$ scheme,~the changing scale ($\Lambda^{\prime} > \Lambda $) accounts for the change in parameters (which retain the same value as at the $\Lambda $ in the fixed UV-scheme) when the chiral symmetry breaking strengths $h_{x} \text{ and } h_{y}$  are decreased away from the physical point.~The problem of fixed-UV scheme is resolved by construction in the fixed-$f_{\pi}$ scheme because the condensates being intimately related to the fixed $f_{\pi}$ do not drop to zero and hence the SCSB does not get lost.~The fixed $f_{\pi}$ scheme suffers from the problem of a heuristic determination of the initial effective action \cite{Resch},~even though it is physically reasonable.~More accurate methods of fixing the model parameters unambiguously will be employed below where one exploits the ChPT predicted scaling relations for the decay  constants $f_{\pi}, f_{K} $ and $M_{\eta}^2 = m_{\eta}^2 + m_{\eta^{\prime}}^2$ when the $m_{\pi}  \ \text{and} \ m_{K} \rightarrow 0$ and therefore the explicit chiral symmetry breaking strengths $h_{x}$ and $h_{y}$ approach zero in the RQM model as one moves away from the physical point towards the chiral limit in the Columbia plot.


The chiral symmetry of the QCD Lagrangian gets enhanced to the $ U_{L}(3) \times U_{R}(3) $ when the quark loops responsible for the $U_{A}(1)$ anomaly get suppressed by taking the large $N_{c}$ limit.~Using the nonet of the Goldstone bosons where the $\eta'$ meson becomes the ninth addition to the $SU_{A}(3)$ octet of ($\pi,K.\eta$),~one can construct the effective Lagrangian with $U(3)$ chiral symmetry.~The systematic expansion of the Green functions of the U(3) ChPT in powers momenta,~quark masses and the $ \mathcal{O}(p^2)$ parameter $\frac{1}{N_{c}}$,~was introduced in the Ref. \citep{gasser} and it got firmly established in Refs.\cite{LeutI, KaisI, herrNPB, herrPLB, Escribano}.~The studies in Refs.~\cite{borasoyI, borasoyII, Beisert} have argued that the large $N_{c}$ arguments are not necessary for constructing  the effective Lagrangian with the singlet field as the additional $\frac{1}{N_{c}}$ counting scheme is imposed only to ensure that loops with an $\eta'$ are suppressed by powers of $\frac{1}{N_{c}}$.~They have pointed out that the mass of the $\eta'$ particle that introduces an additional low energy scale of about 1  GeV,~ is proportional to  $\frac{1}{N_{c}}$.~Hence the $\eta'$ can be treated perturbatively,~for its systematic inclusion in the  ChPT without using the $\frac{1}{N_{c}}$ counting rules.~This alternative method of treating massive fields got firmly established in the Ref. \cite{Becher} where the loop integrals are evaluated using a modified regularization scheme,~the so-called infrared regularization,~in which  Lorentz and chiral invariance are kept at all stages \cite{borasoyI, borasoyII}.~Discussing the inclusion of loops and renormalization issues,~the authors in the Ref. \cite{borasoyII},~have argued that the $\eta'$ loops do not contribute (at the fourth order Lagrangian) in infrared regularization and the $\eta'$ can be treated as a background field while the contribution of Goldstone boson ($\pi,K,\eta$) loops  get evaluated.~The above discussed framework is different from the large $N_c$ standard U(3) ChPT approach taken in the Refs. \citep{herrNPB, herrPLB, Escribano} where the authors have treated the $\eta'$ mesons on the same footing as the original Golstone Bosons and included the $\eta'$ in loops.

~The quark mass and condensate dependent expressions of the decay constants $f_{\pi}, f_{K} $ and the masses $m_{\pi}, \ m_{K}$ and  $M_{\eta}^2$ given by the infrared regularized U(3) ChPT \cite{borasoyI, borasoyII, Beisert} approach are described below for performing the chiral limit study in the RQM-I model.~The $m_{\pi},\ m_{K}$ dependent scaling relations for the  $f_{\pi}, \ f_{K} \text{ and } M_{\eta}^2 $ given by the large $N_c$ standard $U(3)$ ChPT \cite{herrNPB, herrPLB, Escribano} for performing the chiral limit study in the RQM-S model are presented in the subsection~\ref{subsec:SChpt}.~The RQM-I and RQM-S model results of the chiral limit studies and Columbia plots  have been compared exhaustively by varying the mass of the scalar $\sigma$ meson as $m_{\sigma}=400,\ 500\text{ and }600$ MeV.~The Columbia plots have been drawn both in the pion-kaon mass plane and the light-strange quark mass plane.~The  $m_{\pi}-m_{K}$ and $m_{ud}-m_{s}$ planes of the RQM-S model Columbia plots have been drawn also when the $\sigma$ mass becomes very high as $m_{\sigma}=750\text{ and }800$ MeV.~The RQM-S model results of chiral limit studies have been compared with other results available in the current literature.

\subsection*{1 : $ \bf M_{\eta}^2$ from infrared regularized U(3) ChPT inputs}
\label{subsec:IRChpt}

The infrared regularized U(3) ChPT prescription of treating the $\eta'$ meson,~gives a particular expression of the $M_{\eta}^2$ that constitutes the main ingredient of this ChPT based parameter fixing procedure.~It was  proposed originally in the Ref.\cite{Herpay:05} for the linear sigma model studied in the optimized perturbation theory framework.~In this approach,~it is sufficient to use the $ SU_{L}(3) \times SU_{R}(3) $ ChPT framework for knowing the functional forms $f_{\pi} (m_\pi, m_K ) \   \text{and}\ f_{K}(m_\pi, m_K )$.
~The One-loop ChPT calculations in the Ref.~\cite{gasser} give the following expressions of $m_\pi^2, m_K^2, f_{\pi}, f_{K}$ with $ \mathcal{O}(\frac{1}{f^2})$ accuracy in terms of the eight parameters $f, A, q, M_{0}, L_{4}, L_{5}, L_{6}, L_{8}$.
\bqa
\label{mpi2}
\nonumber
m_{\pi}^2&=&2A\left[1+\frac{1}{f^2}\biggl\{ \mu_{\pi}-\frac{\mu_{\eta}}{3} +16A(2 L_{8}-L_{5}) \right. \\ 
&&\left.+16A(2+q)(2L_{6}-L_{4})\biggr\}\right]\;. \\ \nonumber
\label{mk2}
m_{K}^2&=&A(1+q)\left[1+\frac{1}{f^2}\biggl\{\frac{2}{3} \mu_{\eta} +8A(1+q)(2 L_{8}-L_{5}) \right. \\ 
&&\left.+16A(2+q)(2L_{6}-L_{4})\biggr\}\right]\;. 
\eqa
\bqa
\label{fpiol}
f_{\pi}&=&f+\frac{1}{f} \left[-2\mu_{\pi}-\mu_{K}+8AL_{5}+8A(2+q) L_{4}\right]. \\ \nonumber
\label{fkol}
f_{K}&=&f \left[1+\frac{1}{f^2}\biggl\{-\frac{3}{4}(\mu_{\pi}+\mu_{\eta}+2\mu_{K}) \right. \\  
&&\left.\ \ \ +8A(2+q) L_{4}  +4A(1+q) L_{5} \biggr\} \right]\;. 
\eqa
The chiral logarithms $\mu_{\text{\tiny{PS}}}=\frac{m_{\text{\tiny{PS}}}^2}{32 \pi^2} \ln(\frac{m_{\text{\tiny{PS}}}^2}{M_{0}^2})$ are evaluated  at the scale $M_{0}$ in the above expressions   where the leading order squared mass of the corresponding meson in the pseudo-scalar octet,~is substituted for the $ m_{\text{\tiny{PS}}}^2$.~One should  note that the chiral constants $L_i$ do not vary with the pseudo-scalar masses.~The quark mass and quark condensate dependence of the above one-loop ChPT expressions of the $m_\pi^2, m_K^2, f_{\pi} \ \text{and} \ f_{K}$ are governed by the parameters  $q=2 m_s /(m_u+m_d)=m_s/m_{ud}$ and $A=B \ (m_u+m_d)/2=Bm_{ud}$ where the chiral limit value of the quark condensate $<\bar{u} u> $ determines the $B$.~Inverting the Eqs.~(\ref{mpi2}) and (\ref{mk2}) to order $\mathcal{O}(\frac{1}{f^2})$ accuracy,~the $q$ and $A$ can be expressed below in terms of the masses $m_\pi, m_K, m_{\eta}$  and chiral constants $L_i$.
\bqa
\label{Aqpim}
\nonumber
A&=&\frac{m_{\pi}^2}{2}\left[1-\frac{1}{f^2}\biggl\{ \mu_{\pi}-\frac{\mu_{\eta}}{3} +8m_{\pi}^2(2 L_{8}-L_{5}) \right. \\ 
&&\left.+8(2m_{K}^2+m_{\pi}^2)(2L_{6}-L_{4})\biggr\}\right]\;. \\ \nonumber
\label{qmr}
q+1&=&\frac{2m_{K}^2}{m_{\pi}^2}\left[1-\frac{1}{f^2}\biggl\{\mu_{\eta} -\mu_{\pi} \right. \\ 
&&\left.+8(m_K^2-m_{\pi}^2)(2 L_{8}-L_{5})\biggr\}\right]\;.
\eqa
The leading order relations of the above two equations are sufficient to find the following $\mathcal{O}(\frac{1}{f^2})$ accurate $m_{\pi},  m_{K}$-dependence of the $f_{\pi}, f_{K}$ from Eqs.~(\ref{fpiol}) and (\ref{fkol}). 
\bqa
\label{fpi}
f_{\pi}&=&f-\frac{1}{f} \biggl[2\mu_{\pi}+\mu_{K}-4m_{\pi}^2(L_{4}+L_{5})-8m_{K}^2 L_{4}\biggr].  \hspace {.6 cm} \\  \nonumber
\label{fk}
f_{K}&=&f \left[1-\frac{1}{f^2}\biggl\{\frac{3}{4}(\mu_{\pi}+\mu_{\eta}+2\mu_{K})-4m_{\pi}^2 L_{4} \right. \\  
&&\left.\ \ \  -4m_{K}^2(2L_{4}+L_{5}) \biggr\} \right]\;.  
\eqa
 With the input $f_\pi$=93 MeV, $f_K$=113 MeV,  $m_\pi$=138 MeV, $m_K$=495.6 MeV,$m_\eta$=547.8 MeV and $M_{0}=4 \pi f_{\pi}\equiv 1168$ MeV, f=88 MeV, one gets the constants $L_4$ and $L_5$ at the physical point as :
\bqa
L_4=-0.7033 \times 10^{-3} \ ; \ \ \ \  L_5=0.3708 \times 10^{-3}.
\eqa 
The values of $A$ and $q$ that one takes at the physical point, control the constants $L_6$ and $L_8$.~The strange to average non-strange quark mass ratio has the value $q=24.9$ similar to the Ref.\cite{Herpay:05} as it is close to the lattice determination and compatible with the range 
$ 20 \le q \le 34$ indicated by the PDG listing \cite{Eidel}.~Choosing leading order ChPT value of $A$ in the physical point $A=A^{(0)}$ and then using the phenomenological values of $m_{\pi}^2, m_{K}^2$ with the Gell-Mann-Okubo formula for $m_{\eta}^2$ in the $\mathcal{O}(\frac{1}{f^2})$ accurate expressions of $A$ and $q$,~one gets :
\bqa
L_6=-0.3915 \times 10^{-3} \ ; \ \ \ \  L_8=0.511 \times 10^{-3}.
\eqa
The values of $M_{0}$,~$f$ and $L_i$  can be used further for the continuation of $A$ and $q$ from the physical point to an arbitrary point in the $m_{\pi}-m_K$ plane.

One should note that the physical $\eta$ and $\eta^{\prime}$ states are obtained from  the mixtures of the states $\eta_{0}$ and $\eta_{8}$.~Apart from the values of $m_{\pi}^2$ and $m_{K}^2$,~one needs the numerical input for the $m_{\eta,00}^2+m_{\eta,88}^2$ also in order to fix the quartic coupling $\lambda_{2}$ in the LSM/QM model \cite{Lenagh,Schaefer:09,Mao,TiPQM3F}.~After finding the value of $\lambda_{2}$,~the 't Hooft cubic coupling $c$ gets easily fixed.~The experimental masses $m_{\eta}$ and $m_{\eta^{\prime}}$ are known only at the physical point.~Since the numerical value of only the total sum $m_{\eta}^2+m_{\eta^{\prime}}^2=m_{\eta,00}^2+m_{\eta,88}^2=M_{\eta}^2$ is required for finding the $\lambda_{2}$,~even if one does not know the individual values of $m_{\eta}^2 \ (m_{\pi},m_{K}) $ and $m_{\eta^{\prime}}^2 \ (m_{\pi},m_{K}) $,~the knowledge of the sum $M_{\eta}^2  \  (m_{\pi},m_{K})$ as input is necessary and sufficient for fixing the $\lambda_{2}$ when one moves towards the chiral limit away from the physical point.

In order to find the functional form of the $M_{\eta}^2  \  (m_{\pi},m_{K})$,~one needs the application of the infrared regularized $ U_{L}(3) \times U_{R}(3) $ ChPT that proceeds through similar  steps as in the  previous description but the mixing in the masses of the $\eta_{0}, \eta_{8}$  makes it a little involved.~The Ref. \cite{Herpay:05} has given the collection of the relevant formulas spread over  several papers \cite{herrNPB,borasoyI,borasoyII}.~The values of experimental masses $m_{\eta}$,~$m_{\eta^{\prime}}$ together with the choice of mixing angle $\theta_{\eta}=-20^{0}$,~give the values of $m_{\eta_{00}}^2, m_{\eta_{88}}^2, m_{\eta_{08}}^2 $ at the  physical point.~When the ChPT expressions of $m_{\eta_{00}}^2, m_{\eta_{88}}^2, m_{\eta_{08}}^2 $,~are made equal to their respective numerical values at the physical point,~the four chiral constants $L_{7}, v_{0}^{(2)}, v_{2}^{(2)}, v_{3}^{(1)} $ appearing in those expressions,~get restricted by the three relations.~One should note that the constant $ v_{0}^{(2)}$ takes into account the $U_{A}(1)$ anomaly contribution to the $\eta'$ mass that gets determined primarily by the topological characteristics of the gluonic configurations.~Hence it would be insensitive against the quark mass variations.~Choosing the  large $N_c$ relation for  $ v_{0}^{(2)}=-29.3 f^2$,~the remaining three chiral constants as reported in Ref. \cite{Herpay:05} are  :
\bqa
\hspace{-.6 cm }L_7=-0.2272 \times 10^{-3} ; \ v_{3}^{(1)}=0.095 ; \ v_{2}^{(2)}=-0.1382. 
\eqa
The sum of Eqs. (B10) and (B11) in Ref. \cite{Herpay:05} gives the following ($m_{\pi},m_{K}$) dependence of the $M_{\eta}^2 $.	 
\bqa
\label{Meta}
\nonumber
M_{\eta}^2&=&2m_{K}^2-3v_{0}^{(2)}+2(2m_{K}^2+m_{\pi}^2)\
\{ 3v_{2}^{(2)}-v_{3}^{(1)} \}+  \\ \nonumber 
&&\hspace {-.4 cm}\frac{1}{f^2} \biggl[8v_{0}^{(2)}(2m_{K}^2+m_{\pi}^2)(3L_{4}+L_{5})+m_{\pi}^2(\mu_{\eta}-3\mu_{\pi})  \\ \nonumber
&&\hspace {-.4 cm}-4m_{K}^2 \mu_{\eta}+\frac{16}{3}\biggl\{(6L_{8}-3L_{5}+8L_{7})(m_{\pi}^2-m_{K}^2)^2+\\
&&\hspace {-.4 cm}2L_{6}(m_{\pi}^4-2m_{K}^4+m_{\pi}^2m_{K}^2)+L_{7}(m_{\pi}^2+2m_{K}^2)^2\biggr\} \biggr]\;.     \\ \nonumber
\eqa
The $\eta$ mass occurs at the leading order in chiral logarithm i.e.  $(m_{\eta}^{(0)})^{2}=(4 m_{K}^2-m_{\pi}^2)/3$,~hence the functions $f_{\pi}(m_{\pi},m_{K})$, $f_{K} (m_{\pi},m_{K})$ and  $M_{\eta}^2 (m_{\pi},m_{K})$ are applicable only when $4 m_{K}^2 >m_{\pi}^2 $ .~Eqs.(\ref{Meta}) together with the Eqs.(\ref{fpi}) and (\ref{fk}) enable the calculation of the RQM/QM model couplings $\lambda_{2}, c $,$m^{2}$ and $\lambda_{1}$ in the $m_{\pi}-m_{K}$ plane.

The chiral symmetry breaking strengths for  the chiral limit studies are reduced by reducing the $m_{\pi}$ and $m_{K}$ away from the physical point.~The results obtained in the  $m_{\pi}-m_{K}$ plane can be mapped to the corresponding results in the light-strange quark mass $m_{us}-m_{s}$ plane after using the Eqs.~(\ref{Aqpim}) and (\ref{qmr}) whose multiplication gives  $\mathcal{O}(1/f^2)$ expression of  $A(q+1)=B(m_{s}+m_{ud})$ in terms of the $\pi,K$ masses and chiral constants.~The constant $B$ gets determined from the Eq.~(\ref{Aqpim}) for the $m_{\pi}=138$ MeV as $A=B \ m_{ud}$ after taking the  $m_{ud}=4$ MeV at the physical point.~The quark masses away from the physical point can be evaluated easily after 
knowing the constant $B$ and  the Columbia plot in the $m_{ud}-m_{s}$ plane can be drawn.

\subsection{$ \bf f_{\pi},f_{K},m_{\pi},m_{K}  \text{ and }  M_{\eta}^2 \text{ from large } N_{c}$ U(3) ChPT}
\label{subsec:SChpt}

The large $N_c$ standard U(3) ChPT prescription as described in the Refs.~\cite{herrPLB,Escribano} gives 
the following expressions of the pion,~kaon decay constants and masses $f_{\pi},f_{K} \text{ and }m_{\pi}^2 , m_{K}^2$.

\bqa
\label{fpiU3}
f_{\pi}&=&f \Biggl( 1+4\frac{L_{5}}{f^2} m_{\pi}^{2} \Biggr)\;. \\ 
\label{fkU3}
f_{K}&=&f \Biggl( 1+4\frac{L_{5}}{f^2} m_{K}^{2} \Biggr)\;. \\
\label{mpi2U3}
m_{\pi}^2&=&2\ m_{ud} B\left[1+\frac{16  m_{ud} B}{f^2}(2L_{8}-L_{5})\right]\;. \\ 
\label{mk2U3}
\hspace*{-.8cm}m_{K}^2&=&m_{ud}B(1+q)\left[1+\frac{8  m_{ud} B}{f^2}(1+q)(2 L_{8}-L_{5})\right]\;.
\eqa

Note that only the dominant low energy constants $L_{5}$ and $L_{8}$ are appearing in the above expressions as they  are the $\mathcal{O}(N_{c})$ while the other constants $L_{4,6,7}$ are sub leading and hence suppressed in the large $N_{c}$ standard U(3) ChPT.~The chiral limit  value of the quark condensate $<\bar{u} u> $ is used to fix the parameters $q=m_s/m_{ud}$  and $B$.~Taking $f_{\pi}=93$ MeV with $f_{K}=1.223 \ f_{\pi}$ and using  Eqs.~(\ref{fpiU3}) and (\ref{fkU3}) for the physical $m_{\pi}=138$ MeV and $m_{K}=496$ MeV,~one finds $f=91.2599$ MeV and $L_{5}=2.084659\times10^{-3}$.~The expressions of the octet $\eta_{8}$ and singlet $\eta_{0}$ masses \cite{herrPLB,Escribano} are written as the following.
\bqa
\label{et88}
m_{88}^2&=&\frac{(4m_{K}^2-m_{\pi}^2)}{3}+\frac{4}{3}(m_{K}^2-m_{\pi}^2)\Delta_{M}\;. \\ \nonumber
\label{et00}
m_{00}^2&=&\frac{(2m_{K}^2+m_{\pi}^2)}{3}(1-2\Delta_{N})+\frac{2}{3}(m_{K}^2-m_{\pi}^2)\Delta_{M}\\  &&-3v_{02}\;.\\
\label{Delm}
\Delta_{M}&=&\frac{8}{f^2}(m_{K}^2-m_{\pi}^2)\ (2L_{8}-L_{5})\;.\\
\label{Deln}
\Delta_{N}&=&3v_{31}-\frac{12}{f^2}v_{02} L_{5}\;.\\ \nonumber
\label{MetaII}
M_{\eta}^2&=&m_{00}^2+m_{88}^2=2m_{K}^2-\frac{2}{3}(m_{K}^2-m_{\pi}^2)y \\
&&+2(m_{K}^2-m_{\pi}^2)\Delta_{M}-\frac{2}{3}(2m_{K}^2+m_{\pi}^2)\Delta_{N}
\eqa
The $\frac{2(m_{K}^2-m_{\pi}^2)}{3}y=3v_{02}$ in Eq.(\ref{MetaII}) defines $y$.~The $\eta'$ mass $m_{\eta'}^2=(M_{\eta}^2-m_{\eta}^2$) with the following mass for physical $\eta$.
\bqa
\label{meta}
\nonumber
m_{\eta}^2&=&m_{K}^2-\frac{(m_{K}^2-m_{\pi}^2)}{3}[y+(\sqrt{9+2y+y^2}\ ) ]+\Delta_{M} \\ \nonumber
&&(m_{K}^2-m_{\pi}^2)\biggl(1-\frac{9+y}{3\sqrt{9+2y+y^2}}\biggr)-\frac{\Delta_{N}}{3}\biggl\{2m_{K}^2 \\ 
&&+m_{\pi}^2-\frac{3(2m_{K}^2-3m_{\pi}^2)-y(2m_{K}^2+m_{\pi}^2)}{\sqrt{9+2y+y^2}}\biggr\}
\eqa
~The five parameters $v_{02},v_{31},L_{8}$ together with the quark mass parameters $m_{ud}B$ and $q$ can be expressed in terms of five independent observables $m_{\pi},m_{K},m_{\eta},m_{\eta'}$ and the $\eta-\eta'$ mixing angle $\theta$.~Thus large $N_{c}$ ChPT is not predictive at this level.~In order to get predictions from the large $N_{c}$ ChPT, Ref.~\cite{herrPLB} imposed that $\mathcal{O}(\delta)$ corrections are not too large so that the large $N_c$ expansion makes sense \cite{Escribano}.~Keeping the mixing angle in the range $20^{\circ}<\theta<24^{\circ}$, the Ref.\cite{herrPLB} gives the following range of the parameters fit for the physical observables.
\bqa
\nonumber
0.980 \le \frac{2m_{ud}B}{m_{\pi}^2}\le 0.988\; , \\ \nonumber \\
\nonumber
18.3 \le (q-1) \le 20.9\; \\ \nonumber \\
\nonumber
26 \le \frac{-v_{02}}{f^2} \le 29\; , \\ \nonumber \\
\nonumber
-0.164 \le v_{31} \le -0.161\;, \\ \nonumber \\ 
1.35 \times 10^{-3} \le L_{8} \le 1.57 \times 10^{-3}
\eqa
The experimental masses $m_{\eta}=547.5$ MeV and $m_{\eta'}=957.78$ MeV give $M_{\eta}=(\sqrt{m_{\eta}^2+m_{\eta^{\prime}}^2})=1103.22$ MeV.~With this same  numerical value of the $M_{\eta}$ ,~when the pole masses $m_{\eta}=528.48$ MeV and $m_{\eta'}=968.41$ MeV are used as input in the RQM model (With the $f_{\pi},f_{K}$ as above) for calculating the $\eta,\eta'$ self energy corrections and the renormalized parameters,~the same pole masses for the $\eta$ and $\eta'$ are reproduced in the output  after adding the corresponding self energy corrections to their respective masses calculated from the new set of renormalized parameters.~For getting the $M_{\eta}=1103.22$ MeV at the physical point from the ChPT expression given in the Eq.(\ref{MetaII}),~one finds the following numerical values of the U(3) ChPT parameters.  
\bqa
L_{8}=1.44 \times 10^{-3}, v_{31} = -0.1637 \text{ and } v_{02}=-29.1f^2
\eqa
It is important to note that with the above set of parameter values at the physical point,~the ChPT expressions in Eqs. (\ref{meta}) and (\ref{MetaII}) give the $\eta$ and $\eta'$ masses as $m_{\eta}=527.82$ MeV and $m_{\eta'}=968.77$ which are almost same as the corresponding masses obtained in the RQM model.~Note that the QM model parameters give  $m_{\eta}=538.96$ MeV and $m_{\eta'}=962.62$ MeV in the output when the  $M_{\eta}=1103.22$ MeV at the physical point.

One can find the light and strange quark masses corresponding to the reduced masses of pions and kaons from the Eqs.(\ref{mpi2U3}) and (\ref{mk2U3}).~Solving the quadratic Eq.(\ref{mpi2U3}) and
taking the positive root,~one gets $m_{ud}B$ whose value for the $m_{\pi}=138$ MeV  will give the constant B if one takes $m_{ud}=4$ MeV at the physical point.~The positive root of the quadratic Eq.(\ref{mk2U3}) gives  numerical values of $Bm_{ud}(q+1)=B(m_{ud}+m_{s})$ for different kaon masses.~Thus knowing B,~One can know the $m_{ud}$ and $m_{s}$ for different values of the $m_{\pi}$ and $m_{K}$ as one moves away from the physical point in the Columbia plot.

\begin{table*}[!htbp]
\caption{Chiral limit is approached by defining the reduced $\pi,\ K$ meson starred masses as  $\frac{m_{\pi}^*}{m_{\pi}}= \frac{m_{K}^*}{m_{K}}=\beta \le 1$ such that $\frac{m_{\pi}^*}{m_{K}^*}=\frac{m_{\pi}}{m_{K}}$ where $m_{\pi}=138$ and $m_{K}=496$ MeV at the physical point \cite{Schaefer:09}.~The  $f_{\pi}, \ f_{K}$ for any $\beta$ in the RQM-S model using the large  $N_c$ standard U(3) ChPT,~are obtained by putting the corresponding  $m_{\pi}^*$, $m_{K}^*$ in the Eq.~(\ref{fpiU3}) and Eq.~(\ref{fkU3}).~The parameters and chiral constants of the standard U(3) ChPT at the physical point which reproduce the experimental $M_{\eta}$:Expt=1103.22 MeV corresponding to the experimental values $(m_{\eta},m_{\eta^{\prime}}) = (547.5,957.8)$ MeV,~give the $(m_{\eta},m_{\eta^{\prime}}) = (527.82,968.77)$ MeV from the Eqs.(\ref{MetaII}) and (\ref{meta}) for the input set S.~The Eq.~(\ref{fpi}) and Eq.~(\ref{fk}) with the $m_{\pi}^*$, $m_{K}^*$  for any $\beta$,~give the corresponding  $f_{\pi}, \ f_{K}$ for the infrared regularized U(3) ChPT input set I 
where the corresponding $M_{\eta}$ gets calculated using the  Eq.~(\ref{Meta}).~The infrared regularized U(3)  ChPT expression in the  Eq.~(\ref{Meta}) gives $M_{\eta}$-I=1035.55 MeV for the physical $m_{\pi}, m_{K}$.~The  RQM model parameters $\lambda_{20},c_{0},\lambda_{10},m_{0}^2, h_{x0}^{*} \text{ and } h_{y0}^{*}$ for the physical point $\beta=1$ and critical end point ratio $\beta_{\cep}$,~are presented for both the ChPT input sets S and I when the $m_{\sigma}=400,500 \text{ and }600$ MeV and only the input set S when the $m_{\sigma}=750 \text{ and }800$ MeV.} 
\label{tab:table1}
\begin{tabular}{p{0.04\textwidth}|p{0.14\textwidth}| p{0.07\textwidth}|  p{0.07\textwidth}| p{0.09\textwidth}| p{0.09\textwidth}| p{0.08\textwidth}|p{0.07\textwidth} | p{0.10\textwidth} | p{0.10\textwidth}| p{0.10\textwidth}} 
\hline
$m_{\sigma} $ &  $\frac{m_{\pi}^*}{m_{\pi}}= \frac{m_{K}^*}{m_{K}}=\beta  $ & $f_{\pi} (\text{MeV})$ & $f_{K} (\text{MeV})$ & $\text{M}_{\eta}$ $(\text{MeV}$) & $\lambda_{20}$ & $ c_{0} \ (\text{MeV})$&$\lambda_{10}$  &$m_{0}^2$ $(\text{MeV}^2)$&  $ h^*_{x0}  (\text{MeV}^3)$ & $ h^*_{y0} (\text{MeV}^3)$  \\
\hline
 &1(S)&93.0&113.7389&1103.22&33.522&7330.557&1.907 & $(443.341)^2 $  &$(119.804)^3$&$(323.531)^3$  \\
 &1(I) &92.9737&113.2635 & 1035.55 & 35.47&7390.50&1.27 & $(447.71)^2 $  &$(119.79)^3$&$(324.39)^3$  \\
400&$\beta_{\cep}$=0.38738(S) &91.521&94.6332&889.14&31.148&8056.86&7.225&$(175.035)^2 $ & $(63.859)^3$ &$(167.657)^3$  \\
&$\beta_{\cep}$=0.37963(I) &91.571& 97.915&846.25&-24.468&7828.558&25.119&$(159.898)^2 $ & $(63.019)^3$ &$(163.679)^3$  \\
\hline
 &1(S)&93.0&113.7389&1103.22&33.522&7330.557&4.072 & $(397.817)^2 $  &$(119.804)^3$&$(323.531)^3$  \\
&1(I)&92.9737&113.2635 & 1035.55 & 35.47&7390.50&3.46&$(402.433)^2 $ & $(119.79)^3$ &$(324.39)^3$  \\
500&$\beta_{\cep}$=0.37665(S) &91.507&94.449&887.13& 31.05  &8069.107&10.541&$-(114.749)^2 $ & $(62.676)^3$ &$(164.468)^3$  \\
&$\beta_{\cep}$=0.37159(I) &91.481&97.632&845.21&-25.271&7836.42&28.372&$-(125.371)^2 $ & $(62.11)^3$ &$(161.307)^3$  \\
\hline
 &1(S)&93.0&113.7389&1103.22&33.522&7330.557&9.0398 & $(265.265)^2 $  &$(119.804)^3$&$(323.531)^3$  \\
&1(I)&92.9737&113.2635&1035.55 &35.47&7390.50&8.46&$( 272.31)^2 $ & $(119.79)^3$ &$(324.39)^3$  \\
600&$\beta_{\cep}$=0.33634(S)&91.457&93.8028&880.12&30.689&8112.955&17.512&$-(322.909)^2 $ & $(58.126)^3$ &$(152.253)^3$  \\
&$\beta_{\cep}$=0.32445(I) &90.946&95.985 &839.74&-30.081&7880.442&37.757&$-(345.951)^2 $ & $(56.646)^3$ &$(147.077)^3$  \\
\hline
 &1(S)&93.0&113.7389&1103.22&33.522&7330.557&24.492 & $-(540.551)^2 $  &$(119.804)^3$&$(323.531)^3$  \\
750&$\beta_{\cep}$=0.25013(S)&91.369&92.666&867.82& -30.081&8193.523&33.834&$-(558.347)^2 $ & $(47.720)^3$ &$(124.582)^3$  \\
\hline
&1(S) &93.0&113.7389&1103.22&33.522&7330.557&29.826&$-(545.342)^2 $ & $(119.804)^3$ &$(323.531)^3$  \\
800&$\beta_{\cep}$=0.23316(S)&91.354&92.482&865.83&29.86&8207.002&37.957&$-(602.887)^2 $ & $(45.538)^3$ &$(118.818)^3$  \\
\hline
\end{tabular}
\end{table*}


\section{Results and Discussion}
\label{secIII}


The RQM model~\cite{vkkr23, skrvkt24} whose parameters are renormalized on-shell after the proper treatment of the quark one-loop vacuum fluctuation,~got refined very recently into a valuable tool of chiral limit studies by enriching the model with the chiral perturbation theory (ChPT) inputs for finding the model parameters away from the physical point~\cite{vkt25I,vkt25II}.~The $ \mathcal{O}(\frac{1}{f^2})$  accurate results of the infrared regularized U(3) ChPT~\cite{Herpay:05, borasoyI, borasoyII, Beisert, Becher},~as the input set I are used for the ($m_{\pi},m_{K}$) dependence of the $f_{\pi}, f_{K} $ and $M_{\eta}^2$ and the first,~second order and crossover chiral transition regions in the Columbia plot are computed in the RQM-I model  for the $m_{\sigma}=400 \text{ and } 530 $ MeV.~The large $N_c$ standard U(3) ChPT scaling relations for the ($m_{\pi},m_{K}$) dependence of the $f_{\pi}, f_{K} $ and $M_{\eta}^2$ ~\cite{gasser, LeutI,  KaisI, herrNPB, herrPLB, Escribano},~termed here as the input set S,~have also been used in the Ref.~\cite{vkt25I} for fixing the RQM model parameters away from the physical point and computing the Columbia plot in the RQM-S model when the $m_\sigma=530$ MeV in the presence of $U_{A}(1)$ anomaly.~The enrichment of the RQM model with the ChPT inputs,~cures the problem of the loss of the spontaneous chiral symmetry breaking ( SCSB) for the mass range $m_{\sigma}=400-800$ MeV when the $m_{\pi},m_{K} \rightarrow  0$  in the the often used method \cite{Ortman, Lenagh, Schaefer:09,  fuku08, berges, Herbst} of chiral limit studies in the QM model called the fixed-ultraviolet (UV) scheme \cite{Resch} where   the light (strange) explicit chiral symmetry breaking strengths  $ h_{x}(h_{y}) $ are changed while all other parameters are kept same as the ones at the physical point.~The chiral limit in the fixed UV scheme can be explored only for very large $m_{\sigma}\ge800$ MeV when the mass parameter $m^2$ turns negative to get the SCSB.~The nonperturbative FRG study of Resch et.al. \cite{Resch} under the local potential approximation (LPA) circumvented the above problem by proposing the ChPT motivated fixed-$f_{\pi}$ scheme where the initial effective action is heuristically adjusted to larger scales ($\Lambda^{\prime} > \Lambda $) for every smaller mass point in the path to the chiral limit such that the $f_{\pi}$ always retains its physical value and the SCSB is not lost for the mass range $m_\sigma=400-600$ MeV.~The change in parameters (which are kept same as at the physical point) gets compensated by changing the scale ($\Lambda^{\prime} > \Lambda $) of the initial action  when the explicit chiral symmetry breaking strengths $h_{x}$ and $h_{y}$ decrease.~The fixed $f_{\pi}$ scheme suffers from the problem of a heuristic determination of the initial effective action,even though it is physically reasonable.~Furthermore the FRG flow  allows for the scalar $\sigma$ masses only in the range $m_{\sigma} \in [400,600]$ MeV~\cite{Resch}.~In contrast to the above,~the RQM model parameter fixing with the ChPT inputs is free from any ambiguity and heuristic adjustment as  the $\pi, K$ meson masses are reduced to perform chiral limit study  away from the physical point.

~The Columbia plots in the presence of the $U_{A}(1)$ anomaly were computed in the RQM model using the two ChPT input sets I and S only for the case of $m_{\sigma}=530$ MeV in the Ref.~\cite{vkt25I},~so that the results can be compared for checking the consistency of chiral limit study.~One gets slightly larger first order region in the RQM-S model.~The RQM-I and S model tricritical lines,~demarcating the first and second order regions in the $\mu-m_{K}$ plane at $m_{\pi}=0$,~are almost overlapping for the $m_{K}<450$ MeV.~But the real difference emerges for the $m_{K}>450$ MeV when the tricritical line obtained from the large $N_c$ standard U(3) ChPT input set S  shows saturation for the range $m_{K}=450-550$ MeV as its slope  becomes zero in the range $m_{K}=520.0-550.0$ MeV in which the chemical potential stays almost constant at $\mu=234.8$ MeV while the tricritical line for the infrared regularized input set I does not show saturation as its average slope increases  to 0.1566 when it goes  from the point ($m_{K}=500.0,\mu=235.79$) MeV to the ($m_{K}=550.0,\mu=243.62$) MeV.~The proper saturation trend shown by the tricritical line in the $\mu-m_{K}$ plane of the RQM-S model Columbia plot for larger $m_{K}$ is an improvement over the corresponding result obtained in the RQM-I model because the tricritical line is expected to be connected to the tricritical point of the two-flavor chiral limit \cite{Resch, hjss} at some higher value of $m_{K}$.~The above discussion,~prompts us to compute the Columbia plots in the RQM-S model for different $m_{\sigma}=400,500 \text{ and } 600$ MeV and compare them with the existing (for the $m_{\sigma}=400$ MeV in~\cite{vkt25II}) and to be computed  RQM-I model Columbia plots for the $m_{\sigma}=500 \text{ and } 600$ MeV.~Bringing out similarities and differences in the RQM-S and RQM-I model Columbia plots for different values of the $m_{\sigma}$ in the range 400-600 MeV,~the present work aims to know if the large $N_c$ standard U(3) ChPT scaling relations for the ($m_{\pi},m_{K}$) dependence of the $f_{\pi}, f_{K} $ and $M_{\eta}^2$,~constitute  an  improved and better prescription for Chiral limit studies.~It is well known that the strength of the first order chiral transition diminishes for larger $m_{\sigma}$ and the critical end point ($\cep$) disappears from  the entire $\mu-T$ plane as the chiral transition becomes crossover all over the  phase diagram computed for the physical point parameters  when the $m_{\sigma} > 600$ MeV~\cite{vkkr23}.~In order to see how the first order  regions get reduced when the second order chiral transition regions of the $\mu-m_{K}$ plane (at $m_{\pi}=0$) and the chiral crossover transition regions of the $m_{\pi}-m_{K}$ plane (at $\mu=0$),~are expanding in response to the increasing $\sigma$ meson mass,~the Columbia plots will be computed and discussed below also for the $m_{\sigma}=750 \text{ and } 800$ MeV in the RQM-S model.~We will compute,~discuss and compare the light-strange quark mass $m_{ud}-m_{s}$ planes  (at $\mu=0$) and the $\mu-m_{s}$ planes (at $m_{ud}=0$) of the Columbia plots also when the $m_{\sigma}=400,500,600$ MeV in both the RQM-S as well as the RQM-I models and the  $m_{\sigma}=750 \text{ and } 800$ MeV in the RQM-S model.~The complete comparative perspective of the critical quantities obtained in the $\mu-m_{K}$ ($\mu-m_{s}$) and $m_{\pi}-m_{K}$ ($m_{ud}-m_{s}$) planes of the Columbia plots of different model scenarios over the whole range of scalar $\sigma$ meson mass $m_{\sigma}=400-800$ MeV together with the relevant LQCD study results (with references) will be presented in a tabular form.

~Note that since the on-shell renormalization of  parameters in the RQM model,~gives rise to a significantly stronger axial $U_{A}(1)$ anomaly and a reduced strange as well as non-strange direction explicit chiral symmetry breaking strengths  $h_{x} \text{ and } h_{y}$,~the softening effect of the quark one-loop vacuum fluctuation on the strength of the chiral transition is moderate in the RQM model~\cite{vkkr23,skrvkt24}.~Therefore one finds a noticeably large first order regions in the $\mu-m_{K}$ ($\mu-m_{s}$) and $m_{\pi}-m_{K}$ ($m_{ud}-m_{s}$) planes of the RQM-S model Columbia plots for the $m_{\sigma}=530$ MeV as reported recently in the Ref.~\cite{vkt25I}.~In contrast to the above the corresponding first order regions are significantly smaller in the  Columbia plots of the curvature mass based parametrized quark meson with vacuum term (QMVT) model~\cite{vkt25II} and the e-MFA FRG QM model work for the $m_{\sigma}=530$ MeV~\cite{Resch} due to very large softening influence of the quark one-loop vacuum fluctuation on the strength of the chiral transition.~The $U_{A}(1)$ anomaly strength $c$ and  $h_{x}  (h_{y})$ do not change in the QMVT model and  the e-MFA FRG work as they are using curvature meson masses to fix the model parameters.~It will also be  interesting to compare the smaller extents of the first order regions that one finds due to the very strong smoothing effect of the quark one-loop vacuum correction in the QMVT model and the e-MFA FRG QM model Columbia plots for the $m_{\sigma}=530$ MeV~\cite{Resch,vkt25II} with the first order regions of the RQM-S model Columbia plots (for larger $m_\sigma$) which become smaller  due to the effect of increasing the mass of   the $\sigma$ meson to $m_{\sigma}=600,750\text{ and } 800$ MeV.~The three important critical quantities to be computed in the Columbia plots are 1) the tricritical points ($\tcp$) $m_{K}^{\tcp}$ ($m_{s}^{\tcp}$) at $\mu=0$ and also in the $\mu-m_{K}$ ($\mu-m_{s}$) plane for the light chiral limit $m_{\pi}=0=m_{ud}$,~2) the pion (light quark) critical mass $m_{\pi}^{c}$ ($m_{ud}^{c}$) beyond which the chiral transition is a smooth crossover on the three flavor SU(3) symmetric chiral limit line ($m_{\pi}=m_{K}$  i.e.  $m_{ud}=m_{s}$ ) in the $m_{\pi}-m_{K}$ ($m_{ud}-m_{s}$ ) plane and 3) the pion (light quark) terminal mass $m_{\pi}^{t}$ ($m_{ud}^{t}$) at which the chiral critical line of Z(2) critical points,~terminates on the strange chiral limit line $h_{y}=0$ ($m_{s}=0$) in the $m_{\pi}-m_{K}$ ($m_{ud}-m_{s}$) plane at $\mu=0$.~The tricritical point is defined by that kaon (s quark) mass $m_{K}^{\tcp}$ ($m_{s}^{\tcp}$) at which the second order chiral phase transition becomes first order.

The $h_{x0}=0$ in the Eq. \  (\ref{hx0}) defines the light chiral limit for the RQM model whereas the $h_{y0}=0$ in Eq.\ (\ref{hy0}) gives the strange chiral limit line.~The RQM model parameters $\lambda_{20},c_{0},\lambda_{10},m_{0}^2$ corresponding to the reduced chiral  symmetry breaking strengths $h_{x0}^*$ and $h_{y0}^*$ for performing the chiral limit study are calculated using the input sets of two different prescriptions of the ChPT.~The reduced light and strange chiral symmetry breaking strengths for the QM model are given by $h_{x}^{*}=(m_{\pi}^{*})^{2} \ f_{\pi}$ and $h_{y}^{*}=\lbrace\sqrt{2} \ f_K \ (m_{K}^{*})^{2}-\frac{f_{\pi}}{\sqrt{2}} \ (m_{\pi}^{*})^{2}\rbrace$.~The RQM model reduced strengths are defined as : $h_{x0}^{*}=(m_{\pi,c}^{*})^{2} \ f_{\pi}$ and $h_{y0}^{*}=\lbrace\sqrt{2} \ f_K \ (m_{K,c}^{*})^{2}-\frac{f_{\pi}}{\sqrt{2}} \ (m_{\pi,c}^{*})^{2}\rbrace$ where the reduced pion,~kaon curvature masses $m_{\pi,c}^{*} \text{ and } m_{K,c}^{*} $ are found by putting the reduced pion,~kaon mass $m_{\pi}^{*} \text{ and } m_{K}^{*} $ values respectively in the expressions Eq.~(\ref{mpicr}) and Eq.~(\ref{kcr}).~The ($m_{\pi},m_{K}$) dependence of the $f_{\pi}, f_{K} $ and $M_{\eta}^2$ for the input set I of the infrared regularized U(3) ChPT prescription described in the section(\ref{subsec:Chpt}-I),~are calculated by substituting the reduced pion,~kaon masses $m_{\pi}^*$ and $m_{K}^*$ respectively in the Eq.~(\ref{fpi}),~Eq.~(\ref{fk}) and  Eq.(\ref{Meta}).~After substituting the reduced masses $m_{\pi}^*$ and $m_{K}^*$ respectively in the Eq.~(\ref{fpiU3}),~Eq.~(\ref{fkU3}) and Eq.(\ref{MetaII}),~one finds the ($m_{\pi},m_{K}$) dependence of the $f_{\pi}, f_{K} $ and $M_{\eta}^2$ for the input set S of the large $N_c$ standard U(3) ChPT framework given in the section(\ref{subsec:Chpt}-II).~The $m_{\pi}^{*}$ and $m_{K}^{*}$ defined by reducing the $\pi$ and $K$ masses in a fixed ratio. i.e. $\frac{m_{\pi}^*}{m_{\pi}}= \frac{m_{K}^*}{m_{K}}=\beta$,~provides us a direct path from the physical point to the chiral limit by tuning only the fraction $\beta$ instead of changing both the $m_\pi$ and $m_K$ randomly.~The $\frac{m_{\pi}^*}{m_{K}^*}$ always remains equal to the fixed physical point ratio $\frac{m_{\pi}}{m_{K}}$ \cite{Schaefer:09}.~The fraction $\beta$ of the above choice,~also facilitates the comparison of our results with those of the Refs.~\cite{Resch,Schaefer:09}.~The RQM model with the input set I and S has been termed respectively as the RQM-I and RQM-S model whose parameters $\lambda_{20},c_{0},\lambda_{10},m_{0}^2,h_{x0},h_{y0}$ for the $f_{\pi},f_{K},M_{\eta}$ corresponding to the physical point ratio $\beta=1$ ( where $(m_{\pi},m_{K})$=(138,496) MeV) and the critical end point ratio $\beta_{\cep}$ when the $m_{\sigma}=400,500 \text{ and } 600$ MeV are presented in the Table \ref{tab:table1}. The RQM-S model parameters for the $\beta=1\text{ and }\beta_{\cep}$ when $m_{\sigma}=750 \text{ and } 800$ MeV,~are also presented in the Table \ref{tab:table1}.~The critical end point ratio $\beta_{\cep}$ is that fraction of $\pi \text{ and } K$ mass at which the chiral crossover transition,~in the $m_{\pi}-m_{K}$ plane (at $\mu=0$),~becomes first order.

\begin{figure*}[htb]
\subfigure[\ Non-strange chiral condensate.]{
\label{fig1a} 
\begin{minipage}[b]{0.49\textwidth}
\centering
\includegraphics[width=\linewidth]{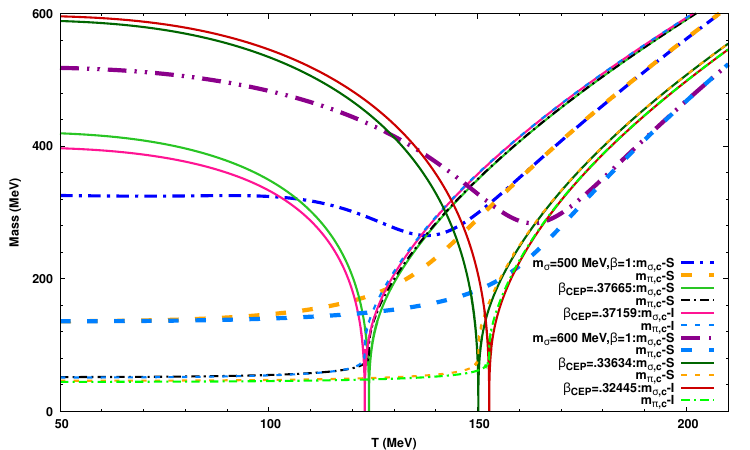}
\end{minipage}}
\hfill
\subfigure[\ Strange chiral condensate.]{
\label{fig1b} 
\begin{minipage}[b]{0.49\textwidth}
\centering 
\includegraphics[width=\linewidth]{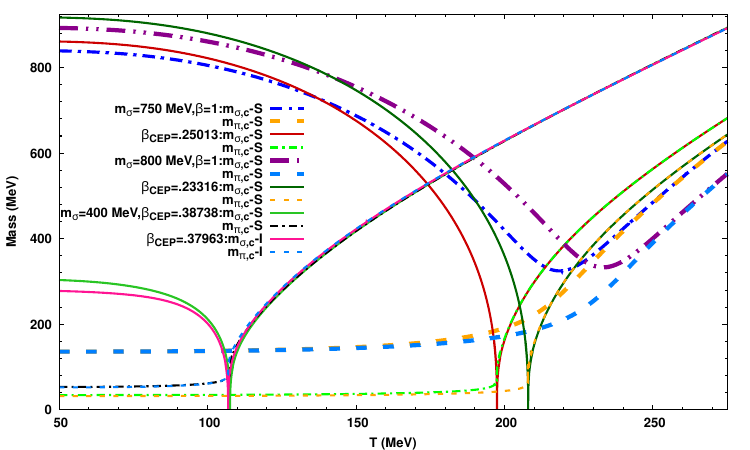}
\end{minipage}}
\caption{The curvature masses of the $\pi \text{ and } \sigma$ mesons in the RQM model when its parameters are fixed using the large $N_c$ standard U(3) ChPT inputs (the infrared regularized U(3) ChPT inputs) are termed  as $m_{\pi,c}-\text{S(I)} \text{ and } m_{\sigma,c}-\text{S(I)}$.~The temperature variations of the $m_{\pi,c}-\text{S} \text{ and } m_{\sigma,c}-\text{S}$ for both the physical point $\beta=1$ and the $\beta_{\cep}$ together with the $m_{\pi,c}-\text{I} \text{ and } m_{\sigma,c}-\text{I}$ temperature variations for the $\beta_{\cep}$ at $\mu=0$ when the $m_{\sigma}=500 \text{ and } 600$ MeV,~are presented in the left panel (a).~The right panel (b) shows the $\mu=0$ temperature variations of the $m_{\pi,c}-\text{S} \text{ and } m_{\sigma,c}-\text{S}$ for the physical point $\beta=1$ and $\beta_{\cep}$ when the $m_{\sigma}=750 \text{ and } 800$ MeV.~The $m_{\pi,c}-\text{S(I)} \text{ and } m_{\sigma,c}-\text{S(I)}$ temperature variations for the $\beta_{\cep}$ when $m_{\sigma}=400$ MeV are also plotted in the right panel where $\mu=0$.~The fraction $\beta=\frac{m_{\pi}^*}{m_{\pi}}= \frac{m_{K}^*}{m_{K}}$ .}
\label{fig:mini:fig1}
\end{figure*}

Using the $M_{\eta}=\sqrt{(m_{\eta}^2+m_{\eta^{\prime}}^2)}=1103.22$ MeV (termed as $M_{\eta}:\text{Expt}$) for  the experimental masses $(m_{\eta},m_{\eta^{\prime}})=(547.5,957.8)$ MeV with the  $f_{\pi} \text{ and } f_{K}$ at $\beta=1$ of the input set I in the Table \ref{tab:table1} for the input,~the renormalized parameters of the RQM-I model give $(m_{\eta},m_{\eta^{\prime}})=(527.82,968.76)$ MeV in the output.~One finds a slightly smaller value of the $M_{\eta}=1035.55$ MeV from the ChPT expression in Eq.(\ref{Meta}) for the input set I at the physical point and with this $M_{\eta}$ in the input,~the RQM-I model give $(m_{\eta},m_{\eta^{\prime}})=(520.62,895.17)$ MeV in the output.~Values taken from the prescribed range for the parameters and chiral constants of the large $N_{c}$ standard U(3) ChPT in the Refs.~\cite{herrPLB,Escribano} as given in the section (\ref{subsec:Chpt}-II),~reproduce the experimental $M_{\eta}$ : Expt=1103.22 MeV at the physical point for the input set S.~The $M_{\eta}$ : Expt=1103.22 MeV with the $f_{\pi},f_{K}$ for $\beta=1$ from the input set S,~give the $(m_{\eta},m_{\eta^{\prime}}) = (527.82,968.77)$ MeV from the standard U(3) ChPT expressions Eqs.~(\ref{MetaII}) and (\ref{meta}) whereas the corresponding RQM model calculation gives almost the same $(m_{\eta},m_{\eta^{\prime}}) = (528.48,968.41)$ MeV.~The above computations are consistent as the calculated $M_{\eta}$ from the 
$(m_{\eta},m_{\eta^{\prime}})$ values obtained in the output  of all the RQM-S(I) model calculations is the same as in the input. 


\subsection{\bf{Comparing temperature variations of the $\pi\text{ and }\sigma$ curvature masses at $\beta=1,\beta_{\cep}$ in RQM-S(I) model for $\bf m_{\sigma}=400,500,600,750 \text{ and }800$ MeV.}}
\label{secIIIA}

The chiral limit study has been performed in the above described set up of the RQM-S and RQM-I model framework.~The explicit chiral symmetry breaking strengths $ h^*_{x0}$ and $ h^*_{y0}$ in the non-strange and strange directions are systematically decreased by reducing the pion and kaon masses when the ratio $\beta$ is tuned to smaller fractions.~The temperature variations of the light and strange condensates $x$ and $y$ ~when $\beta$ is reduced from 1 for different $m_{\sigma}=400,500,600,750 \text{ and } 800$ MeV,~have similar pattern as shown for the RQM-I and S models in the Ref.~\cite{vkt25I} for the case of $m_{\sigma}=530$ MeV.~Becoming sharper in response to the decreasing $\beta$,~the physical point chiral crossover transition turns second order at  the $Z_{2}$ critical point fraction $\beta_{\cep}$ for reduced $\cep$ masses $m_{\pi,\cep} \text{ and }m_{K,\cep}$.~For $\beta < \beta_{\cep}$,~one gets the first order chiral phase transition which becomes successively stronger when $\beta$ approaches zero in the chiral limit where the masses of the octet mesons $\pi,K \text{ and } \eta$ become zero~\cite{vkt25I,vkt25II}.

~The chiral transition is analyzed also by studying the temperature variations of the masses of  $\sigma \text{ and } \pi$ mesons which are the chiral partners.~When the chiral crossover transition turns second order,~the effective potential flattens in the $\sigma$ direction and the associated critical $\sigma$ mode becomes massless.~Hence if the temperature variation of the scalar mass $m_\sigma$,~hits zero on the temperature axis,~one confirms the identification of the  critical end point ($\cep$).~The correlation length  $\xi=1/m_{\sigma}$ diverges at the $\cep$ when mass of the critical mode $m_\sigma \rightarrow 0$.~The Ref.~\cite{Resch} has discussed that when only  the chiral symmetry breaking  source strengths $j_{x} \text{ and } j_{y}$ in the light and strange directions are reduced for performing the chiral limit study in the fixed UV scheme whereas the other parameter values of the model are kept fixed as at the physical point,~the critical sigma mode mass $m_{\sigma}$  increases and  the corresponding correlation length $\xi$ decreases to such an extent that the spontaneous chiral symmetry breaking (SCSB) gets lost.~They proposed the ChPT motivated fixed-$f_{\pi}$ scheme as a  remedy of the above problem and performed the chiral limit study by heuristically adjusting the initial effective action  to the larger scales ($\Lambda^{\prime} > \Lambda $) when the symmetry breaking source strengths $j_{x} \text{ and } j_{y}$,~are reduced for smaller ($m_{\pi}, \ m_{K}$)  such that the $f_{\pi}$ remains fixed to its physical value and the SCSB is not lost when the $m_\sigma=400-600$ MeV.~The change in parameters (which are kept same as at the physical point) for the smaller explicit chiral symmetry breaking source strengths,~are compensated by changing  the scale ($\Lambda^{\prime} > \Lambda $).

When the $f_{\pi},\ f_{K}, \ M_{\eta}$ and all the model parameters change according to the ($m_{\pi}, \ m_{K}$) dependent scaling relations provided by the large $N_c$ standard U(3) ChPT and the infrared regularized U(3) ChPT prescriptions respectively in the RQM-S and RQM-I model,~the consistent chiral limit study can be performed without any loss of SCSB in our work.~Here it is relevant to point out that the FRG flow  allows for the scalar $\sigma$ masses only in the range $m_{\sigma} \in [400,600]$ MeV~\cite{Resch} while the chiral limit study can be accurately performed over the complete range of $m_{\sigma}=400-800$ MeV in our RQM-S model work.~Comparing the Columbia plots of the RQM-S model and RQM-I model for different $m_{\sigma}$ in the range 400-600 MeV,~it will be shown later that RQM-S model gives better and refined results.~Therefore the plots of temperature variations of the $\sigma , \ \pi$ curvature masses $m_{\sigma,c}$ and $m_{\pi,c}$ have been shown in the Fig.~\ref{fig:mini:fig1}.~The plots of $m_{\sigma,c}$ and $m_{\pi,c}$ in the RQM-S model for $\beta=1 \text{ and }\beta_{\cep}$ and RQM-I model for $\beta_{\cep}$ are presented in the  Fig.~\ref{fig1a} where $m_{\sigma}=500\text{ and } 600$ MeV.~The Fig.~\ref{fig1b} presents the $m_{\sigma,c}$ and $m_{\pi,c}$ temperature plots for the RQM-S model at the physical point as well as  $\beta_{\cep}$ when $m_{\sigma}=750\text{ and } 800$ MeV and the RQM-S and  I models at $\beta_{\cep}$ when $m_{\sigma}=400$ MeV.~Since the temperature variations of the $m_{\sigma,c}$ and $m_{\pi,c}$ in the RQM-S model overlap with the corresponding temperature plots in the RQM-I model at the physical point,~the  Fig.~\ref{fig1a} shows the $\beta=1$ temperature plots of the $(m_{\sigma,c},m_{\pi,c})$ only for the RQM-S model when $m_{\sigma}=500,600$ MeV while the $(m_{\sigma,c},m_{\pi,c})$ temperature variations at $\beta=1$ for $m_{\sigma}=400$ MeV,~are not shown in the Fig.~\ref{fig1b}  for either the RQM-S or the RQM-I model.~The temperature plots of the $m_{\sigma,c}$ and $m_{\pi, c}$ in the Fig.~\ref{fig1a} and Fig.~\ref{fig1b} become degenerate in the chiral symmetry restored phase after the chiral transition.

It is worth pointing out that the $\sigma$ meson vacuum curvature mass $m_{\sigma,c}$ for all the cases of  $m_{\sigma}$ in both the RQM-S and RQM-I models,~is lowest at the physical point.~When the temperature variations of the $m_{\sigma,c}$ at the $\beta_{\cep}$ are compared with those for the physical point $\beta=1$ in both the RQM-S and I models,~one finds  noticeably larger values of the $m_{\sigma,c}$  in the beginning of their temperature variations for the $\beta=\beta_{\cep}$ in the Fig.~\ref{fig1a} and Fig.~\ref{fig1b}.~Successively increasing in the path to the chiral limit when the ($m_{\pi},m_{K}$) are reduced by the fraction $\beta$,~the $m_{\sigma,c}$ becomes highest when the chiral limit is reached (not shown here,~it can be seen in the Ref.~\cite{vkt25II}).~The above pattern is opposite of what one finds in the e-MFA-FRG:QM model study for the $m_\sigma=530$ MeV in the Ref.~\cite{Resch} where the $\sigma$ mass in the Fig.2(c) goes through a significant successive decrease when the chiral limit is approached by  decreasing  the source strengths $j_{x} \text{ and } j_{y}$ successively by the fractions $\alpha=\sqrt{\beta}=1.0,0.17,0.04 \text{ and } 0$ and correspondingly adjusting the initial effective action (to start the FRG flow) successively to the higher scales $\Lambda=700,1000,1100 \text{ and } 1143$ MeV.~Furthermore the $m_{\sigma,c}$ temperature variations of the RQM-S model for  $m_{\sigma}=400\text{ and }500$ MeV  respectively in the Fig.~\ref{fig1b} and Fig.~\ref{fig1a},~are lying above the corresponding temperature variations of the RQM-I model.~The above trend gets reversed in the Fig.~\ref{fig1a} for the  case of $m_{\sigma}=600$ MeV where the RQM-S model $m_{\sigma,c}$ temperature plot lies below that of the RQM-I model.~This behavior gets explained when one notes that the $\beta_{\cep}$ for the RQM-I model is slightly smaller than the $\beta_{\cep}$ of the RQM-S model when the $m_{\sigma}=400\text{ and }500$ MeV  and this differences becomes noticeably larger for the $m_{\sigma}=600$ MeV.

For the $m_{\sigma}=400$ MeV,~the crossover transition terminates at the second order $Z_{2}$ critical point masses $(m_{\pi,\cep}^*,m_{K,\cep}^*)=(53.46,~192.14)$ MeV for the $\cep$ ratio $\beta_{\cep}|_{(m_{\sigma}=400 \ \text{MeV} : \text{RQM-S})}=0.38738$ in the RQM-S model and $(m_{\pi,\cep}^*,m_{K,\cep}^*)=(52.39,~188.30)$ MeV for the $\cep$ ratio $\beta_{\cep}|_{(m_{\sigma}=400 \ \text{MeV} : \text{RQM-I})}=0.37963$ in the RQM-I model.~When the $m_{\sigma}=500$ MeV,~one finds a reduced set of masses for the second order $Z_{2}$ critical point at $(m_{\pi,\cep}^*,m_{K,\cep}^*)=(51.98,~186.82)$ MeV for the $\cep$ ratio $\beta_{\cep}|_{(m_{\sigma}=500 \ \text{MeV} : \text{RQM-S})}=0.37665$  in the RQM-S model and $(m_{\pi,\cep}^*,m_{K,\cep}^*)=(51.33,~184.49)$ MeV for the $\cep$ ratio  $\beta_{\cep}|_{(m_{\sigma}=500 \ \text{MeV} : \text{RQM-I})}=0.37159$ in the RQM-I model.~The critical end point masses for the $m_{\sigma}=600$ MeV are obtained at the $(m_{\pi,\cep}^*,m_{K,\cep}^*)=(46.41,~166.82)$ MeV for the $\cep$ ratio $\beta_{\cep}|_{(m_{\sigma}=600 \ \text{MeV} : \text{RQM-S})}=0.33634$ in the RQM-S model and the $(m_{\pi,\cep}^*,m_{K,\cep}^*)=(44.77,~160.93)$ MeV for the $\cep$ ratio $\beta_{\cep}|_{(m_{\sigma}=600 \ \text{MeV} : \text{RQM-I})}=0.32445$ in the RQM-I model.~One finds that for all the $\sigma$ masses $m_{\sigma}=400,500\text{ and } 600$ MeV the RQM-S model $\cep$ ratio is larger than the  $\cep$ ratio of the RQM-I model i.e. $\beta_{\cep}|_{(m_{\sigma}=400,500 \text{ and } 600 \text{MeV} : \text{RQM-S})}> \beta_{\cep}|_{(m_{\sigma}=400,500 \text{ and } 600 \text{MeV} : \text{RQM-I})}$.~It is worth emphasizing that when the chiral limit study results obtained in the RQM-S model are compared with the corresponding results found in the RQM-I model,~one finds that the use of the large $N_c$ standard U(3) ChPT input set S,~generates a stronger approach towards the first order  chiral transition as the chiral crossover transition turns second order and then first order transition is obtained for $\beta<\beta_{\cep}$ by a comparatively smaller reduction in the  masses of $\pi \text{ and } K $ mesons.~Note that since $\beta_{\cep}|_{(m_{\sigma}=400 \ \text{MeV} : \text{RQM-S(I)})}>\beta_{\cep}|_{(m_{\sigma}=500  \text{MeV} : \text{RQM-S(I)})}>\beta_{\cep}|_{(m_{\sigma}=600  \text{MeV} : \text{RQM-S(I)})}$ ,~the expected pattern gets confirmed that lower values of the $\sigma$ meson mass in both the RQM-S and I model scenarios give rise to a stronger first order chiral transition whose strength become successively weaker for the higher $m_{\sigma}$.~Further the $\lbrace{\beta_{\cep}|_{(m_{\sigma}=400\text{MeV})}-\beta_{\cep}|_{(m_{\sigma}=500\text{MeV})}\rbrace}<$ $\lbrace{\beta_{\cep}|_{(m_{\sigma}=500\text{MeV})}-\beta_{\cep}|_{(m_{\sigma}=600\text{MeV})}\rbrace}$ indicates that the chiral transition strength becomes only marginally weak when $m_{\sigma} = 400 \rightarrow 500$ MeV whereas  it gets significantly weaker when $m_{\sigma} = 500 \rightarrow 600$ MeV.  

\begin{figure*}[htb]
	\subfigure[\ Columbia plot: RQM-S model]{
		\label{fig2a} 
		\begin{minipage}[b]{0.49\textwidth}
			\centering
			\includegraphics[width=\linewidth]{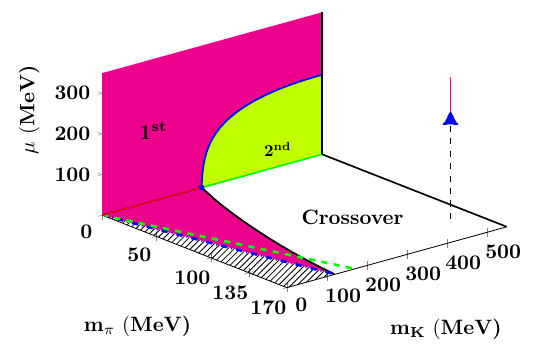}
	\end{minipage}}
	\hfill
	\subfigure[\ Columbia plot: RQM-I model]{
		\label{fig2b} 
		\begin{minipage}[b]{0.49\textwidth}
			\centering 
			\includegraphics[width=\linewidth]{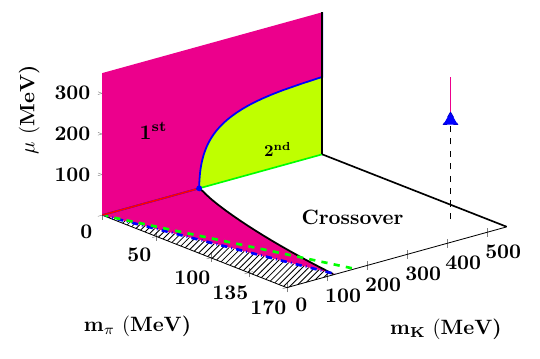}
	\end{minipage}}
	\caption{~The  chiral transitions for the $m_{\pi}=0$ and  $\mu=0$ are depicted respectively in the $\mu-m_{K}$ and $m_{\pi}-m_{K}$ plane.~The results for the RQM-S model when its parameters are fixed using the large $N_c$ standard U(3) ChPT inputs,~are presented in the left panel (a) whereas the right panel (b) shows the results for the RQM-I model where its parameters are fixed using the infrared regularized U(3) ChPT inputs.~The second order $Z(2)$ critical line in solid black color that separates the crossover from the first order transition region in the Fig.(a) and (b),~intersects the $m_{\pi}=m_{K}$ green dash line respectively at the critical pion mass $m_{\pi}^{c}=134.51$ and $m_{\pi}^{c}=134.16$ MeV and it terminates respectively at the terminal pion mass $m_{\pi}^{t}=169.25$ and $m_{\pi}^{t}=168.39$ MeV.~Solid blue line of the tricritical points which starts respectively at the blue dot $m_{K}^{\tcp}=248.8$ and  $m_{K}^{\tcp}=242.6$  MeV in the Fig.(a) and (b),~separates the second  and first order regions.~The $h_{y0}$ is negative in the dashed area below the dash dot blue line for the strange chiral limit $h_{y0}=0$.~The black dashed vertical line at the physical point in the Fig(a) and (b),~shows the crossover transition which ends at the critical end point in blue triangle and the solid red line shows the first order transition.~The scalar $\sigma$ mass $m_{\sigma}=400$ MeV. }
	\label{fig:mini:fig2} 
\end{figure*}

For a large $m_{\sigma}=750$ MeV in the RQM-S model,~one finds the critical end point masses as the $(m_{\pi,\cep}^*,m_{K,\cep}^*)=(34.52,~124.06)$ MeV for the $\cep$ ratio $\beta_{\cep}|_{(m_{\sigma}=750 \ \text{MeV} : \text{RQM-S})}=0.25013$.~When the $\sigma$ mass becomes very large $m_{\sigma}=800$ MeV,~the
 second order $Z_{2}$ critical point masses become significantly small as $(m_{\pi,\cep}^*,m_{K,\cep}^*)=(32.18,~115.65)$ MeV for the smallest $\cep$ ratio of $\beta_{\cep}|_{(m_{\sigma}=800 \ \text{MeV} : \text{RQM-S})}=0.23316$.~Here it would be appropriate to compare our results in the RQM-S model with the 
e-MFA:QM model work of Ref.~\cite{Resch} where the parameter $\alpha$ in their study is related to the $\beta$ in our work as the $\beta=\sqrt{\alpha}$.~The second order $Z_{2}$ critical point in the e-MFA:QM model study for the $m_{\sigma}=530$ MeV is obtained for a significantly smaller $\beta_{\cep}|_{(m_{\sigma}=530  \text{MeV} : \text{e-MFA:FRG})}=\sqrt{\alpha_{c}}=0.2$  which corresponds to quite small pion and kaon masses  $(m_{\pi,\cep}^*,m_{K,\cep}^*)=(27.6,~99.2)$ MeV.~Thus even when the chiral transition gets strongly softened and diluted due to a very high  $\sigma$ mass ($m_{\sigma}=800$ MeV) in the RQM-S model,~its strength remains stronger than the strength of chiral transition in the e-MFA:QM model~\cite{Resch} and the curvature mass based parametrized QMVT model~\cite{vkt25II} study for the moderate $m_{\sigma}=530$ MeV because $\beta_{\cep}|_{(m_{\sigma}=800 \ \text{MeV} : \text{RQM-S})}>\beta_{\cep}|_{(m_{\sigma}=530  \text{MeV} : \text{e-MFA:FRG})}\sim \beta_{\cep}|_{(m_{\sigma}=530  \text{MeV} : \text{QMVT})}$. The diluted strength of the chiral transition in response to the increase of $\sigma$ meson masses in the RQM-S model can be gauged from the $\beta_{\cep}$ values which change from $\beta_{\cep}|_{(m_{\sigma}=530  \text{MeV} : \text{RQM-S})}=0.36778$ \cite{vkt25I} to $\beta_{\cep}|_{(m_{\sigma}=800 \ \text{MeV} : \text{RQM-S})}=0.23316$ when the $m_{\sigma}=530\rightarrow800$ MeV.~The softening of the chiral phase transition due to the effect of  quark one-loop vacuum fluctuations is quite moderate in our RQM model studies whether one uses the ChPT input set S or I.~On account of the very large softening effect of the quark one-loop vacuum correction in the e-MFA:QM model work in the Ref.~\cite{Resch} and the curvature mass based parametrized QMVT model study in the Ref.~\cite{vkt25II},~one finds significantly smaller regions or range of parameters in which the first order chiral transition occurs.~For having the complete picture of the results,~one should also look at the QM model results obtained after neglecting the Dirac's sea contribution under the s-MFA and taking the  $m_{\sigma}=800\text{ MeV}$ such that the SCSB does not get lost in the work of Ref.~\cite{Schaefer:09} where the Fig.~(9) shows that the chiral crossover transition terminates at the second order transition for a significantly larger ratio $\beta_{cep}=.488$ that corresponds to noticeably larger $\pi \text{ and } K$ masses $(m_{\pi,\cep}^*,m_{K,\cep}^*)=(67.34,~242.05)$ MeV.~The Table~(\ref{tab:table2}) presents an exhaustive comparison of the results of the RQM-S and RQM-I model for the $m_{\sigma}=400,500\text{ and } 600$ MeV and the RQM-S model for the $m_{\sigma}=750\text{ and } 800$ MeV in the present work.~The results of the RQM-S model study in Ref.\cite{vkt25I},~the e-MFA:QM model FRG study in Ref.~\cite{Resch} and the QMVT model study in  Ref.~\cite{vkt25II} for the $m_{\sigma}=530$ MeV,~are also shown.~The results of s-MFA QM model study in Ref~\cite{Schaefer:09} for $m_{\sigma}=800$  MeV and  different Lattice QCD results with proper references are also presented in the Table~(\ref{tab:table2}).

\subsection{\bf{Comparison of  RQM-S and I model Columbia plots for $\bf{m_{\sigma}}=400,500,600\text{ MeV}$ and the RQM-S model plot description for $\bf{m_{\sigma}}=750,800\text{ MeV}.$}}
\label{Colplotmeson}

One needs to implement different conditions, other than simply reducing the $m_{\pi} \text{ and } m_{K}$ by tuning the only one parameter $\beta$, for computing different regions of the Columbia plot.~The tricritical line is defined as the locus of tricritical points in the vertical  $\mu-m_{K}$ plane where the second order chiral phase transition ends to become the first order transition. This line is computed by implementing  the condition of light chiral limit ($m_{\pi}=0\implies h_{x0}=0$).~The  $Z(2)$ critical end points ($\cep$) are located after fixing the $m_{\pi,\cep}$ and  finding the corresponding $m_{K,\cep}$ at which the crossover transition  for the $m_{K}>m_{K,\cep}$,~turns second order to become first order for the $m_{K}<m_{K,\cep}$.~Demarcating the first order transition from the crossover transition regions, the chiral critical line is constituted by the critical end points in the horizontal $m_\pi-m_{K}$ plane at the $\mu=0$. This line has been depicted by the solid black line in the figures to follow. The implementation of strange chiral limit  $h_{y0}=0=h_{y}$ condition,~gives the blue dash dotted lines in the figures below. The strange direction chiral symmetry breaking strength $h_{y}$ is negative in 
the shaded regions of the figures. The $SU(3)$ symmetric chiral limit lines for the $m_{\pi}=m_{K}$,~are plotted by the green dash line in the figures below. 

\begin{figure*}[htb]
\subfigure[\ Columbia plot: RQM-S model]{
\label{fig3a} 
\begin{minipage}[b]{0.49\textwidth}
			\centering
			\includegraphics[width=\linewidth]{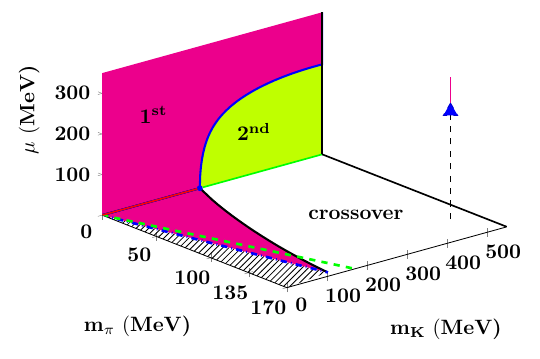}
	\end{minipage}}
	\hfill
	\subfigure[\ Columbia plot: RQM-I model]{
		\label{fig3b} 
		\begin{minipage}[b]{0.49\textwidth}
			\centering 
			\includegraphics[width=\linewidth]{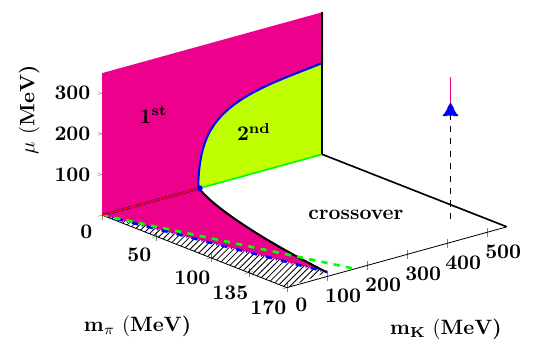}
	\end{minipage}}
	\caption{~The chiral transitions in the $\mu-m_{K}$ plane for the $m_{\pi}=0$ and the $m_{\pi}-m_{K}$ plane for the $\mu=0$ when the $m_{\sigma}=500$ MeV are presented in the left panel (a) for the RQM-S model and the right panel (b) for the RQM-I model.~The points,~lines,~crossover, first and second order chiral  transition regions and features are defined similar to the Fig.(\ref{fig:mini:fig2}).~The critical quantities ($m^{\tcp}_{K},\ m_{\pi}^{t}, \ m_{\pi}^{c}$) in the RQM-S and RQM-I model are (244.1, 164.65, 130.71) MeV and (240.1, 164.02, 130.84) MeV in respective order.} 
\label{fig:mini:fig3} 
\end{figure*}

The s-MFA QM model study,~where the quark one-loop vacuum term is dropped,~shows first order chiral transition in the light chiral limit ($m_{\pi}=0$) independent of the value of $m_{K}$ and  the strength of the $U_A(1)$ anomaly  \cite{Schaefer:09}.~Confirming the predictions of the Ref. \cite{rob}, the Chiral  phase transition for the light chiral limit turns out to be second order when the  $m_{K} \ge 496$ MeV due to the effect of the quark one-loop vacuum fluctuations in the present RQM-S and I model work as in the e-MFA:FRG QM model \cite{Resch} or the QMVT model \cite{vkt25II} studies.~The second order chiral phase transition that occurs on the points of the $m_{K}$ axis  in the light chiral limit for $m_{K}<496$ MeV,~belongs to the O(4) universality class.~This O(4) second order line  terminates at the  tricritical point $m_{K}^{\tcp}$ where the transition turns first order and becomes stronger till the chiral limit is reached when $m_{K}=0$.~The tricritical point $m_{K}^{\tcp}$ is marked by  the blue dot  in the figures below.~The solid black color chiral critical line terminates on the strange chiral limit line at the terminal 
pion mass $m_{\pi}^{t}$ beyond which the transition is a smooth crossover everywhere  in the $m_{\pi}-m_{K}$ plane.~The  intersection point of the  SU(3) symmetric $m_{\pi}=m_{K}$ chiral limit path depicted by the green dash line with the chiral critical line in the solid black color,~gives the critical pion mass $m_{\pi}^{c}$ where the boundary of first order region ends in the $m_{\pi}-m_{K}$ plane.

\begin{figure*}[htb]
\subfigure[\ Columbia plot: RQM-S model]{
\label{fig4a} 
\begin{minipage}[b]{0.49\textwidth}
			\centering
			\includegraphics[width=\linewidth]{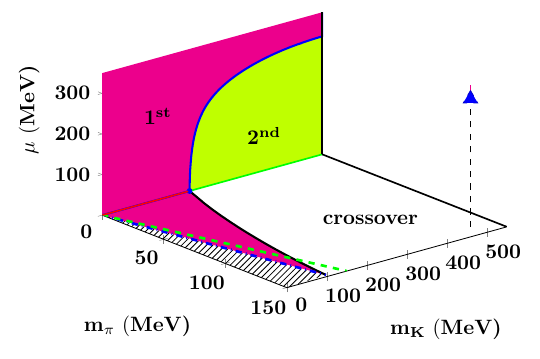}
	\end{minipage}}
	\hfill
	\subfigure[\ Columbia plot: RQM-I model]{
		\label{fig4b} 
		\begin{minipage}[b]{0.49\textwidth}
			\centering 
			\includegraphics[width=\linewidth]{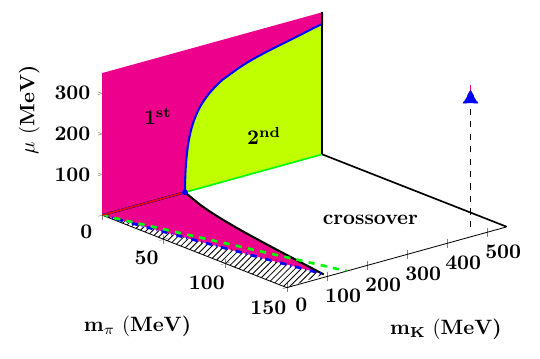}
	\end{minipage}}
	\caption{~Chiral transitions in the $\mu-m_{K}$ plane at $m_{\pi}=0$ and the $m_{\pi}-m_{K}$ plane at $\mu=0$  for the $m_{\sigma}=600$ MeV,~are presented in the left panel (a) for the RQM-S model and the right panel (b) for the RQM-I model.~The points,~lines,~crossover, first and second order chiral  transition regions and features are defined similar to the Fig.(\ref{fig:mini:fig2}).~The critical quantities ($m^{\tcp}_{K},\ m_{\pi}^{t}, \ m_{\pi}^{c}$) are respectively (218.5, 147.66, 117.26) MeV and (207.05, 146.35, 118.01) MeV in the RQM-S and RQM-I model.}
\label{fig:mini:fig4} 
\end{figure*}

The  identification of the $\tcp$ at $m_{K}^{\tcp}$,~serves the two fold  quantitative measure. The maximum extent of the first order transition region below the crossover region in the horizontal $m_{\pi}-m_{K}$ plane of the Columbia plot at $\mu=0$ and the minimum extent of the first order above the second order transition region in the vertical $\mu-m_{K}$ plane of the Columbia plot at $m_{\pi}=0$,~both are quantified by the numerical value $m_{K}^{\tcp}$ of the $\tcp$.~The terminal pion mass $m_{\pi}^{t}$  quantifies the maximum spread of the first order transition on the $m_{\pi}$ axis between the crossover region above and the unphysical $h_{y}=0$ region below it while the critical pion mass $ m_{\pi}^{c}$ serves as the quantifying measure of the maximum spread of the first order transition region on the $m_{\pi}$ axis covered between the crossover region above and the SU(3) symmetric  $m_{\pi}=m_{K}$ chiral limit line below it in the horizontal $m_{\pi}-m_{K}$ plane of the Columbia plot at $\mu=0$.~The set of critical quantities $(m_{K}^{\tcp},~m_{\pi}^{t},~m_{\pi}^{c}) \equiv (248.8,~169.25,~134.51)$ MeV in the Fig.~\ref{fig2a} and $(m_{K}^{\tcp},~m_{\pi}^{t},~m_{\pi}^{c}) \equiv (242.6,~168.39,~134.16)$ MeV in the  Fig.~\ref{fig2b} obtained respectively in the RQM-S model and RQM-I model computations,~signify the largest spread of the first order regions  in both the horizontal and vertical planes of the Columbia plots for the respective model scenarios when the scalar $\sigma$ meson mass is taken as the $m_{\sigma}=400$ MeV.~Note that the extent of the first order regions in the vertical and horizontal planes of the Columbia plot are slightly larger in the RQM-S model than what is obtained in the corresponding RQM-I model computations.~When the scalar $\sigma$ meson mass is sequentially increased to $m_{\sigma}=500 \text{ and }600$ MeV,~the sets of the critical quantities are obtained respectively as $(m_{K}^{\tcp},m_{\pi}^{t},m_{\pi}^{c}) \equiv (244.1,164.65,130.71)$ $\lbrace{(240.1,164.02,130.84) \rbrace} $ MeV in the Fig.~\ref{fig3a} $\lbrace{\text{Fig.~\ref{fig3b}}\rbrace}$ and $(m_{K}^{\tcp},~m_{\pi}^{t},~m_{\pi}^{c}) \equiv (218.5,147.66,117.26)$ $\lbrace{(207.05,146.35,118.01)\rbrace}$ MeV in the Fig.~\ref{fig4a} $\lbrace{\text{ Fig.~\ref{fig4b}} \rbrace}$ for the RQM-S model $\lbrace{ \text{RQM-I model} \rbrace}$.~On increasing the $m_{\sigma}$ from $400\rightarrow500$ MeV,~the first order regions in the $m_{\pi}-m_{K}$ and $\mu-m_{K}$ planes of Columbia plots get slightly reduced as indicated by the marginal reduction of the critical quantities by $(\Delta m_{K}^{\tcp},~\Delta m_{\pi}^{t},~\Delta m_{\pi}^{c}) \equiv (4.7,~4.6,~3.8)$ MeV in the  RQM-S model and $(\Delta m_{K}^{\tcp},~\Delta m_{\pi}^{t},~\Delta m_{\pi}^{c}) \equiv (2.5,~4.37,~3.32)$ MeV in the  RQM-I model.~When the $m_{\sigma}$ changes from  $500 \rightarrow 600$ MeV,~the ability to produce the first order transition over the large regions in the $m_{\pi}-m_{K}$ and $\mu-m_{K}$ planes of the RQM-S and I model Columbia plots,~gets noticeably compromised in comparison to the case of  $m_{\sigma}=400 \rightarrow 500$.~One finds moderate but noteworthy reduction of $(\Delta m_{K}^{\tcp},~\Delta m_{\pi}^{t},~\Delta m_{\pi}^{c}) \equiv (25.6,~16.99,~13.45)$ MeV for the critical quantities computed in the RQM-S model while the RQM-I model critical quantities suffer from a larger reduction of $(\Delta m_{K}^{\tcp},~\Delta m_{\pi}^{t},~\Delta m_{\pi}^{c}) \equiv (33.05,~17.67,~12.83)$ MeV when the $m_{\sigma}$ goes from the $500 \rightarrow 600$ MeV.~In comparison to the RQM-S Columbia plot in the Fig.~\ref{fig4a},~the horizontal ($m_{\pi}-m_{K}$) and vertical ($\mu-m_{K}$) planes,~show distinctly smaller regions of the first order chiral transition in the Fig.~\ref{fig4b} for the RQM-I Columbia plot when the $m_{\sigma}=600$ MeV.

The first order region near the chiral limit in the $\mu-m_{K}$ plane at the $m_{\pi}=0$,~which is very large when the $m_{\sigma}=400\text{ and }500$ MeV  respectively in the Fig.~\ref{fig2a} (Fig.~\ref{fig2b}) and Fig.~\ref{fig3a} (Fig.~\ref{fig3b})  and though smaller but still large when the  $m_{\sigma}=600$ MeV in the Fig.~\ref{fig4a}(Fig.~\ref{fig4b}) for the RQM-S model (RQM-I model),~increases when the chemical potential $\mu$ is increased.~The tricritical points chiral line  lying below the first order and above the second order region,~starts at $\mu=0$ from the  $m_{K}^{\tcp}=$248.8(242.6) MeV with a large positive slope in the Fig.~\ref{fig2a} (Fig.~\ref{fig2b})  for the RQM-S(RQM-I) model when the $m_{\sigma}=400$ MeV and its successively decreasing  positive slope for the larger $\mu$ and $ m_{K}$,~ gradually approaches zero showing good saturation (near saturation) for the $m_{K}>500$ MeV.~The $\mu \text{ rises from }191.12 (184.7 ) \rightarrow 195.09 (188.25)$ MeV on the  tricritical line of the Fig.~\ref{fig2a} (Fig.~\ref{fig2b}) when the $m_{K}$ goes from $450\rightarrow 500$ MeV in the RQM-S (RQM-I) model giving 0.079 (0.071) as its slope which becomes very small to .016 (small to 0.049) when the $\mu=195.09(188.25)$ MeV at $m_{K}=500$ MeV becomes $\mu=195.91(190.7)$ MeV at $m_{K}=550$ MeV.~Since the slope of the RQM-S model tricritical line in the Fig.~\ref{fig2a}  becomes zero in the range $m_{K}=530 \text{ to }550$ MeV  where $\mu$ remains constant as $\mu=195.91$ MeV,~it shows better saturation than the tricritical line of the RQM-I model in the Fig.~\ref{fig2b}.~On increasing the $\sigma$ mass to the $m_{\sigma}=500$ MeV,~the RQM-S (RQM-I) model tricritical line   starting with large positive slope from the $m_{K}^{\tcp}=244.1(240.1)$ MeV at $\mu=0$ in the Fig.~\ref{fig3a} (Fig.~\ref{fig3b}) shows (does not show) the saturation pattern like the one that we get for the $m_{\sigma}=400$ MeV in the RQM-S (RQM-I) model.~The slope 0.1634(0.168) between the points $(\mu,m_{K})=(209.11,400.0)\lbrace{(204.6,400.0)\rbrace}$ MeV and $(\mu,m_{K})=(217.28,450.0)\lbrace{(213.0,450.0)\rbrace}$ MeV on the tricritical line of the RQM-S (RQM-I) model
in the Fig.~\ref{fig3a} (Fig.~\ref{fig3b}), decreases to the 0.0764(0.1222) when one goes from the point $(\mu,m_{K})=(217.28,450.0)\lbrace{(213.0,450.0)\rbrace}$ MeV to the point $(\mu,m_{K})=(221.1,500.0)\lbrace{(219.11,500.0)\rbrace}$ MeV and it becomes 0.011(0.1158) between the point $(\mu,m_{K})=(221.1,500.0)\lbrace{(219.11,500.0)\rbrace}$ and $(\mu,m_{K})=(221.65,550.0)\lbrace{(224.9,550.0)\rbrace}$ MeV.~With the zero slope for the range $m_{K}=520 \text{ to }550$ MeV  where $\mu$ is very close to the constant $\mu=221.7$ MeV,~the RQM-S model tricritical line saturates properly.~The RQM-I model tricritical line does not show saturation having a noticeable non-zero slope 0.1158 close to $m_{K}=550$ MeV.

When the $m_{\sigma}$ is increased to 600 MeV,~the RQM-S model tricritical line that begins in the Fig.~\ref{fig4a} with large positive slope from the $m_{K}^{\tcp}=218.5$ MeV at $\mu=0$,~repeats the saturation trend that one observes in the RQM-S model for the case of $m_{\sigma}=400\text{ and }500$ MeV
respectively in the Fig.~\ref{fig2a} and \ref{fig3a}.~The RQM-S model tricritical line slope 0.1864 in the Fig.~\ref{fig4a} between the points $(\mu,m_{K})=(274.13,400.0)\text{ and }(283.45,450.0)$ MeV  decreases noticeably to 0.1042 when the $(\mu,m_{K})$ shifts from the point $(283.45,450.0) \rightarrow (288.66,500.0)$ MeV and the slope becomes very small to 0.0408 when the $\mu \text{ changes from }288.66\rightarrow290.7$ MeV corresponding to the shift of $m_{K}\text{ from }500\rightarrow550$ MeV.~The RQM-S model tricritical line shows proper saturation as its slope 0.01 is negligible in the range $m_{K}=530 \text{ to }550$ MeV  where  $\mu=290.2\rightarrow290.7$ MeV.~In complete contrast to the RQM-S model,~the tricritical line of the RQM-I model,~that starts in the Fig.~\ref{fig4b} with large positive slope from a noticeably smaller $m_{K}^{\tcp}=207.05$ MeV at $\mu=0$,~not only does not show the saturation trend but it becomes distinctly and significantly divergent for the $m_{K}>400$ MeV.~The quite large slope 0.229 of the RQM-I model tricritical line between the points $(\mu,m_{K})=(287.82,400.0)$ MeV and (299.27,450.0) MeV in the Fig.~\ref{fig4b},~decreases only marginally to 0.2188 when one goes from the point  $(\mu,m_{K})=(299.27,450.0)\rightarrow(310.21,500.0)$ MeV and the slope increases  again to 0.2228 when the $\mu \text{ changes from }310.21\rightarrow321.35$ MeV for the increasing $m_{K} \text{ from }500\rightarrow550$ MeV.


Here,~it is relevant to point out the comparison of RQM-S and RQM-I model Columbia plots as reported recently in Ref~\cite{vkt25I}  for the case of $m_{\sigma}=530$ MeV.~The use of input set $\text{M}_{\eta}$-I and the input set $\text{M}_{\eta}$-II in the Ref~\cite{vkt25I} for performing the chiral limit study,~has been termed in the present study as the RQM-I model and the RQM-S model.~The tricritical lines in the $\mu-m_{K}$ plane of the RQM-I and RQM-S model Columbia plots,~start respectively from the $m_{K}^{\tcp}=235.7$ MeV in the Fig.(4a) and $m_{K}^{\tcp}=239.58$ MeV  in the Fig.(5) of the Ref.~\citep{vkt25I}.~The RQM-I and RQM-S model tricritical lines,~are almost overlapping for the $m_{K}<450$ MeV.~Showing noticeable  difference after the $m_{K}\ge450$,~the RQM-S model tricritical line shows saturation for the range $m_{K}=450-550$ MeV while the RQM-I model tricritical line turns somewhat divergent.~The RQM-S model tricritical line has the slope of only 0.078 in going from the point $(m_{K},\mu)=(450.0,230.22) \rightarrow (500.0,234.12)$ MeV whereas the slope 0.1478 of the RQM-I model tricritical line between the point $(m_{K},\mu)=(450.0,228.4)$ MeV and $(500.0,235.79)$ MeV,~is nearly  double of the above slope of the RQM-S model line.~The slope of RQM-I model line slightly increases  to 0.1566 when it moves  from the point $(m_{K},\mu)=(500.0,235.79)\rightarrow(550.0,243.62)$ MeV whereas the decreasing average slope between the point $(m_{K},\mu)=(500.0,234.12)\text{ and }(550.0,234.8)$ MeV is only 0.0136 for the RQM-S model tricritical line whose slope turns zero in the range $m_{K}=520.0-550.0$ MeV as the  chemical potential is almost constant at $\mu=234.8$ MeV  in this range.~Thus the  divergence that was observed in the Ref.~\cite{vkt25I} for the RQM-I model tricritical line for the case of $m_{\sigma}=530$ MeV,~becomes noticeable and distinctly large divergence when the $m_{\sigma}$ becomes 600 MeV as discussed in the last paragraph.

\begin{figure*}[htb]
\subfigure[\ Columbia plot: RQM-S model for $m_{\sigma}=750$ MeV]{
\label{fig5a} 
\begin{minipage}[b]{0.49\textwidth}
			\centering
			\includegraphics[width=\linewidth]{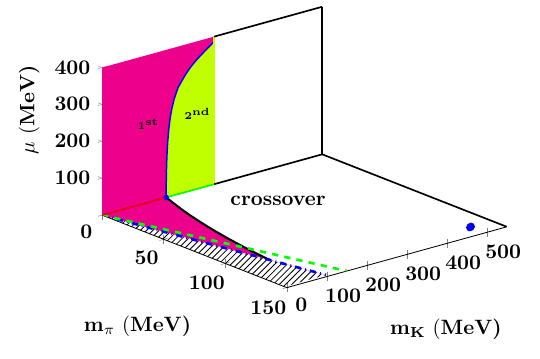}
	\end{minipage}}
	\hfill
	\subfigure[\ Columbia plot: RQM-S model for $m_{\sigma}=800$ MeV]{
		\label{fig5b} 
		\begin{minipage}[b]{0.42\textwidth}
			\centering 
			\includegraphics[width=\linewidth]{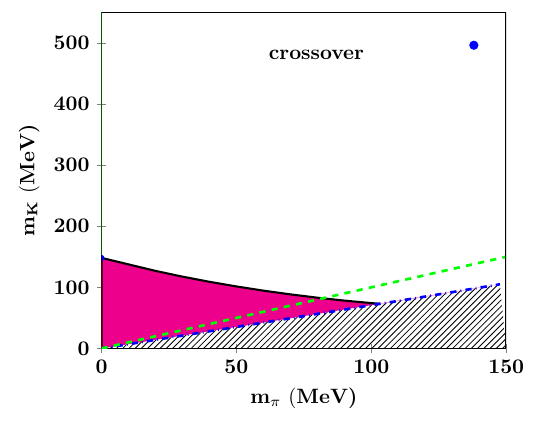}
	\end{minipage}}
\caption{~The left panel (a) depicts the RQM-S model chiral transition in the horizontal plane $m_{\pi}-m_{K}$ at $\mu=0$ and the vertical $\mu-m_{K}$ plane at $m_{\pi}=0$ when the $m_{\sigma}=750$ MeV.~~The horizontal plane $m_{\pi}-m_{K}$ at $\mu=0$ in the right panel (b) presents the  chiral transition in RQM-S model for the case of $m_{\sigma}=800$ MeV.~~The points,~lines,~crossover,~first and second order chiral  transition regions and features are defined similar to the Fig.(\ref{fig:mini:fig2}).~The tricritical line demarcating the first and second order colored regions,~is computed only upto the point $(m_{K},\mu)=(280,386.8)$ MeV in the Fig.~(a) because the chemical potential becomes very high for a significantly  smaller kaon mass in the $\mu-m_{K}$ plane.~The RQM-S model critical quantities ($m^{\tcp}_{K},\ m_{\pi}^{t}, \ m_{\pi}^{c}$) are respectively (159.85, 110.40, 87.90) MeV and (148.2, 103.19, 82.07) MeV when $m_{\sigma}=750 \text{ and } 800$ MeV.}
\label{fig:mini:fig5} 
\end{figure*}

The large initial positive slope at the beginning of the tricritical lines of the RQM-S and RQM-I model respectively in the Fig.~\ref{fig2a}, \ref{fig3a}, \ref{fig4a}, and  the Fig.~\ref{fig2b}, \ref{fig3b}, \ref{fig4b},~is consequence of the fact that the  chiral critical surface,~in the three dimensions of the $m_{\pi}-m_{K}-\mu$ for the RQM model set up,~has a positive curvature whose intersection with the $\mu-m_{K}$ plane at the $m_{\pi}=0$,~gives the tricritical line.~The chiral critical lines of the $Z(2)$ critical end points which make the chiral critical surface,~are its intersection with the  $\mu-m_{K}$ planes for different  $m_{\pi}\neq0$.~Due to its positive curvature~\cite{forcrd2},~when the chiral critical surface intersects the dashed black vertical line for the chiral crossover transition at the physical point,~the existence of a critical end point gets marked  by a solid blue arrow in the Figs.~(\ref{fig:mini:fig2}),(\ref{fig:mini:fig3}) and (\ref{fig:mini:fig4}).~The physical point $\cep$ for the $m_{\sigma}=400,500\text{ and }600$ MeV,~lies respectively at the successively higher $\mu_{\cep}=243.77,265.62 \text { and } 315.85$ MeV in the Fig.~\ref{fig2a}, \ref{fig3a} and \ref{fig4a} for the RQM-S model and $\mu_{\cep}=243.29, 265.3 \text{ and } 315.74$ MeV in the Fig.~\ref{fig2b}, \ref{fig3b} and  \ref{fig4b} for the RQM-I model.~With nearly similar shape for the  $m_{\sigma}=400$ MeV,~the RQM-S and RQM-I model chiral critical surfaces are positioned in comparatively lower chemical potential ranges where the  tricritical lines in the $\mu-m_{K}$ plane  of both the models,~show the similar saturation pattern after $m_{K}>400$ MeV,~though the saturation behavior of the RQM-I model tricritical line is somewhat approximate.~While shifting  to the higher chemical potentials for the $m_{\sigma}=500 \text{ and }530$ MeV,~the chiral critical surface for the RQM-I model gets lifted to relatively higher $\mu$ near ($m_{\pi}=0,m_{K}>400$     MeV) as its tricritcal line picks up the diverging trend after the $m_{K}>400$ MeV while the RQM-S model chiral critical surface gets lifted up evenly in the $\mu$ direction since its tricritcal line shows the consistent saturation pattern  after the $m_{K}>400$ MeV.

~It is worth emphasizing that even though the RQM-S and RQM-I model  $\cep$ coordinates,~for the physical point and $m_{\sigma}=600$ MeV,~respectively at the ($\mu_{\cep},\text{T}_{\cep}$)=(315.85,21.96) and (315.74,22.1) MeV have almost no difference,~the RQM-I model chiral critical line at the physical point is part of such a chiral critical surface that gets a noticeably larger $\mu$ direction lift up in the vicinity of the ($m_{\pi}=0,m_{K}>400$ MeV) in its overall  positioning  at significantly higher $\mu$.~The above mentioned  tilting in the critical surface,~ends up giving the RQM-I model tricritical line that diverges significantly after the $m_{K}>400$ MeV in the $\mu-m_{K}$ plane at $m_{\pi}=0$ whereas the chiral critical surface of the RQM-S model has a proper evenly lifted up shape in the $\mu$ direction since it connects its chiral critical line at the physical point to the tricritical line which shows the expected saturation pattern for the $m_{K}>400$ MeV in the $\mu-m_{K}$ plane at the $m_{\pi}=0$.~Note that the saturation behavior for the tricritical line in the $\mu-m_{K}$ plane (for whatever be the value of $m_{\sigma}$),~is consistent and desirable on the  physical grounds as the 2+1 flavor tricritical line is expected to be connected to  the tricritical point of the two flavor chiral limit \cite{hjss} at some higher value of the $\mu$ and $m_{K}$.~Therefore it is reasonable to conclude that when one uses the large $N_{c}$ standard U(3) ChPT scaling relations to determine the ($m_{\pi},m_{K}$) dependence of the $f_{\pi}, f_{K} $ and $M_{\eta}^2$ away from the physical point,~the RQM-S model gives the improved and more refined results for the chiral limit studies for all values of the $\sigma$ mass in the range 400-600 MeV.

The Columbia plots are drawn after increasing the $m_{\sigma}$ successively to the $400,500 \text{ and }600$ MeV in order to see how the shifting critical end points present in the $\mu-\text{T}$ plane of the RQM model phase diagrams for the physical point parameters~\cite{vkkr23,skrvkt24},~are associated with and influenced by the change in the chiral critical surface and the tricritcal line for the light chiral limit.~The chiral crossover transition line gets extended and the first order chiral transition line shrinks in the $\mu-\text{T}$ plane when the $\sigma$ mass $m_{\sigma}$ increases and the position of the $\cep$ shifts to the right lower corner of the phase diagrams.~The strength of the chiral transition becomes so weak that it turns crossover everywhere in the $\mu-\text{T}$ plane and the $\cep$  altogether disappears from the phase diagrams for the physical point when the $m_{\sigma}>600$ MeV.~Since the RQM-S model  gives better and accurate results for the chiral limit studies,~it is worthwhile to draw the Columbia plots for the $m_{\sigma}=750$ and 800 MeV.~It will be interesting to compute the tricritical line for $m_{\pi}=0$ in the $\mu-m_{K}$ plane for the $m_{\sigma}=750$ MeV.~The above tricritical line will give us a picture of shrinkage in the shape of the chiral critical surface that exists only for a small range of the $m_{\pi}-m_{K}$ and   terminates well before the $m_{\pi}=138$ MeV,~hence does not have any chiral critical line that lies above the physical point on its other end  since the chiral transition is crossover all over.

\begin{figure*}[htb]
\subfigure[Vertical plane $\mu-m_{K}$ for $m_{\sigma}$=400,500,600 and 750 MeV]{
\label{fig6a} 
\begin{minipage}[b]{0.49\textwidth}
			\centering
			\includegraphics[width=\linewidth]{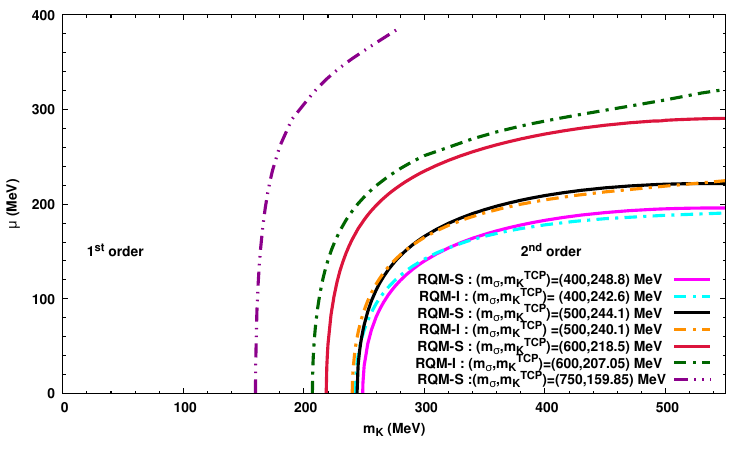}
	\end{minipage}}
	\hfill
	\subfigure[ Horizontal plane $m_{\pi}-m_{K}$ for $m_{\sigma}$=400,500,600,750,800 MeV]{
		\label{fig6b} 
		\begin{minipage}[b]{0.49\textwidth}
			\centering 
			\includegraphics[width=\linewidth]{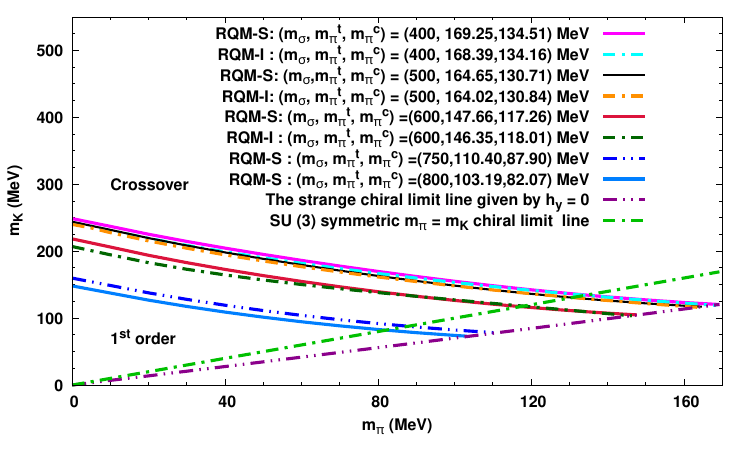}
	\end{minipage}}
\caption{Left panel (a) (Right panel (b)) presents the comparison of lines and regions in the vertical $\mu-m_{K}$, $m_{\pi}=0$ (horizontal $m_{\pi}-m_{K}$, $\mu=0$) planes of the three dimensional RQM-S and RQM-I model Columbia plots for the $m_{\sigma}=400,500,600,750 \text{ and } 800$ MeV in the Fig.~(\ref{fig2a}),~(\ref{fig2b}),~(\ref{fig3a}),~(\ref{fig3b}),~(\ref{fig4a}),~(\ref{fig4b}) and ~(\ref{fig5a}),~(\ref{fig5b}).~RQM-S and  RQM-I model tricritical lines for the $m_{\sigma}=400,500,600 \text{ and } 750$ MeV,~as marked and explained in the Fig.(a),~start from the $m_{K}^{\tcp}$ on the $m_{K}$ axis and demarcate the first order region above them on their left hand side  from the second order region below them on their right hand side.~RQM-S and RQM-I model Z(2) critical lines for the $m_{\sigma}=400,500,600,750 \text{ and } 800$ MeV,~as marked and explained in the Fig.(b), separating the crossover transition region above them from the first order region that lies below,~intersect the SU(3) symmetric $m_{\pi}=m_{K}$ chiral limit line at the critical pion mass $m_{\pi}^{c}$ and terminate on the strange chiral limit line satisfying  the $h_{y0}=0$,~at the terminal pion mass $m_{\pi}^{t}$.}
\label{fig:mini:fig6} 
\end{figure*}

Even when the strength of chiral transition is significantly weaker for the large
$m_{\sigma}=750$ MeV in the RQM-S model Columbia plot of the Fig.~\ref{fig5a},~the corresponding critical quantities  $(m_{K}^{\tcp},~m_{\pi}^{t},~m_{\pi}^{c}) \equiv (159.85,~110.40,~87.90)$ MeV,~are comparable to the critical quantities $(m_{K}^{\tcp},~m_{\pi}^{t},~m_{\pi}^{c}) \equiv (169,~110,~86)$ MeV obtained for the $m_{\sigma}=530$ MeV in the e-MFA QM model study done after switching off the quantum and thermal fluctuations of the mesons in the  FRG framework  of the Ref.\cite{Resch}.~The extent of the first order region in the horizontal $m_{\pi}-m_{K}$ plane of the Fig.~\ref{fig5a},~is similar to the area of the first order region seen in the $m_{\pi}-m_{K}$ plane of the Fig.(3a) in the Ref.~\cite{Resch} but the tricritical line in the vertical $\mu-m_{K}$ plane of the e-MFA:FRG QM model Columbia plot
shows proper saturation while the corresponding  tricritical line for the RQM-S model when $m_{\sigma}=750$ is strongly divergent.~The tricritcal line in the vertical $\mu-m_{K}$ plane begins from a smaller $m_{K}^{\tcp}=159.85$ MeV at $m_{\pi}=0$ in the Fig.~\ref{fig5a} with a very large positive slope but it shows a very large divergence and a smaller bending because the chemical potential becomes very high $\mu=386.8$ MeV for a significantly smaller kaon mass $m_{K}=280$ MeV.~Thus the chiral critical surface can be constructed only for the range $m_{K}^{\tcp}=159.85\rightarrow m_{K}=280$ MeV on the $m_{K}$ axis and the range  $m_{\pi}=0\rightarrow m_{\pi}^{t}=110.4$ MeV on the $m_{\pi}$ axis.~The quite smaller and shrunk chiral critical surface would lie far from the region that could  have possibly covered the physical point for some non-zero temperature and a large chemical potential,~hence one does not find any critical end point $\cep$ in the $\mu-\text{T}$ plane of the RQM-S model for physical point when the $m_{\sigma}=750$ MeV.~Since the tricritical lines in the $m_{\pi}=0$ vertical $\mu-m_{K}$ planes of the RQM-S model Columbia plots for the $m_{\sigma}=400,500\text{ and } 600$ MeV show consistent and proper saturation trend,~the other end of the corresponding chiral critical surfaces contain the chiral critical lines which pass over and cover the physical point,~hence the corresponding $\cep$ exists.~One finds a very large reduction of $(\Delta m_{K}^{\tcp},~\Delta m_{\pi}^{t},~\Delta m_{\pi}^{c}) \equiv (58.65,~37.26,~29.36)$ MeV for the critical quantities computed in the RQM-S model when the $m_{\sigma}$ goes from the $600 \rightarrow 750$ MeV.

The $m_{\pi}-m_{K}$ plane of the Columbia plot for the $\mu=0$ has been computed and presented in the Fig.~\ref{fig5b}  for the case of $m_{\sigma}=800$ MeV when the strength for producing the first order chiral transition in the RQM-S model becomes very weak.~The critical quantities $(m_{K}^{\tcp},~m_{\pi}^{t},~m_{\pi}^{c}) \equiv (148.2,~103.19,~82.07)$ MeV obtained in the RQM-S model when the $m_{\sigma}=800$ MeV,~indicate that the area of the first order region is smallest while the crossover transition region has the largest area in the $m_{\pi}-m_{K}$ when the results are compared with the corresponding RQM-S model results for the case of $m_{\sigma}=400,500,600\text{ and }750$ MeV.~When the $m_{\sigma}$ goes from the $750 \rightarrow 800$ MeV,~the computed critical quantities get reduced by the amount $(\Delta m_{K}^{\tcp},~\Delta m_{\pi}^{t},~\Delta m_{\pi}^{c}) \equiv (11.65,~7.21,~5.83)$ MeV.~Here it 
is relevant to mention that the Schaefer et. al. are finding the critical pion mass for the $SU(3)$ 
chiral limit as $m_{\pi}^{c}\equiv150$ MeV in the Ref. \cite{Schaefer:09} for the case of 
$m_{\sigma}=800$ MeV where the quark one-loop vacuum fluctuations are neglected  under the 
s-MFA.~Further as mentioned in the Ref. \cite{pisarski24},~the chiral matrix model study using 
the mean field analysis gives $m_{\pi}^{c}\equiv110$ MeV.

\begin{table*}[!htbp] 
\caption{The  quantities $m_{K}^{\tcp},\ m_{s}^{\tcp}, \ m_{\pi}^{t}, \ m_{ud}^{t}, \ m_{\pi}^{c},\ m_{ud}^{c}$ and the $\beta_{\cep}:m_{\pi,\cep}^*,m_{K,\cep}^* $ are presented in this table for the RQM-S model whose parameters away from the physical point,~are fixed using the large $N_c$ standard U(3) ChPT inputs and the RQM-I model whose parameters are fixed using the infrared regularized U(3) ChPT inputs.~Very recent results of these quantities computed in the curvature mass based parametrization of the quark meson model with quark one-loop vacuum term (QMVT-I) where the infrared regularized U(3) ChPT inputs are used for parameter fixing~\cite{vkt25II},~are also presented.~The results for some of these quantities obtained in the e-MFA:QM model FRG study for $m_{\sigma}=530$ MeV and the QM model s-MFA study for $m_{\sigma}=800$ MeV,~are also shown for comparison.~In the QM model s-MFA study,~the TCP and second order chiral transition are absent in the  light chiral limit $m_\pi=0$.~The change of the quantities $\Delta m_{K}^{\tcp},\ \Delta m_{s}^{\tcp}, \ \Delta m_{\pi}^{t}, \ \Delta m_{ud}^{t}, \ \Delta m_{\pi}^{c},\ \Delta m_{ud}^{c}$ and the $ \Delta \beta_{\cep}: \Delta m_{\pi,\cep}^*,\Delta m_{K,\cep}^* $ are also presented when the $m_{\sigma}$ sequentially changes from 400$\rightarrow$500 and 500$\rightarrow$600 (600$\rightarrow$750 and 750$\rightarrow$800) in the RQM-S and RQM-I model
( RQM-S model).~The SU(3) chiral limit results of the critical pion (light quark) mass $m_{\pi}^{c}$ ($m_{ud}^{c}$) obtained from different LQCD studies for three degenerate flavors at $\mu=0$,~are presented in the last two rows where the corresponding fermion implementation and the number  of time slices $N_t$ are also shown.}
	\label{tab:table2}
	\begin{tabular}{p{0.115\textwidth}| p{0.169\textwidth} |p{0.07\textwidth}| p{0.06\textwidth}|p{0.06\textwidth}| p{0.07\textwidth}|p{0.07\textwidth} |p{0.08\textwidth}| p{0.085\textwidth} | p{0.085\textwidth}| p{0.085\textwidth}}
		\hline 
		$m_{\sigma}$&Model&$m_{K}^{\tcp}$&$m_{s}^{\tcp}$&$m_{\pi}^{t}$ &$m_{ud}^{t}$ &$m_{\pi}^{c}$&$m_{ud}^{c}$&$\beta_{\cep}$& $m_{\pi,\cep}^{*} $ & $m_{K,\cep}^{*}$\\
		\hline 
		400 &RQM-S&248.8& 25.23 &169.25& 5.97 &134.51&3.80 &0.38738 & 53.46& 192.14\\
		    &RQM-I\cite{vkt25II}&242.6& 25.53 &168.39& 6.13 &134.16&3.84 &0.37963 & 52.39& 188.30\\
		    &QMVT-I\cite{vkt25II}&137.2& 8.06 &98.15& 2.05 &78.45&1.31 &0.2215 & 30.57& 109.88\\
		\hline
	500 & RQM-S &244.1&24.33&164.65&5.66 & 130.71&3.59&0.37665 &51.98 &186.82\\
	    & RQM-I &240.1&25.00&164.02&5.79 & 130.84&3.66&0.37159 &51.33 &184.49\\
	    \hline
	    600 & RQM-S &218.5&19.65&147.66& 4.57 & 117.26&2.90 &0.33634 &46.41 &166.82\\
	    & RQM-I &207.05&17.70&146.35&4.59 & 118.01&2.97&0.32445 &44.77 &160.93\\
		\hline
		750 & RQM-S & 159.85 & 10.65 &110.40&2.57 & 87.90&1.64 &0.25013 &34.52 &124.06\\
		\hline
		800 & RQM-S &148.2& 9.21 &103.19&2.25 & 82.07&1.43 &0.23316 &32.18 &115.65\\
		\hline
	 530	& RQM-S\cite{vkt25I} &239.58&23.47&161.35&5.44 & 128.04&3.45&0.36778 &50.9 &182.96\\
	    & RQM-I\cite{vkt25I} &235.7&24.08&160.75&5.58 & 128.38&3.51&0.3642 &50.26 &180.64\\
		&e-MFA:QM-FRG\cite{Resch}&169&- & 110&3.0 &86 &-& 0.20 & 27.6 &99.2\\
		&QMVT-I\cite{vkt25II}&124.7&6.64&90.03&1.72 & 72.12&1.1&0.2028 &27.98 &100.58\\
		\hline
		800&s-MFA:QM\cite{Schaefer:09}&No TCP&--& 177&- &150 &- & 0.488 & 67.34 &242.05\\
		\hline %
	\end{tabular}
\begin{tabular}{p{0.115\textwidth}| p{0.169\textwidth} |p{0.07\textwidth}| p{0.06\textwidth}|p{0.06\textwidth}| p{0.07\textwidth}|p{0.07\textwidth} |p{0.08\textwidth}| p{0.085\textwidth} | p{0.085\textwidth}| p{0.085\textwidth}}
		\hline 
		Change of $m_{\sigma}$&Model&$\Delta m_{K}^{\tcp}$&$\Delta m_{s}^{\tcp}$&$\Delta m_{\pi}^{t}$ &$\Delta m_{ud}^{t}$ &$\Delta m_{\pi}^{c}$&$\Delta m_{ud}^{c}$&$\Delta \beta_{\cep}$& $\Delta m_{\pi,\cep}^{*} $ & $\Delta m_{K,\cep}^{*}$\\
\hline 
400$\rightarrow$500 & RQM-S &4.7&0.9&4.6&0.31 & 3.8&0.21&0.01073 &1.48 &5.32\\
	    & RQM-I &2.5&0.53&4.37&0.34 & 3.32&0.18&0.00804 &1.06 &3.81\\
		\hline 
500$\rightarrow$600 & RQM-S &25.6&4.68&16.99&1.09 & 13.45&0.69&0.04031 &5.57 &20.0\\
	    & RQM-I &33.05&7.3&17.67&1.2 & 12.83&0.69&00.4714 &6.56 &23.56\\
		\hline 
600$\rightarrow$750 & RQM-S &58.65&9.0&37.26&2.0 & 29.36&1.26&0.08621 &11.89 &42.76\\
		\hline
750$\rightarrow$800 & RQM-S &11.65&1.44&7.21&0.32 & 5.83&0.21&0.01697 &2.34 &8.41\\
\hline
\end{tabular}		
\begin{tabular}{p{0.05\textwidth}| p{0.135\textwidth} |p{0.11\textwidth}| p{0.135\textwidth}|p{0.11\textwidth}| p{0.11\textwidth}|p{0.11\textwidth} |p{0.11\textwidth}| p{0.11\textwidth}}   
LQCD Study&$N_{t}=4$,Standard Staggered $m_{\pi}^c$ &$N_{t}=4$, p4 Staggered $m_{\pi}^c$&$N_{t}=6$,Standard Staggered $m_{\pi}^c$&$N_{t}=6$,Stout Staggered $m_{\pi}^c$&Wilson Clover $N_{t}=6$, $m_{\pi}^c$&$N_{t}=6$, HISQ Staggered $m_{\pi}^c$&Wilson Clover $N_{t}=8$, $m_{\pi}^c$& $m_{ud}^c$ : Möbius Domain Wall \\
\hline
Ref.&$\sim$290\ \cite{karsch2}&$\sim$67\ \cite{karsch3}&$\sim$150 \ \cite{forcrd}&$\sim$0 \ \cite{varnho}&$\sim$300\ \cite{jin}&$\le$50 \ \cite{Bazav}&$\le$170\ \cite{jin2}&$\le$4 \ \cite{zhang24}\\
\hline 
\end{tabular}
\end{table*}

The complete comparative perspective of change in the area of first order and second order chiral transition regions of the RQM-S and RQM-I model when the scalar $\sigma$ mass changes 
from $m_{\sigma}=400\rightarrow 750$ MeV,~have been presented in the Fig.~(\ref{fig6a}) after re-plotting the tricritical lines of the  vertical $\mu-m_{K}$ planes at $m_{\pi}=0$ of the RQM-S $\lbrace{\text{RQM-I}\rbrace}$ model Columbia plots in the Fig.~(\ref{fig2a}) $\lbrace{\text{(\ref{fig2b})}\rbrace}$,~(\ref{fig3a}) $\lbrace{\text{(\ref{fig3b})}\rbrace}$,~(\ref{fig4a})$\lbrace{\text{(\ref{fig4b})}\rbrace}$ and ~(\ref{fig5a}) respectively for the $m_{\sigma}=400,500,600 \text{ and } 750$ MeV.~One can clearly see that as the $m_{\sigma}$ increases,~the tricritical lines shift to higher chemical potentials  in general for both the models.~The RQM-S and RQM-I model tricritical lines are nearly overlapping when the $m_{\sigma}=400\text{ and }500$ MeV and one can see that both lines become almost flat for $m_{K}>450$ MeV showing saturation for the $m_{\sigma}=400$ MeV.~Picking up small divergence from below when the $m_{K}>450$,~the RQM-I tricritical line for the $m_{\sigma}=500$ MeV crosses the properly saturating RQM-S tricritical line  near the $m_{K} \sim 500$ MeV.~Being quite distinct and separate,~the RQM-I model tricritical line,~that starts from a smaller $m_{K}^{\tcp}=207.05$ MeV for the $m_{\sigma}=600$ MeV,~becomes quite divergent after $m_{K}>400$ MeV while the flattening RQM-S model tricritical line shows the proper saturation trend.~The RQM-S model tricritcal line for the case of $m_{\sigma}=750$ MeV that starts from a significantly smaller $m_{K}^{\tcp}=159.85$ MeV,~shows a very strong divergence and a smaller bending because one needs to take very high chemical potential  $\mu=386.8$ MeV such that the second order transition turns first order even when the kaon mass $m_{K}=280$ MeV  is significantly smaller,~the value upto which the tricritical line can be calculated.~In order to appreciate the proper context of the above discussed tricritical lines,~one has to recall that the $\cep$ does not exist in the RQM-S model phase diagram when the $m_{\sigma}=750$ MeV,~since the chiral transition is crossover all over in the $\mu-\text{T}$ plane whereas one finds $\cep$ in both the scenarios of  RQM-S and I model when the  $m_{\sigma}=400,500\text{ and } 600$ MeV.

The chiral critical solid black  lines of the $Z(2)$ critical end points in the horizontal $m_{\pi}-m_{K}$ planes (at $\mu=0$) of the RQM-S $\lbrace{\text{RQM-I}\rbrace}$ model Columbia plots in the Fig.~(\ref{fig2a})$\lbrace{\text{(\ref{fig2b})}\rbrace}$,~(\ref{fig3a})$\lbrace{\text{(\ref{fig3b})}\rbrace}$,~(\ref{fig4a})$\lbrace{\text{(\ref{fig4b})}\rbrace}$,~(\ref{fig5a}) and (\ref{fig5b}) respectively for the $m_{\sigma}=400, 500, 600, 750 \text{ and } 800$ MeV,~have been redrawn in one frame of the Fig.~(\ref{fig6b}) in order to show the relative shrinking in the area of first order chiral transition regions in response to the expanding chiral crossover transition regions in the RQM-S and RQM-I model when the scalar $\sigma$ mass changes from $m_{\sigma}=400\rightarrow 800$ MeV.~The SU(3) symmetric chiral limit lines for $m_{\pi}=m_{K}$ and strange chiral limit $h_{y}=0$ line are also plotted.~The line types of the RQM-S(I) model chiral critical lines for different $m_{\sigma}$,~chiral crossover transition and first order transition regions are marked and explained in the  Fig.~(\ref{fig6b}) where the numerical values of the critical quantities $m_{\pi}^{t}\text{ and }m_{\pi}^{c}$ in the RQM-S(I) model for different $m_{\sigma}$  are also given.

The $m_{\sigma}=400 (500)$ MeV case RQM-S model chiral critical line,~which lies  marginally above the corresponding critical line of the RQM-I model when the $m_{\pi}<80$ MeV~overlap completely with it in the range $m_{\pi}=80-169.25(164.65)$ MeV.~The numerical values of the critical quantities $m_{\pi}^{c},\ m_{\pi}^{t}$ and the extent of the first order and crossover chiral transition regions in the $m_\pi-m_{K}$ plane,~are nearly the same in the RQM-S and the RQM-I model for the respective cases of the $m_{\sigma}=400\text{ and }500$ MeV.~While lying noticeably above the critical line of the RQM-I model in the range $m_{\pi}=0\rightarrow80$ MeV,~the RQM-S model chiral critical line for the case of $m_{\sigma}=600$ MeV,~overlaps with it completely in the range $m_{\pi}=80-147.66$ MeV.~The first order region of the chiral transition in the $m_{\pi}-m_{K}$ plane gets reduced by a proportion through which the corresponding chiral crossover transition region increases when the $\sigma$ mass increases from $400\rightarrow600$ MeV in both the RQM-S and I model.~The RQM-S chiral critical lines are enclosing comparatively larger first order transition regions for all the $m_{\sigma}$.~The RQM-S model first order regions go through significant reduction in the $m_{\pi}-m_{K}$ plane for the significantly large  $m_{\sigma}=750\text{ and }800$ MeV,~hence one finds that the  critical quantities $m_{\pi}^{c},\ m_{\pi}^{t}$ are also getting  reduced accordingly.

First order chiral phase transitions occur over a significantly large and wide area in the vertical and horizontal planes of the RQM-S and I model Columbia plots when the $m_{\sigma}=400\text{ and }500$ MeV.~Even after the noticeable reduction for the case of larger  $m_{\sigma}=600$ MeV,~the extents of first order chiral transition regions in the RQM-S and RQM-I model Columbia plots,~are significantly larger than the first  order regions obtained in the  e-MFA:FRG QM model Columbia plots for the $m_{\sigma}=530$ MeV in Ref.~\cite{Resch} and  QMVT model Columbia plots for the $m_{\sigma}=400\text{ and }530$ MeV in the Ref.~\cite{vkt25II}.~The critical quantities $m_{\pi}^{t} \text{ and } m_{\pi}^{c}$ in the horizontal $m_{\pi}-m_{K}$ planes of the RQM-S model for the $m_{\sigma}=750\text{ and } 800$ MeV,~when the strength of producing the first order chiral transition is very weak,~are respectively comparable to and larger than the values of the $m_{\pi}^{t}\text{ and }m_{\pi}^{c}$ obtained  respectively in the $m_{\pi}-m_{K}$ planes of the e-MFA:FRG QM model Columbia plot for the $m_{\sigma}=530$ MeV and the QMVT model Columbia plots for the case of $m_{\sigma}=400\text{ and }530$ MeV.~The e-MFA:FRG QM model work ~\cite{Resch} and the QMVT model studies~\cite{schafwag12, vkkr12, vkkt13} treat the quark one-loop vacuum correction term using the $\overline{\text{MS}}$ renormalization scheme and curvature masses of mesons are used to fix the model parameters.~The above method of treating quark one-loop vacuum fluctuations,~generates a very large smoothing effect on the strength of the chiral transition,~hence one finds significantly smaller first order regions in the Columbia plots of these studies~\cite{vkt25I,vkt25II}.

Different effective theory approaches are actively pursuing the Columbia plot studies in the light of the most recent LQCD results.~The very recent work of the Ref.\cite{Giacosa} has discussed a scenario in the Columbia plot of its Fig.1,~ where the area of first order region is maximum when the sixth order anomalous coupling is zero and the first order region decreases corresponding to the increasing sixth order coupling strength while the cubic coupling strength decreases.~They have shown a possibility where the chiral phase transition in the whole $m_{\pi}-m_{K}$ plane of the Columbia plot becomes second order,~even though  the $U_{A}(1)$ anomaly is still broken because of the large strength of the  sixth order anomalous coupling term while the cubic coupling strength is zero.~The FRG QM model study under the LPA in the Ref.~\cite{Resch},~has shown that,~the critical pion mass $m_{\pi}^{c}\equiv 86$ MeV obtained in the e-MFA of the QM model,~becomes quite small $m_{\pi}^{c}\equiv 17$ MeV indicating much reduced first order regions in the Columbia plot when the thermal and vacuum quantum fluctuations of meson loops are also included in their study with the full FRG flow.~However the Ref.~\cite{pisarski24} has pointed out that the approximation used in the above work is known to overestimate the masonic fluctuations~\cite{PawlRen} that tend to soften the chiral transition.~Further one should also note that it has been cautioned in the Ref.~\cite{fejoHastuda} that the LPA in the FRG studies,~completely neglects the wave function renormalization which might affect the final result.~Including the nonperturbative mesonic loop corrections in the Hartree approximation of LSM using the Cornwall-Jackiw-Tomboulis (CJT)  two-particle (2PI) irreducible effective action formalism,~the beyond perturbation method studies~\cite{Rischke:00,Lenagh} reported a large first order region in the Columbia plot when the $U_A(1)$ anomaly is absent and the results show drastic $\sigma$ mass dependence in the presence of the $U_A(1)$ anomaly where the heavier $m_{\sigma}$ values give the first order regions which disappear for the realistic  $m_{\sigma}$.~Using the symmetry improved CJT called the SICJT formalism~\cite{Sicjt} in the LSM,~the very recent study in the Ref.~\cite{Tomiya},~finds a stable first-order regime with a definite location of the $\tcp$ at $m_{s}^{\tcp}/m_{s}^{\text{\tiny{Phys}}}=0.696$ which shows small variation on increasing the $m_{\sigma / f_{0}(500)}$ from 672.4 MeV to 797.2 MeV whereas the critical pion mass reported
in their work is  $m_{\pi}^{c}$=52.4 MeV for the $m_{\sigma / f_{0}(500)}$=672.4 MeV.~They have clarified that the enhanced first order regions found with the conventional CJT formalism,~are merely an artifact due to the lack of manifest invariance of the Nambu Goldstone and low energy theorems at finite temperature.
~The relative position of the RQM-S model $\tcp$ at $m_{K}^{\tcp}/m_{K}^{\text{\tiny{Phys}}}$ ($m_{s}^{\tcp}/m_{s}^{\text{\tiny{Phys}}}$ ),~with respect to the value of kaon (s quark) mass at the physical point,~shifts from the 0.5016 to 0.4405 (0.2927 to 0.2280) when the $m_{\sigma}=400\rightarrow600$ MeV and one finds a significantly large shift in  the $m_{K}^{\tcp}/m_{K}^{\text{\tiny{Phys}}}$ ($m_{s}^{\tcp}/m_{s}^{\text{\tiny{Phys}}}$ ) from the  0.4405 to 0.2988 (0.2280 to 0.1068 ) when the $m_{\sigma}=600\rightarrow800$ MeV.~In a similar fashion the $m_{\pi}^{c}(m_{ud}^{c})$ changes from 134.51 (3.8) to 117.26 (2.9) MeV when the $m_{\sigma}=400\rightarrow600$ whereas the change in the $m_{\pi}^{c}(m_{ud}^{c})$ from 117.26 (2.9) to 82.07 (1.43) MeV is significantly large when the $m_{\sigma}=600\rightarrow800$ MeV.
~It is important to note that we are finding stable first order regions in the  RQM model Columbia plots even though its area gets reduced for very high $m_{\sigma}=800$ MeV.~Furthermore the relatively larger first order regions in the RQM-S model Columbia plots are caused by the fact that the $U_{A}(1)$ anomaly strength $c$ which contains a condensate dependent part,~ gets significantly enhanced when the meson self energies due to quark loops are evaluated using the meson pole masses and parameters are fixed on-sell where the smoothing effect of the quark one-loop vacuum fluctuations become moderate.

\begin{figure*}[htb]
\subfigure[\ Columbia plot: RQM-S model]{
\label{fig7a} 
\begin{minipage}[b]{0.49\textwidth}
			\centering
			\includegraphics[width=\linewidth]{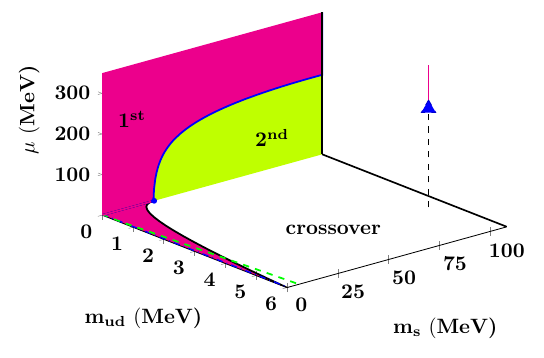}
	\end{minipage}}
	\hfill
	\subfigure[\ Columbia plot: RQM-I model ]{
		\label{fig7b} 
		\begin{minipage}[b]{0.49\textwidth}
			\centering 
			\includegraphics[width=\linewidth]{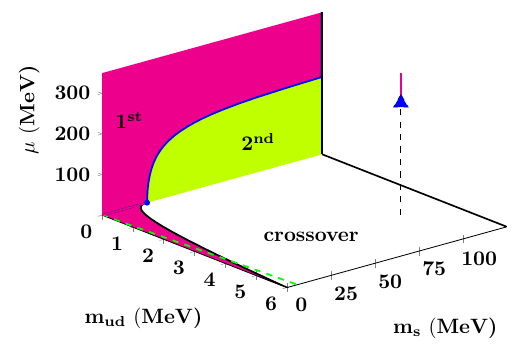}
	\end{minipage}}
	\caption{~The  chiral transitions for the $m_{ud}=0$ and  $\mu=0$ are depicted respectively in the $\mu-m_{s}$ and $m_{ud}-m_{s}$ plane.~The results for the RQM-S model when its parameters are fixed using the large $N_c$ standard U(3) ChPT inputs,~are presented in the left panel (a) whereas the right panel (b) shows the results for the RQM-I model where its parameters are fixed using the infrared regularized U(3) ChPT inputs.~The second order $Z(2)$ critical line in solid black color that separates the crossover from the first order transition region in the Fig.(a) and (b),~intersects the $m_{ud}=m_{s}$ green dash line respectively at the critical light quark mass $m_{ud}^{c}=3.80$ and $m_{ud}^{c}=3.84$ MeV and it terminates respectively at the terminal light quark mass $m_{ud}^{t}=5.97$ and $m_{ud}^{t}=6.13$ MeV.~Solid blue line of the tricritical points which starts respectively at the blue dot $m_{s}^{\tcp}=25.23$ and  $m_{s}^{\tcp}=25.53$  MeV in the Fig.(a) and (b),~separates the second  and first order regions.~The black dashed vertical line at the physical point in the Fig(a) and (b),~shows the crossover transition which ends at the critical end point in blue triangle and the solid red line shows the first order transition.~The scalar $\sigma$ mass $m_{\sigma}=400$ MeV.}
\label{fig:mini:fig7}
\end{figure*}

\begin{figure*}[htb]
\subfigure[\ Columbia plot: RQM-S model]{
\label{fig8a} 
\begin{minipage}[b]{0.49\textwidth}
			\centering
			\includegraphics[width=\linewidth]{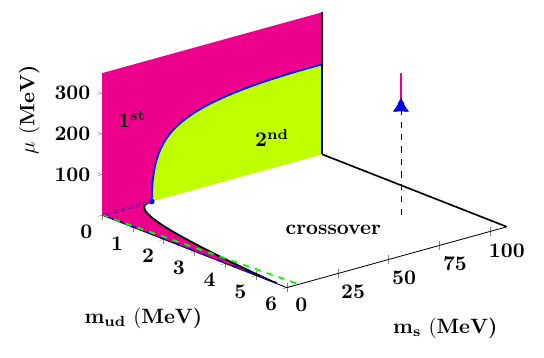}
	\end{minipage}}
	\hfill
	\subfigure[\ Columbia plot: RQM-I model ]{
		\label{fig8b} 
		\begin{minipage}[b]{0.49\textwidth}
			\centering 
			\includegraphics[width=\linewidth]{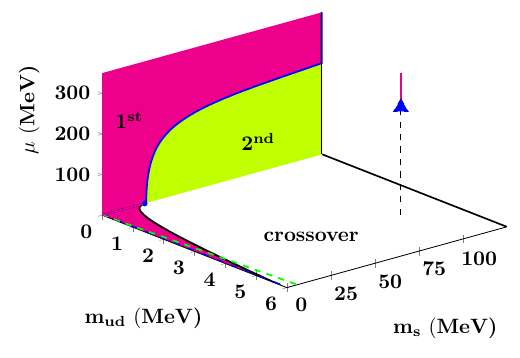}
	\end{minipage}}
	\caption{~~The chiral transitions in the $\mu-m_{s}$ plane for the $m_{ud}=0$ and the $m_{ud}-m_{s}$ plane for the $\mu=0$ when the $m_{\sigma}=500$ MeV are presented in the left panel (a) for the RQM-S model and the right panel (b) for the RQM-I model.~The points,~lines,~crossover, first and second order chiral  transition regions and features are defined similar to the Fig.(\ref{fig:mini:fig7}).~The critical quantities ($m^{\tcp}_{s},\ m_{ud}^{t}, \ m_{ud}^{c}$) in the RQM-S and RQM-I model are (24.33, 5.66, 3.59) MeV and (25.0, 5.79, 3.66) MeV in respective order.}
	\label{fig:mini:fig8} 
\end{figure*}

\subsection{\bf{Comparing the RQM-S and I model Columbia plots in the quark mass $\bf {m_{ud}-m_{s}}$ and $\bf {\mu-m_{s}}$ planes}}
\label{Colplotquark}

The Columbia plots in the  $m_{\pi}-m_{K}$ plane at $\mu=0$ and the $\mu-m_{K}$ plane at $m_{\pi}=0$ can be redrawn in the light-strange quark mass $m_{ud}-m_{s}$ plane at $\mu=0$ and the chemical potential-strange quark mass $\mu-m_{s}$ plane at $m_{ud}=0$ by using the Eqs.~(\ref{Aqpim}) and (\ref{qmr}) for the infrared regularized U(3) ChPT input set for the RQM-I model and the Eqs.(\ref{mpi2U3}) and (\ref{mk2U3}) for the large $N_c$ standard U(3) ChPT input set for the RQM-S model.~Columbia plots in the quark mass planes would make the results of chiral limit studies in effective model approach,~available for comparison with LQCD  results and studies in other approaches.~In order to map the results from the $m_{\pi}-m_{K}$  to the $m_{ud}-m_{s}$ plane,~the light quark mass has been fixed to  $m_{ud}=4$ MeV at the physical point.~With the $m_{ud}=4$ MeV,~the solutions of Eqs.~(\ref{Aqpim}) and (\ref{qmr}) give the physical point value of the strange quark mass as ($m_{s}^\text{\tiny{Phys}}=99.7$ MeV)  for the RQM-I model whereas the solutions of Eqs.~(\ref{mpi2U3}) and (\ref{mk2U3}) give the physical point value of  the strange quark mass for the RQM-S model as ($m_{s}^\text{\tiny{Phys}}=86.2$ MeV).~The zero light quark mass $m_{ud}=0$ gives the light chiral limit.~The strange quark mass physical point value for the physical kaon mass $m_{K}=496$ MeV,~gives the second order chiral phase transition on the $m_{s}$ axis when $m_{ud}=0$.~The second order phase transition turns first order chiral transition at the tricritical point of strange quark mass $m_{s}^{\tcp}$ corresponding to the kaon mass $m_{K}^{\tcp}$ when the $m_{s} (m_{K})$ is reduced keeping the $m_{ud}=0$ and $\mu=0$.~For different non-zero values of $\mu$ on the chemical potential axis,~the tricritical point value of the strange quark mass shifts up on the $m_{s}$ axis and the resulting tricritical line in the $\mu-m_{s}$ plane at $m_{ud}=0$,~depicted by blue line in the figures below,~starts from the $m_{s}^{\tcp}$ at the $\mu=0$ (shown by solid blue dot in the figures) and demarcates the first order transition region above it from the second order transition region lying below.~The chiral critical line separating the chiral crossover transition region from the first order region in the $m_{ud}-m_{s}$ plane at $\mu=0$,~is obtained from the corresponding line in the $m_{\pi}-m_{K}$ plane and it has been depicted here also by the solid black color.~The chiral critical line terminates on the $m_{ud}$ axis at the terminal light quark mass $m_{ud}^{t}$ (obtained from the $m_{\pi}^{t}$) beyond which the transition is a smooth crossover everywhere in the $m_{ud}-m_{s}$ plane.~The  intersection point of the  SU(3) symmetric $m_{ud}=m_{s}$ chiral limit path,~depicted by the green dash line in the figures,~with the chiral critical line in the solid black color,~gives the critical light quark mass $m_{ud}^{c}$ (obtained from the $m_{\pi}^{c}$) where the boundary of first order region ends in the $m_{ud}-m_{s}$ plane.

\begin{figure*}[htb]
	\subfigure[\ Columbia plot: RQM-S model]{
		\label{fig9a} 
		\begin{minipage}[b]{0.49\textwidth}
			\centering
			\includegraphics[width=\linewidth]{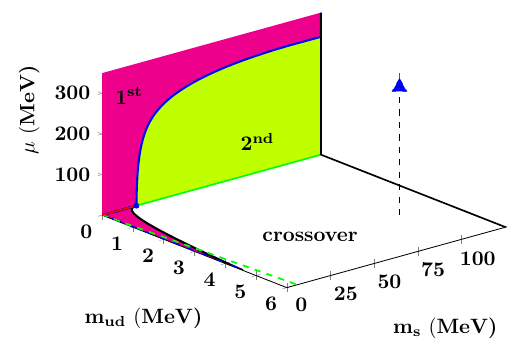}
	\end{minipage}}
	\hfill
	\subfigure[\ Columbia plot: RQM-I model]{
		\label{fig9b} 
		\begin{minipage}[b]{0.49\textwidth}
			\centering 
			\includegraphics[width=\linewidth]{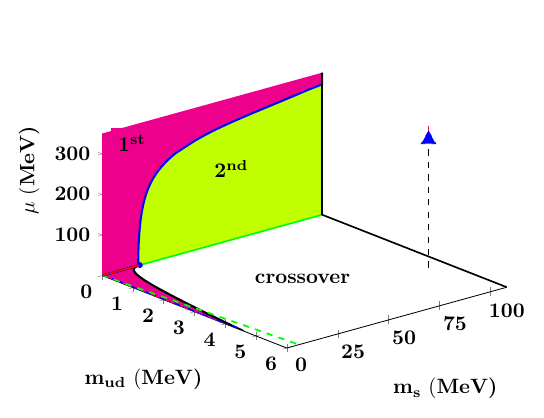}
\end{minipage}}
\caption{~Chiral transitions in the $\mu-m_{s}$ plane at $m_{ud}=0$ and the $m_{ud}-m_{s}$ plane at $\mu=0$  for the $m_{\sigma}=600$ MeV,~are presented in the left panel (a) for the RQM-S model and the right panel (b) for the RQM-I model.~The points,~lines,~crossover, first and second order chiral  transition regions and features are defined similar to the Fig.(\ref{fig:mini:fig7}).~The critical quantities ($m^{\tcp}_{s},\ m_{ud}^{t}, \ m_{ud}^{c}$) are respectively (19.65, 4.57, 2.90) MeV and (17.70, 4.59, 2.97) MeV in the RQM-S and RQM-I model.}
	\label{fig:mini:fig9} 
\end{figure*}

The  horizontal $m_{ud}-m_{s}$ and vertical $\mu-m_{s}$ quark mass planes of the RQM-S model Columbia plot for the $m_{\sigma}=400 $ MeV show the largest extent of the first order chiral transition regions in the  Fig.~\ref{fig7a} where the obtained critical quantities are ($m^{\tcp}_{s},\ m_{ud}^{t}, \ m_{ud}^{c}$)=(25.23, 5.97, 3.80) MeV corresponding to the $(m_{K}^{\tcp},~m_{\pi}^{t},~m_{\pi}^{c}) \equiv (248.8,~169.25,~134.51)$ MeV.~Even though the RQM-I model first order regions in the $m_{\pi}-m_{K}$
and $\mu-m_{K}$ planes of the Fig.~\ref{fig2b} are marginally smaller than the corresponding RQM-S model first order regions in the Fig.~\ref{fig2a},~the first order regions in the $m_{ud}-m_{s}$ and $\mu-m_{s}$ quark mass planes of the RQM-I model Columbia plot for the $m_{\sigma}=400 $ MeV in the Fig.~\ref{fig7b}
are equivalent and negligibly larger than the corresponding first order regions in the Fig.~\ref{fig7a} for the RQM-S model because the strange quark mass value ($m_{s}^\text{\tiny{Phys}}=99.7$ MeV) for the RQM-I model  at the physical point,~is larger than its RQM-S model value ($m_{s}^\text{\tiny{Phys}}=86.2$ MeV).~Hence the RQM-I model critical quantities in the Fig.~\ref{fig7b} are obtained as ($m^{\tcp}_{s},\ m_{ud}^{t}, \ m_{ud}^{c}$)=(25.53, 6.13, 3.84) MeV corresponding to the $(m_{K}^{\tcp},~m_{\pi}^{t},~m_{\pi}^{c}) \equiv (242.6,~168.39,~134.16)$ MeV.~The extents of first order regions in the quark mass planes of the RQM-S and RQM-I Columbia plots for the $m_{\sigma}=500$ MeV respectively in the Fig.~\ref{fig8a} and Fig.~\ref{fig8b},~get reduced only marginally when compared to the case of $m_{\sigma}=400$ MeV.~After slight reduction by ($\Delta m^{\tcp}_{s},\Delta m_{ud}^{t}, \Delta m_{ud}^{c}$)=(0.9, 0.31, 0.21) MeV when the $m_{\sigma}=400\rightarrow500$ MeV,~the RQM-S model critical quantities in the Fig.~\ref{fig8a} are obtained as ($m^{\tcp}_{s},\ m_{ud}^{t}, \ m_{ud}^{c}$)=(24.33, 5.66, 3.59) MeV corresponding to the $(m_{K}^{\tcp},m_{\pi}^{t},m_{\pi}^{c}) \equiv (244.1,164.65,130.71)$ MeV whereas with the corresponding marginal reduction of ($\Delta m^{\tcp}_{s},\Delta m_{ud}^{t}, \Delta m_{ud}^{c}$)=(0.53, .34, 0.18) MeV,~the RQM-I model critical quantities in the Fig.~\ref{fig8b} are obtained as ($m^{\tcp}_{s},\ m_{ud}^{t}, \ m_{ud}^{c}$)=(25.0, 5.79, 3.66) MeV corresponding to the $(m_{K}^{\tcp},m_{\pi}^{t},m_{\pi}^{c}) \equiv (240.1,164.02,130.84) $ MeV when  the $m_{\sigma}=500$ MeV.

In contrast to the results obtained when the $m_{\sigma}=400\rightarrow500$ MeV,~the first order regions shrink noticeably when the $m_{\sigma}=500\rightarrow600$ MeV as evident from the regions of the first order transition witnessed in the quark mass planes of the RQM-S and RQM-I Columbia plots for the $m_{\sigma}=600$ MeV respectively in the Fig.~\ref{fig9a} and Fig.~\ref{fig9b}.~After moderately large reduction of ($\Delta m^{\tcp}_{s},\Delta m_{ud}^{t}, \Delta m_{ud}^{c}$)=(4.68, 1.09, 0.69) MeV  when the $m_{\sigma}=500\rightarrow600$ MeV,~the RQM-S model critical quantities in the Fig.~\ref{fig9a} are obtained as ($m^{\tcp}_{s},\ m_{ud}^{t}, \ m_{ud}^{c}$)=(19.65, 4.57, 2.90) MeV corresponding to the $(m_{K}^{\tcp},m_{\pi}^{t},m_{\pi}^{c}) \equiv (218.5,147.66,117.26)$ MeV.~On account of the significantly large reduction of ($\Delta m^{\tcp}_{s},\Delta m_{ud}^{t}, \Delta m_{ud}^{c}$)=(7.3, 1.2, 0.69) MeV in the RQM-I model critical quantities when the $m_{\sigma}=500\rightarrow600$ MeV,~one finds noticeably smaller areas of first order regions in the Fig.~\ref{fig9b} when compared to the first order regions of the RQM-S model in the Fig.~\ref{fig9a} even though the physical point value of the strange quark mass ($m_{s}^\text{\tiny{Phys}}=99.7$ MeV) is larger for the RQM-I model.~With the above mentioned reduction,~the RQM-I model critical quantities in the Fig.~\ref{fig9b} are obtained as ($m^{\tcp}_{s},\ m_{ud}^{t}, \ m_{ud}^{c}$)=(17.70, 4.59, 2.97) MeV corresponding to the $(m_{K}^{\tcp},m_{\pi}^{t},m_{\pi}^{c}) \equiv (207.05,146.35,118.01) $ MeV when  the $m_{\sigma}=600$ MeV.

\begin{figure*}[htb]
	\subfigure[\ Columbia plot: RQM-S model for $m_{\sigma}=750$ MeV]{
		\label{fig10a} 
		\begin{minipage}[b]{0.49\textwidth}
			\centering
			\includegraphics[width=\linewidth]{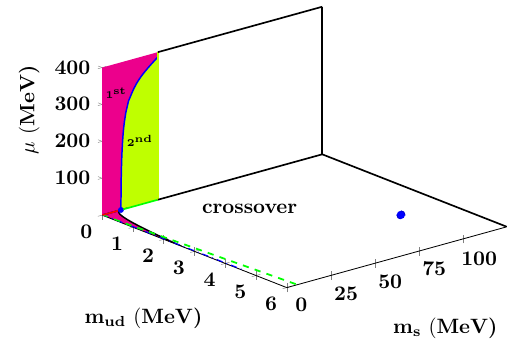}
	\end{minipage}}
	\hfill
	\subfigure[\ Columbia plot: RQM-S model for $m_{\sigma}=800$ MeV]{
		\label{fig10b} 
		\begin{minipage}[b]{0.42\textwidth}
			\centering 
			\includegraphics[width=\linewidth]{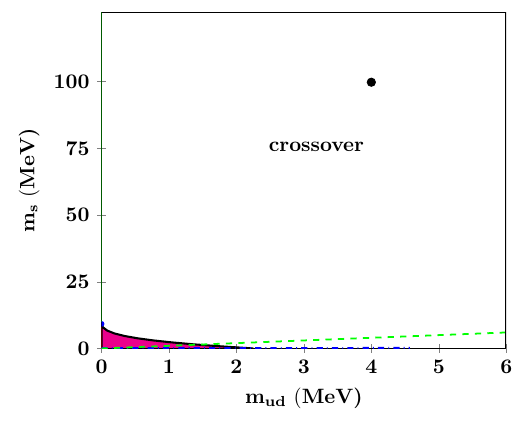}
\end{minipage}}
\caption{The left panel (a) depicts the RQM-S model chiral transition in the horizontal plane $m_{ud}-m_{s}$ at $\mu=0$ and the vertical $\mu-m_{s}$ plane at $m_{ud}=0$ when the $m_{\sigma}=750$ MeV.~~The horizontal plane $m_{ud}-m_{s}$ at $\mu=0$ in the right panel (b) presents the  chiral transition in RQM-S model for the case of $m_{\sigma}=800$ MeV.~~The points,~lines,~crossover,~first and second order chiral  transition regions and features are defined similar to the Fig.(\ref{fig:mini:fig7}).~The tricritical line demarcating the first and second order colored regions,~is computed only upto the point $(m_{s},\mu)=(31.61,386.8)$ MeV in the Fig.~(a) because the chemical potential becomes very high for a significantly  smaller strange mass in the $\mu-m_{s}$ plane.~The RQM-S model critical quantities ($m^{\tcp}_{s},\ m_{ud}^{t}, \ m_{ud}^{c}$) are respectively (10.65, 2.57, 1.64) MeV and (9.21, 2.25, 1.43) MeV when $m_{\sigma}=750 \text{ and } 800$ MeV.}
\label{fig:mini:fig10} 
\end{figure*}

All the tricritical lines in the vertical $\mu-m_{s}$ planes at $m_{ud}=0$,~in the Fig.~\ref{fig7a},~\ref{fig8a} and ~\ref{fig9a} of the RQM-S model and Fig.~\ref{fig7b},~\ref{fig8b} and ~\ref{fig9b} of the RQM-S model respectively for the $m_{\sigma}=400,500\text{ and }600$ MeV,~begin 
with a large positive slope from their origin at the $m_{s}^{\tcp}$ and $\mu=0$ and suffer a large bending for higher chemical potentials.~The RQM-S model tricritical lines show proper saturation for all the cases of the the $m_{\sigma}=400,500\text{ and }600$ MeV as each of them become flat and horizontal with almost a zero slope in the strange quark mass range $m_{s}=91.5-108$ MeV that corresponds to the kaon mass range of $m_{K}=500-550$ MeV.~When the $m_{\sigma}=400$ MeV in the RQM-I model,~its tricritical line shows a saturation pattern close to that of the RQM-S model as its slope approaches nearly zero in the strange quark mass range $m_{s}=105.22-124.66$ MeV for the $m_{K}=500-550$ MeV.~But the RQM-I model tricritical line for the $m_{\sigma}=500$ MeV,~picks up a divergence trend after $m_{s}>69.1$ MeV for the $m_{K}>400$ MeV.~Showing a significantly large divergence after the $m_{s}>69.1$ MeV,~the RQM-I model tricritical in the Fig.~\ref{fig9b} for the $m_{\sigma}=600$ MeV,~looks very different from the corresponding RQM-S model tricritical line in the Fig.~\ref{fig9a}.

\begin{figure*}[htb]
\subfigure[\ Vertical plane $\mu-m_{s}$ for $m_{\sigma}$=400,500,600 and 750 MeV.]{
\label{fig11a} 
\begin{minipage}[b]{0.49\textwidth}
\centering
\includegraphics[width=\linewidth]{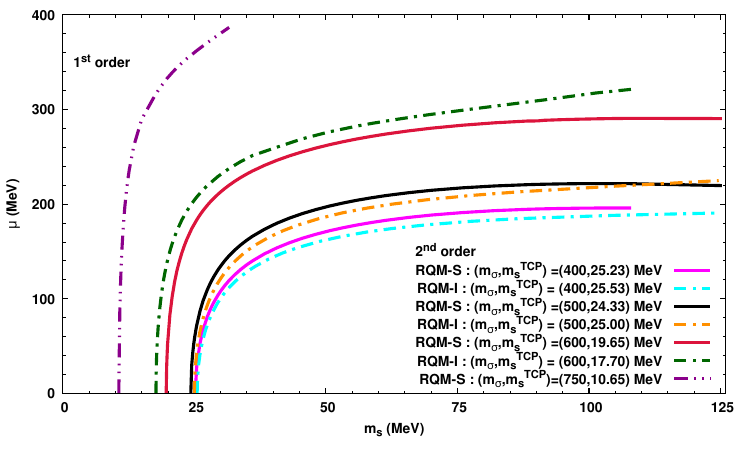}
\end{minipage}}
\hfill
\subfigure[\ Horizontal plane $m_{ud}-m_{s}$ for $m_{\sigma}$=400,500,600,750,~800 MeV.]{
\label{fig11b} 
\begin{minipage}[b]{0.49\textwidth}
\centering 
\includegraphics[width=\linewidth]{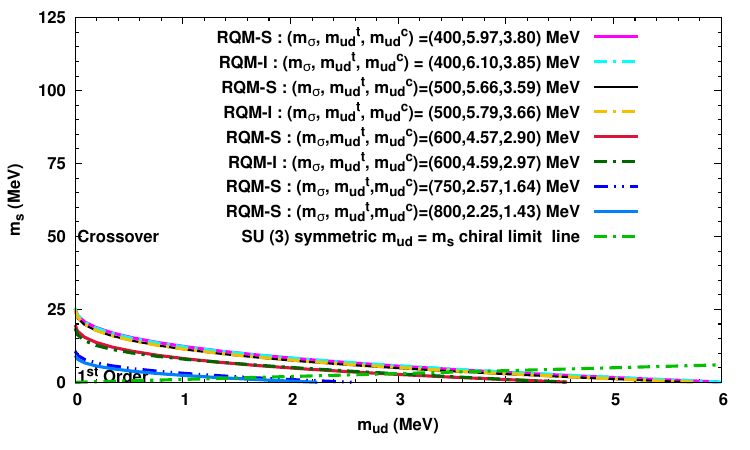}
\end{minipage}}
\caption{~Left panel (a) (Right panel (b)) presents the comparison of lines and regions in the vertical $\mu-m_{s}$, $m_{ud}=0$ (horizontal $m_{ud}-m_{s}$, $\mu=0$) planes of the three dimensional RQM-S and RQM-I model Columbia plots for the $m_{\sigma}=400,500,600,750 \text{ and } 800$ MeV in the Fig.~(\ref{fig7a}),~(\ref{fig7b}),~(\ref{fig8a}),~(\ref{fig8b}),~(\ref{fig9a}),~(\ref{fig9b}) and ~(\ref{fig10a}),~(\ref{fig10b}).~RQM-S and  RQM-I model tricritical lines for the $m_{\sigma}=400,500,600 \text{ and } 750$ MeV,~as marked and explained in the Fig.(a),~start from the $m_{s}^{\tcp}$ on the $m_{s}$ axis and demarcate the first order region above them on their left hand side  from the second order region below them on their right hand side.~RQM-S and RQM-I model Z(2) critical lines for the $m_{\sigma}=400,500,600,750 \text{ and } 800$ MeV,~as marked and explained in the Fig.(b), separating the crossover transition region above them from the first order region that lies below,~intersect the SU(3) symmetric $m_{ud}=m_{s}$ chiral limit line at the critical light quark mass $m_{ud}^{c}$ and terminate on the $m_{ud}$ axis at the terminal light quark mass of $m_{ud}^{t}$.}
\label{fig:mini:fig11} 
\end{figure*}

~The critical quantities suffer a very large  reduction of ($\Delta m^{\tcp}_{s},\Delta m_{ud}^{t}, \Delta m_{ud}^{c}$)=(9.0, 2.0, 1.26) MeV when the $m_{\sigma}=600\rightarrow750$ MeV in the RQM-S model.~
One finds very small areas of first order chiral transition regions in the vertical $\mu-m_{s}$ and horizontal $m_{ud}-m_{s}$ planes of the  RQM-S model Columbia plot in the Fig.~\ref{fig10a} when the $m_{\sigma}=750$ MeV as also signified by the observed critical quantities ($m^{\tcp}_{s},\ m_{ud}^{t}, \ m_{ud}^{c}$)=(10.65, 2.57, 1.64) MeV.~The tricritical line in the vertical $\mu-m_{s}$ plane is very strongly divergent since it suffers quite a small bending and it could be drawn only upto $m_{s}=31.6$ MeV 
for $m_{K}=280$ MeV because even for such a smaller $m_{s}$ value,~one needs to increase the chemical potential to very high $\mu=386.83$ MeV for finding the $\tcp$ since strength of  the chiral transition    becomes very weak for the $m_{\sigma}=750$ MeV.~The divergent tricritical line is associated with a chiral critical region whose spread is significantly smaller in the Columbia plot,~hence the $\cep$ does not exist at the physical point co-ordinates in the $m_{ud}-m_{s}$ plane.~The critical quantities change by ($\Delta m^{\tcp}_{s},\Delta m_{ud}^{t}, \Delta m_{ud}^{c}$)=(1.44, 0.32, 0.21) MeV when the $m_{\sigma}=750\rightarrow800$ MeV in the RQM-S model.~The Fig.~\ref{fig10b} shows the horizontal $m_{ud}-m_{s}$ plane of the Columbia plot where  the  chiral critical line of $Z(2)$ points underneath the largest chiral crossover area,~covers the smallest area of first order regions which gets indicated also by the values of the critical quantities ($m^{\tcp}_{s},\ m_{ud}^{t}, \ m_{ud}^{c}$)= (9.21, 2.25, 1.43) MeV  when the $m_{\sigma}= 800$ MeV in the RQM-S model.

The tricritical lines in the  chemical potential-strange quark mass vertical $\mu-m_{s}$ planes at $m_{ud}=0$ of the RQM-S $\lbrace{\text{RQM-I}\rbrace}$ model Columbia plots in the Fig.~(\ref{fig7a}) $\lbrace{\text{(\ref{fig7b})}\rbrace}$,~(\ref{fig8a}) $\lbrace{\text{(\ref{fig8b})}\rbrace}$,~(\ref{fig9a})$\lbrace{\text{(\ref{fig9b})}\rbrace}$ and ~(\ref{fig10a}) respectively for the $m_{\sigma}=400,500,600 \text{ and } 750$ MeV have been redrawn in a single frame of the Fig.~(\ref{fig11a}) in order to show 
how the comparative shifting of the tricritical lines to higher chemical potentials,~indicate the decreasing first order regions while the corresponding area of second order transitions increase in the  $\mu-m_{s}$ plane  when the $m_{\sigma}$ increases from $400\rightarrow750$ MeV.~Turning flat with a zero slope in the range  $m_{s}=91.5 \text{ to }  108$ MeV of strange quark mass,~the tricritical lines for the  RQM-S model,~show proper saturation for all the cases of the $m_{\sigma}=400,500\text{ and } 600$ MeV.~While showing a saturation pattern in the range $m_{s}=105.22 \text{ to } 124.66$ MeV for the $m_{\sigma}=400$ and picking up a small divergence for the $m_{\sigma}=500$ MeV,~the RQM-I model tricritical line for the case of $m_{\sigma}=600$ MeV,~becomes significantly divergent after the $m_{s}>69.1$ MeV and looks completely different from the RQM-S model tricritical line.~Owing to its very strong divergence for the large $m_{\sigma}=750$ MeV,~the RQM-S model tricritical line could be drawn only upto $(m_{s},\mu)=(31.6,386.83)$ MeV since the strength of  the chiral transition  becomes very weak.

The horizontal $m_{ud}-m_{s}$ planes (at $\mu=0$) of the RQM-S $\lbrace{\text{RQM-I}\rbrace}$ model Columbia plots together with the chiral critical solid black lines and the SU(3) symmetric chiral limit lines for the $m_{ud}=m_{s}$ in the Fig.~(\ref{fig7a})$\lbrace{\text{(\ref{fig7b})}\rbrace}$,~(\ref{fig8a})$\lbrace{\text{(\ref{fig8b})}\rbrace}$,~(\ref{fig9a})$\lbrace{\text{(\ref{fig9b})}\rbrace}$,~(\ref{fig10a}) and (\ref{fig10b}) respectively for the $m_{\sigma}=400, 500, 600, 750 \text{ and } 800$ MeV,~have been plotted again at one place in the Fig.~(\ref{fig11b}) for depicting the relative shrinkage 
in the first order chiral transition regions when the areas of chiral crossover transition in the RQM-S and RQM-I model,~expand corresponding to the increase of scalar $\sigma$ meson mass in the range $m_{\sigma}=400 \text{ to } 800$ MeV.~The different line types and regions are marked and explained in the  Fig.~(\ref{fig11b}) where the values of critical quantities  $m_{ud}^{t}\text{ and }m_{ud}^{c}$ are also given.~The RQM-S and RQM-I model $Z(2)$ chiral critical lines for the $m_{\sigma}=400 \text{ and  } 500$ MeV are nearly overlapping.~In order to understand the above mentioned results,~one has to recall that when the light quark mass is fixed to  $m_{ud}=4$ MeV at the physical point,~the solutions of Eqs.~(\ref{Aqpim}) and (\ref{qmr}) for the RQM-I model give larger strange quark mass $m_{s}^\text{\tiny{Phys}}=99.7$ MeV while the solutions of the Eqs.(\ref{mpi2U3}) and (\ref{mk2U3}) for the RQM-S model give moderate strange quark mass $m_{s}^\text{tiny{Phys}}=86.2$ MeV at the physical point.~If one tries to match the strange quark mass value at the physical point for both the input sets I and S,~by taking  $m_{s}^\text{\tiny{Phys}}=86.2$ MeV at the physical point for the infrared regularized ChPT input set I,~one needs to take a lower value of light quark mass  $m_{ud}=3.46$ MeV at the physical point such that the $m_s$ becomes 86.2 MeV and if that is done,~the critical quantities defined in terms of light and strange quark masses will have a smaller value in the RQM-I model~\cite{vkt25I}.~One should also note that the chiral constants and parameters of the large $N_c$ standard U(3) ChPT input set-S,~give $q=\frac{m_{s}}{m_{ud}}=21.56$ at the physical point and this value lies in its appropriate range given in the Ref.~\cite{herrPLB,Escribano,vkt25I}.~The first order region under the chiral critical line is very small for the $m_{\sigma}=750$ MeV and it is smallest for the $m_{\sigma}=800$ MeV.~In the RQM-S model,~the largest value of the critical light quark mass is obtained  as $ m_{ud}^{c}=3.8$ MeV when the $m_{\sigma}=400$ MeV while its smallest value  $ m_{ud}^{c}=1.43$ MeV  is found for a very large  $m_{\sigma}=800$ MeV.~It is relevant to recall that the quark one-loop vacuum fluctuation exerts a very strong smoothing effect on the strength of chiral transition in the curvature mass parametrized QMVT model where one finds significantly smaller areas of first order transition even for the $m_{\sigma}=400\text{ and } 530 $ MeV as shown in a very recent work in ref.~\cite{vkt25II} where one finds the smallest $ m_{ud}^{c}=1.1$ MeV for the $m_{\sigma}=530$ MeV.~It is important to see that the range $m_{ud}^c=3.8\text{ to } 1.43$ MeV of critical light quark mass values (for different $m_{\sigma}$)    found in our present RQM-S model work,~lies well within the range of $m_{ud}^c\le$4 MeV given by  recent  LQCD study using Möbius domain wall fermions in Ref.~\cite{zhang24}.~The critical pion mass $m_{\pi}^c$ and the critical light quark mass $m_{ud}^c$ obtained from the different LQCD study results,~are presented in the Table(\ref{tab:table2}) for a comparative view.

\section{Summary}
\label{secIV}


The problem of the loss of the spontaneous chiral symmetry breaking ( SCSB ) in the quark meson (QM) model 
for the $m_{\sigma}=400-800$ MeV when the chiral limit studies are performed ~\cite{Ortman, Lenagh, Schaefer:09,  fuku08, berges, Herbst} in the often used fixed-ultraviolet (UV) scheme~\cite{Resch},~where 
the light (strange) explicit chiral symmetry breaking strengths $ h_{x}(h_{y}) $ are reduced 
and the other model parameter values are kept fixed as at the physical point,~got cured very recently in the on-shell renormalized 2+1 flavor quark meson (RQM) model~\cite{vkt25I,vkt25II} when its parameter fixing away from the physical point for the reduced $\pi \text{ and } K $ masses (as the $m_{\pi},m_{K} \rightarrow  0$),~was refined by using the $ (m_{\pi},m_{K}) $ dependent scaling relations for the $f_{\pi}, f_{K} \text{ and } M_{\eta}^2$ given by the $ \mathcal{O}(\frac{1}{f^2})$  accurate results of the large $N_{c}$ standard U(3) chiral perturbation theory (ChPT)~\cite{gasser, LeutI,  KaisI, herrNPB, herrPLB, Escribano} in the RQM-S model and the infrared regularized U(3) ChPT~\cite{Herpay:05, borasoyI, borasoyII, Beisert, Becher} in the RQM-I model.~The Columbia plots of the RQM-S model for the $m_{\sigma}=400,500,600,750 \text{ and } 800$ MeV and the RQM-I model for the $m_{\sigma}=500\text{ and }600$ MeV are computed in the present work.~Bringing out similarities and differences after a thorough and comprehensive comparison of the RQM-S and RQM-I model Columbia plots for the $m_{\sigma}=400,500\text{ and } 600$ MeV,~it has been shown conclusively in the present work that the large $N_c$ standard U(3) ChPT scaling relations for the ($m_{\pi},m_{K}$) dependence of the $f_{\pi}, f_{K} $ and $M_{\eta}^2$,~constitute the  improved and better prescription for Chiral limit studies.~We emphasize that the RQM-S model has given us the framework in which the chiral limit studies can be consistently performed over the complete range of $m_{\sigma}=400\rightarrow800$ MeV without any ambiguity or heuristic adjustment in its parameter fixing whereas the e-MFA:FRG QM model chiral limit study,~where the FRG flow allows for the scalar $\sigma$ masses only in the range $m_{\sigma} \in [400,600]$ MeV~\cite{Resch},~could avoid the loss of SCSB only by heuristically adjusting the initial effective action successively to larger scales ($\Lambda^{\prime} > \Lambda $ (for starting the FRG flow) at each step of reducing the symmetry breaking strengths such that the $f_{\pi}$ does not change in their ChPT motivated fixed $f_{\pi}$ scheme.

The fraction $\beta=\frac{m_{\pi}^*}{m_{\pi}}= \frac{m_{K}^*}{m_{K}}$ provides us a direct path from the physical point to the chiral limit as it defines the reduced $(\pi,K)$ masses $(m_{\pi}^* ,m_{K}^* )$ such that the $\frac{m_{\pi}^*}{m_{K}^*}$ = physical point ratio $\frac{m_{\pi}}{m_{K}}$ where the ChPT parameter ratio $q=\frac{2 \ m_s}{m_u+m_d}$ also gets fixed.~Turning sharper as the $\beta$ is decreased,~ the chiral crossover transition for the physical point  $\beta$=1 becomes second order transition at the $Z_{2}$ critical point fraction $\beta_{\cep}$.~The changing $\beta_{\cep}$ from the $0.38738 \  \lbrace{0.37963\rbrace}\rightarrow0.37665 \ \lbrace{0.37159\rbrace}\rightarrow0.33634 \     \lbrace{0.32445\rbrace}$ for the changing $m_{\sigma}$ as $400\rightarrow500\rightarrow600$ MeV gives us the correspondingly changed reduced $\cep$ masses $(m_{\pi,\cep}^*,m_{K,\cep}^*)$ as $(53.46,~192.14) \  \lbrace{(52.39,~188.30)\rbrace}  \rightarrow(51.98,186.82)\ \lbrace{(51.33,184.49)\rbrace}\rightarrow(46.41,166.82)$ $ \lbrace{ (44.77,160.93)\rbrace}$ MeV in respective order for which the chiral crossover transition becomes second order transition in the RQM-S $\lbrace{\text{RQM-I}\rbrace}$ model.~The relative decrease in $\beta_{\cep}$ indicates that the chiral transition strength becomes only marginally weak when $m_{\sigma} = 400 \rightarrow 500$ MeV whereas  it gets noticeably weaker when $m_{\sigma} = 500 \rightarrow 600$ MeV.~Note that since $\beta_{\cep}|_{(m_{\sigma}=400,500,600 \text{MeV} : \text{RQM-S})}> \beta_{\cep}|_{(m_{\sigma}=400,500,600 \text{MeV} : \text{RQM-I})}$,~the use of input set-S gives rise to a stronger strength of transition in the RQM-S model as the first order  chiral transition is obtained by a smaller reduction in the masses of $\pi \text{ and } K $ mesons.~The $\beta_{\cep}$ changes from the $0.33634\rightarrow 0.25013 \rightarrow 0.23316   $ for the changing $m_{\sigma}$ in the RQM-S model as $600\rightarrow750\rightarrow800$ MeV .~Even though the chiral transition gets strongly  diluted due to a very high $m_{\sigma}=800$ MeV,~its strength in the RQM-S model remains stronger than what is found in the e-MFA:FRG QM model~\cite{Resch} and the curvature mass based parametrized QMVT model~\cite{vkt25II} study for the moderate $m_{\sigma}=530$ MeV because the $\beta_{\cep}|_{(m_{\sigma}=800 \ \text{MeV} : \text{RQM-S})}=0.23316>\beta_{\cep}|_{(m_{\sigma}=530  \text{MeV} : \text{e-MFA:FRG})}=0.2  \sim \beta_{\cep}|_{(m_{\sigma}=530  \text{MeV} : \text{QMVT})}=0.2028$.


The second order transition turns first order on the kaon (s quark) mass axis of the Columbia plot for the light chiral limit $m_{\pi}=0$ ($m_{ud}=0$),~at the tricritical point at $m_{K}=m_{K}^{\tcp}$ ($m_{s}=m_{s}^{\tcp}$).~The chiral critical line,~of the second order $Z(2)$ universality points that demarcates the crossover from the first order region in the $m_{\pi}-m_{K}$ ($m_{ud}-m_{s}$) plane at the $\mu=0$,~terminates on the strange chiral limit line  $h_{y0}=0$ ($m_{s}=$0) at the terminal $\pi$  (light quark) mass $m_{\pi}^{t}(m_{ud}^t)$ beyond which the transition is a smooth crossover everywhere and this line intersects the  SU(3) symmetric $m_{\pi}=m_{K}$ ($m_{ud}=m_{s}$) chiral limit line also at the critical pion mass $m_{\pi}^{c}$ (light quark mass $m_{ud}^{c}$) where the boundary of first order region ends.~The values of the critical quantities $(m_{K}^{\tcp},m_{\pi}^{t},m_{\pi}^{c})$ $\lbrace{(m^{\tcp}_{s},m_{ud}^{t}, m_{ud}^{c}) \rbrace}$ in the meson mass $(m_{\pi}-m_{K})$  $\lbrace{ \text{quark mass }(m_{ud}-m_{s}) \rbrace}$ planes of the RQM-S model and RQM-I model Columbia plots are  (248.8,169.25,134.51) $\lbrace{(25.23,5.97,3.80)\rbrace}$ MeV and (242.6,168.39,134.16) $\lbrace{ (25.53,6.13,3.84)\rbrace}$ MeV in the respective order of models for the $m_{\sigma}=400$ MeV and (244.1,164.65,130.71) $\lbrace{ (24.33, 5.66, 3.59)\rbrace}$ MeV and  (240.1,164.02,130.84)  $\lbrace{25.0,5.79,3.66 \rbrace}$ MeV in the respective order of models for the $m_{\sigma}=500$ MeV.~The largest first order regions are found in the Columbia plot of the RQM-S model when the $m_{\sigma}=400$ MeV.~The RQM-S and RQM-I model first order regions,~get only slightly reduced when the $m_{\sigma}=400\rightarrow500$ MeV.~Even though the $m_{\sigma}=400\text{ and } 500 $ MeV case first order regions in the $m_{\pi}-m_{K}$ and $\mu-m_{K}$ planes  of the RQM-S model are slightly larger than those of the RQM-I model,~the first order regions in the $m_{ud}-m_{s}$ and $\mu-m_{s}$ quark mass planes of the RQM-I model Columbia plot look equivalent and negligibly larger than those of the RQM-S model because the strange quark mass ($m_{s}^\text{\tiny{Phys}}=99.7$ MeV) for the RQM-I model  at the physical point,~is larger than its RQM-S model value ($m_{s}^\text{\tiny{Phys}}=86.2$ MeV).~When the $m_{\sigma}=600$ MeV,~the critical quantities $(m_{K}^{\tcp},m_{\pi}^{t},m_{\pi}^{c})$ $\lbrace{(m^{\tcp}_{s},m_{ud}^{t}, m_{ud}^{c}) \rbrace}$ computed in the meson $\lbrace{\text{quark}\rbrace}$ mass planes,~are  (218.5,147.66,117.26) $\lbrace{ (19.65, 4.57, 2.90)\rbrace}$ MeV in the RQM-S model and (207.05,146.35,118.01)  $\lbrace{(17.70, 4.59, 2.97)\rbrace}$ MeV in the RQM-I model.~In contrast to a marginal change in first order regions of both the RQM-S and RQM-I models for the $m_{\sigma}=400\rightarrow500$ MeV,~the shrinking of first order regions in the RQM-I model Columbia plots,~in comparison to that of the RQM-S model,~is noticeably larger when the $m_{\sigma}=500\rightarrow600$ MeV.~Even if the physical point strange quark mass $m_{s}^\text{\tiny{Phys}}=99.7$ MeV is larger in the RQM-I model,~its quark mass planes for the $m_{\sigma}=600$ MeV case,~show noticeably smaller first order regions.

The  tricritical lines in the vertical $\mu-m_{K},m_{\pi}=0$ ($\mu-m_{s},m_{ud}=0$) planes of the RQM-S and RQM-I model Columbia plots,~start with a large positive slope from the $m_{K}^{\tcp}$ ($m_{s}^{\tcp}$) at $\mu=0$ and  bend strongly for higher $\mu$.~Turning flat with almost a zero slope at higher $\mu$ in the kaon (s quark) mass range of $m_{K}=500-550$ ($m_{s}=91.5-108$) MeV,~the RQM-S model tricritical lines for all the cases of the $m_{\sigma}=400,500\text{ and }600$,~show proper saturation.~The RQM-I model tricritical line 
shows a saturation pattern close to that of the RQM-S model only when the $m_{\sigma}=400$ MeV.~Since the slope  does not decrease after the $m_{K}>400$ ($m_{s}>69.1$ ) MeV,~the RQM-I model tricritical lines for the $m_{\sigma}=500\text{ and }600$ MeV,~pick up a moderate and significantly large divergent  trend, respectively.~The strongly divergent RQM-I model tricritical line looks very different from the RQM-S model tricritical line for the $m_{\sigma}=600$ MeV.~The RQM-S and RQM-I model $Z(2)$ chiral critical lines are likely to have similar shapes in that $\mu-m_{K}$ plane which corresponds to the $m_{\pi}=138$ MeV since the $\cep$  co-ordinates are very nearly identical in $\mu-T$ planes of the RQM-S and I model phase diagrams computed for the physical point parameters.~Even though  the chiral critical surfaces of the  RQM-S and RQM-I models are similar and convergent on their corresponding chiral critical lines at the physical point,~the RQM-I model critical surface gets lifted higher up in the $\mu$ direction near the ($m_{\pi}=0,m_{K}>400$ MeV),~hence the tricritical line on its other end at the $m_{\pi}=0$ becomes significantly divergent.~The RQM-S model critical surface seems to have  evenly leveled up rise in the  $\mu$ direction for all $m_{\pi}$ as its tricritical line at the $m_{\pi}=0$ shows proper saturation that is consistent and desirable on the  physical grounds also since the 2+1 flavor tricritical line is expected to be connected to  the tricritical point of the two flavor chiral limit \cite{hjss} at higher $\mu$ and $m_{K}$.~Our first goal of comprehensively comparing the RQM-I and RQM-S model Columbia plots has clearly established that the RQM-S model with the large $N_{c}$ standard U(3) ChPT inputs,~constitutes the improved and better framework for the Chiral limit studies.

The RQM-S model critical quantities $(m_{K}^{\tcp},m_{\pi}^{t},m_{\pi}^{c})$ $\lbrace{(m^{\tcp}_{s},m_{ud}^{t}, m_{ud}^{c}) \rbrace}$ in the $(m_{\pi}-m_{K})$  $\lbrace{(m_{ud}-m_{s}) \rbrace}$ plane,~get strongly reduced when the $\sigma$ mass changes from $600\rightarrow750\text{ and }800$ MeV to give significantly smaller values as (159.85,~110.40,~87.90) $\lbrace{ (10.65, 2.57, 1.64)\rbrace}$ MeV for the $m_{\sigma}=750$ MeV and (148.2,~103.19,~82.07) $\lbrace{ (9.21, 2.25, 1.43)\rbrace}$ MeV for the $m_{\sigma}=800$ MeV.~The relative position of the RQM-S model $\tcp$ at $m_{K}^{\tcp}/m_{K}^{\text{\tiny{Phys}}}$ ($m_{s}^{\tcp}/m_{s}^{\text{\tiny{Phys}}}$ ),~with respect to the value of kaon (s quark) mass at the physical point,~shifts from the 0.5016 to 0.4405 (0.2927 to 0.2280) when the $m_{\sigma}=400\rightarrow600$ MeV and one finds a significantly large shift in  the $m_{K}^{\tcp}/m_{K}^{\text{\tiny{Phys}}}$ ($m_{s}^{\tcp}/m_{s}^{\text{\tiny{Phys}}}$ ) from the  0.4405 to 0.2988 (0.2280 to 0.1068 ) when the $m_{\sigma}=600\rightarrow800$ MeV.~In a similar fashion the $m_{\pi}^{c}(m_{ud}^{c})$ changes from 134.51 (3.8) to 117.26 (2.9) MeV when the $m_{\sigma}=400\rightarrow600$ whereas the change in the $m_{\pi}^{c}(m_{ud}^{c})$ from 117.26 (2.9) to 82.07 (1.43) MeV is significantly large when the $m_{\sigma}=600\rightarrow800$ MeV.~Note that the very recent study in the Ref.~\cite{Tomiya} in the SICJT formalism~\cite{Sicjt} in the LSM,~finds a stable first-order regime with a definite location of the $\tcp$ at $m_{s}^{\tcp}/m_{s}^{\text{\tiny{Phys}}}=0.696$ which shows small variation on increasing the $m_{\sigma / f_{0}(500)}$ from 672.4 MeV to 797.2 MeV whereas the critical pion mass reported in their work is  $m_{\pi}^{c}$=52.4 MeV for the $m_{\sigma / f_{0}(500)}$=672.4 MeV.~The RQM-S model tricritical line for the case of $m_{\sigma}=750$ MeV is very strongly divergent in the $\mu-m_{K}(\mu-m_{s})$  plane as it suffers a small bending since the chemical potential becomes very high  $\mu=386.8$ MeV for quite a small kaon mass $m_{K}=280$ MeV.~Hence the associated chiral critical surface would exist in a very reduced range of $m_{K}^{\tcp}=159.85\text{ to } m_{K}=280$ MeV on the $m_{K}$ axis and $m_{\pi}=0\text{ to } m_{\pi}^{t}=110.4$ MeV on the $m_{\pi}$ axis.~Apart from demonstrating that the RQM-S model critical quantities go through a significantly large relative shifts when the $m_{\sigma}$ becomes very high,~our second goal of computing the Columbia plots for the higher $m_{\sigma}=750$ MeV in the RQM-S model,~has given us the idea of boundaries of a significantly smaller and shrunk shape of chiral critical surface that would be terminating well before the region under which lies the physical point,~hence the associated critical end point $\cep$ would not exist in the $\mu-\text{T}$ plane of the phase diagram.

~It is important to note that even a weak chiral transition strength in the RQM-S model for $m_{\sigma}=750$ MeV gives (159.85,~110.40,~87.90) MeV for the values of critical quantities $(m_{K}^{\tcp},m_{\pi}^{t},m_{\pi}^{c})$ which  are comparable to their values (169,~110,~86) MeV reported in the e-MFA:FRG QM model study~\cite{Resch} where $m_{\sigma}=530$ MeV and the spreads of first order regions in the horizontal $m_{\pi}-m_{K}$ plane,~look similar in both the studies.~Furthermore,~even the significantly diluted strength of chiral transition,~for a very high $m_{\sigma}=800$ MeV in the RQM-S model,~gives the  critical quantities $(m_{K}^{\tcp},~m_{\pi}^{t},~m_{\pi}^{c}) \equiv (148.2,~103.19,~82.07)$ MeV which turn out to be larger than the critical quantities $(m_{K}^{\tcp},~m_{\pi}^{t},~m_{\pi}^{c}) =(137.2,~98.15,~78.45)$ MeV reported for the case of small  $m_{\sigma}=400$ MeV in the QMVT model Columbia plot in the Ref.~\cite{vkt25II}.~Thus the third goal of the present work gives us the quantitative evidence and strongly confirms the conclusion that the effect of quark one-loop vacuum correction is quite moderate due to on-shell renormalization of the parameters in the RQM-S model whereas the strength of chiral transition gets strongly softened and diluted by a very large smoothing influence of the quark one-loop vacuum fluctuations when its effect gets treated  in the $\overline{\text{MS}}$ scheme of renormalization and the curvature masses of mesons are used to fix the model parameters.~Here,~it is worthwhile to recall that the critical $m_{\pi}^{c}\equiv150$ MeV and the  chiral transition is first order everywhere on the $m_{K}$ axis for $m_{\pi}=0$ when the quark one-loop vacuum fluctuations are neglected  under the s-MFA in the Ref. \cite{Schaefer:09} for the case of $m_{\sigma}=800$ MeV.~We finally conclude by noting that the range $m_{ud}^c=3.8\text{ to } 1.43$ MeV of critical light quark mass values, for different $m_{\sigma}$, found in our present RQM-S model work,~lies well within the range of $m_{ud}^c\le$4 MeV given by  recent  LQCD study using Möbius domain wall fermions in Ref.~\cite{zhang24}.

\ \ \ \ \ \ \ \ \ \ \ \ \ \ {\bf {ACKNOWLEDGMENTS}} \\ \\
I would like to express my sincere gratitude towards Suraj Kumar Rai,
Akanksha Tripathi and Pooja Kumari for the careful
readings of the manuscript. I am deeply indebted and very
much tankful to Swatantra Kumar Tiwari for making the
three dimensional Columbia plots.



\bibliographystyle{apsrmp4-1}

\end{document}